\documentclass[fleqn,usenatbib]{mnras}
\usepackage{newtxtext,newtxmath}
\usepackage[T1]{fontenc}

\DeclareRobustCommand{\VAN}[3]{#2}
\let\VANthebibliography\thebibliography
\def\thebibliography{\DeclareRobustCommand{\VAN}[3]{##3}\VANthebibliography}

\usepackage{graphicx}	% Including figure files
\usepackage{amsmath}	% Advanced maths commands
\newcommand{\Kepler}{{\em Kepler}}
\newcommand{\MESA}{\textsc{mesa}}
\newcommand{\kic}[1]{KIC~#1}
\newcommand{\TheStar}{\kic{9244992}}

\newcommand{\BV}{Brunt--V\"ais\"al\"a}
\newcommand{\rmr}{\mathrm{r}}
\newcommand{\Br}{B_{\rmr}}
\newcommand{\Brup}{\Br^{\mathrm{UL}}}
\newcommand{\Bphi}{B_{\phi}}
\newcommand{\Btheta}{B_{\theta}}

\newcommand{\Mr}{M_{\rmr}}
\newcommand{\Wrr}{W_{\rmr}}
\newcommand{\Kr}{K_{\rmr}}

\newcommand{\mean}[1]{\langle{#1}\rangle}
\newcommand{\RMS}[1]{\mean{#1^2}^{1/2}}
\newcommand{\BrRMS}{\RMS{\Br}}

\newcommand{\BphiRMS}{\RMS{\Bphi}}

\newcommand{\Brmax}{\Br^{\mathrm{max}}}
\newcommand{\Bmax}{B^{\mathrm{max}}}

\newcommand{\xir}[1]{\xi_{\mathrm{r}; #1}}
\newcommand{\xih}[1]{\xi_{\mathrm{h}; #1}}
\newcommand{\xirn}{\xi_{\mathrm{r}}}
\newcommand{\xihn}{\xi_{\mathrm{h}}}
\newcommand{\nurot}{\nu_{\mathrm{rot}}}
\newcommand{\Prot}{P_{\mathrm{rot}}}

\newcommand{\GB}{\text{G}_{B}}
\newcommand{\G}{\text{G}}

\newcommand{\Ylm}{Y_{\ell}^{m}}
\newcommand{\nabperp}{\nabla_{\perp}}
\newcommand{\nabYlm}{\nabperp\Ylm}

\newcommand{\xupest}{0.5}
\newcommand{\xupestpercent}{50}

\newcommand{\Brest}{3.5\pm 0.1}
\newcommand{\Bphiest}{92 \pm 7}

\newcommand{\Brestone}{4.0\pm 0.1}
\newcommand{\Bphiestone}{175\pm 13}

\newcommand{\nlow}{{-38}}
\newcommand{\nhigh}{{-20}}

\newcommand{\xup}{x_{\text{up}}}
\newcommand{\rup}{r_{\text{up}}}

\newcommand{\rin}{r_{\text{in}}}

\newcommand{\rd}{\text{d}}
\newcommand{\ra}{\text{a}}
\newcommand{\rb}{\text{b}}
\newcommand{\dV}{\rd V}
\newcommand{\dS}{\rd S}
\newcommand{\dr}{\rd r}
\newcommand{\dK}{\rd\mathcal{K}}
\newcommand{\dOmega}{\rd\mathit{\Omega}}
\newcommand{\Vex}{V_{\mathrm{ex}}}

\newcommand{\gla}{{\mathscr{A}}}
\newcommand{\glb}{{\mathscr{K}}}
\newcommand{\glc}{{\psi}}

\newcommand{\Neff}{N_{\mathrm{eff}}}

\newcommand{\bOmega}{\boldsymbol{\Omega}}
\newcommand{\br}{\boldsymbol{r}}
\newcommand{\bB}{\boldsymbol{B}}
\newcommand{\bBex}{\boldsymbol{B}_{\mathrm{ex}}}

\newcommand{\bj}{\boldsymbol{j}}
\newcommand{\bn}{\boldsymbol{n}}

\newcommand{\hbxi}{\hat{\boldsymbol{\xi}}}
\newcommand{\bxia}{\boldsymbol{\xi}\vphantom{\hbxi^*}}

\newcommand{\bxin}{{\boldsymbol{\xi}}}

\newcommand{\be}{\boldsymbol{e}}
\newcommand{\ber}{\be_{\text{r}}}
\newcommand{\betheta}{\be_{\theta}}
\newcommand{\bephi}{\be_{\phi}}
\newcommand{\bepsi}{\be_{\Psi}}

\newcommand{\pdr}[1]{\frac{\partial{#1}}{\partial r}}
\newcommand{\pdtheta}[1]{\frac{\partial{#1}}{\partial\theta}}
\newcommand{\pdphi}[1]{\frac{\partial{#1}}{\partial\phi}}

\newcommand{\tdr}[1]{\frac{\rd{#1}}{\rd r}}
\newcommand{\tdpsi}[1]{\frac{\rd{#1}}{\rd\Psi}}

\newcommand{\Dlm}{D_{\ell}^m}
\newcommand{\Flm}{F_{\ell}^m}
\newcommand{\Glm}{G_{\ell}^m}

\newcommand{\cP}{\mathcal{P}}
\newcommand{\oP}[3]{\cP_{#1}^{\left(#2\right)}\left(#3\right)}

\newcommand{\wtj}[6]{\begin{pmatrix}
 {#1} & {#2} & {#3} \\
 {#4} & {#5} & {#6}
\end{pmatrix}}

\newcommand{\kr}{k_{\text{r}}}

\newcommand{\Sr}{S_{\text{r}}}
\newcommand{\Sh}{S_{\text{h}}}

\newcommand{\RMSr}{{\mathsf{B}_{\text{r}}^{\text{min}}}}
\newcommand{\RMSphi}{{\mathsf{B}_{\phi}^{\text{min}}}}

\newcommand{\kc}{\mathfrak{c}}
\newcommand{\kd}{\mathfrak{d}}

\newcommand{\asymm}{\mathfrak{a}}

\newcommand{\lspeed}{c_{*}} % speed of light

\newcommand{\ABV}{{\mathscr{N}}}

\newcommand{\Bwa}{a_*}

\defcitealias{Saio:2015aa}{S15}

\title[{Internal magnetic field of \TheStar}]%
{%
Asteroseismic detection of 
a predominantly toroidal magnetic field
in the deep interior of
the main-sequence F star
\TheStar}

\author[M. Takata et al.]{%
Masao Takata,$^{1}$\thanks{E-mail: takata@astron.s.u-tokyo.ac.jp (MT)}
Simon J. Murphy,$^{2}$\thanks{E-mail: simon.murphy@usq.edu.au (SJM)}
Donald W. Kurtz,$^{3,4}$
Hideyuki Saio$^{5}$
and
Hiromoto Shibahashi$^{1}$
\\
$^{1}$Department of Astronomy, School of Science, The University of Tokyo, 7-3-1 Hongo, Bunkyo-ku, Tokyo 113-0033, Japan\\
$^{2}$Centre for Astrophysics, University of Southern Queensland, Toowoomba, QLD 4350, Australia\\
$^{3}$Centre for Space Research, North-West University, Dr Albert Luthuli Drive, Mahikeng 2735, South Africa\\
$^{4}$Jeremiah Horrocks Institute, University of Lancashire, Preston, PR2\,2HE, UK\\
$^{5}$Astronomical Institute, Graduate School of Science, Tohoku University, Sendai, Miyagi 980-8578, Japan\\
}

\date{Accepted 2025 November 28. Received 2025 October 17; in original form 2025 August 15}

\pubyear{2025}

\begin{document}
\label{firstpage}
\pagerange{\pageref{firstpage}--\pageref{lastpage}}
\maketitle

\begin{abstract}
An
asteroseismic analysis
has revealed
%a possible signature of
a magnetic field
in the deep interior
of a slowly-rotating main-sequence F star
\TheStar,
which was observed by
the \Kepler\ spacecraft
for four years.
The star shows
clear asymmetry of
frequency splittings
of high-order dipolar gravity modes,
which
cannot be explained by rotation {alone},
but are fully consistent with a  model with
rotation, a magnetic field and a discontinuous structure (glitch).
Careful examination
of the frequency dependence of the asymmetry
allows us to put constraints on
not only the radial component of the magnetic field,
but also its azimuthal (toroidal) component.
The lower bounds of
the root-mean-squares 
of the radial and azimuthal
components
in the radiative region within \xupestpercent\ per cent in radius,
which have the highest sensitivity in the
layers
just outside the convective core
with a steep gradient
of chemical compositions,
are estimated to be
$\RMSr=\Brest\,\text{kG}$
and
$\RMSphi = \Bphiest\,\text{kG}$, respectively.
%
%while
%these estimates change to
%\Brestone\ and \Bphiestone, respectively,
%if the field is confined in
%the layers of the steep gradient 
%of chemical composition.
%
%We also find that
%the radial component is
%stronger on average near the equator
%than in the polar region,
%and that
%Although
%the dominant azimuthal component over the radial component
%is qualitatively consistent with
%the ${\Omega}$ effect of the dynamo process,
%
The much stronger azimuthal component than the radial one is consistent with
the significant contribution of the differential rotation %(${\Omega}$ effect)
although the star has almost uniform rotation at present.
%
%Possible origins of the detected field include dynamo processes in the convective core or in the radiative envelope.
The estimated field strengths are too strong
to be explained by dynamo mechanisms
in the radiative zone associated with
the magnetic Tayler instability.
%%the Tayler--Spruit dynamo.
%
The aspherical glitch is found to be located in the innermost radiative layers where there is a steep gradient of chemical composition.
%
%If we take
%the evolutionary effect
%into account,
%the field strength
%is 
%more or less
%consistent with
%those in red giants,
%which have been reported recently.
%
%This %asteroseismic 
%detection 
%of the interior field
%follows
%that in three red giant stars,
%which has recently been reported.
%
%The present result 
The first detection of magnetic fields
in the deep interior of a main-sequence star
%based on its oscillation frequencies
sheds new light
on the problem of stellar magnetism,
for which there remain many uncertainties.
%poorly been understood.
%
% a single paragraph not more than 250 words (200 words for Letters)
\end{abstract}


%%%%%%%%%%%%%%%%%%%%%%%%%%%%%%%%%%%%%%%%%%%%%%%%%%

%%%%%%%%%%%%%%%%% BODY OF PAPER %%%%%%%%%%%%%%%%%%

\section{Introduction}

%The problem of stellar magnetism
%Hale, Babcock
%Sun, Ap stars
%Mestel

The study of magnetic fields in stars
has been a major subject in astrophysics
since \citet{Hale:1908tu} first detected them in sunspots.
The Sun has not only strong magnetic fields 
of the order of $10^3$ G
confined
in sunspots but also a global-scale weak field
of the order of $1$ G,
which extends from the photosphere to the corona.
The number of sunspots 
\citep{Schwabe:1844aa}
and their latitudinal position
\citep[Sp\"orer's law; ][]{1863oss..book.....C}
both change over a period of about eleven years,
which actually corresponds to a half of the magnetic cycle of about twenty-two years \citep{Hale:1919aa}.
Understanding the mechanism
of how the magnetic fields are generated and maintained 
is the central problem of
the activity of the Sun and
low-mass main-sequence stars
\citep[e.g.][]{Hathaway:2015aa}.

On the other hand, 
large-scale nearly dipolar magnetic fields 
with typical strengths of $10^3$ G
are detected at the surface of 
a small fraction of early-type main-sequence stars that
do not have a thick convective envelope
{\citep{1982ApJ...258..639L}}.
In early studies
\citet{1947ApJ...105..105B}
first reported the detection of magnetic fields 
mainly in A and B type stars
with chemical peculiarities,
whereas recent surveys have extended
the detections to O and B type stars
\citep[e.g.][]{Briquet:2015aa,Wade:2016aa,Shultz:2019aa,Petit:2025aa}.
Unlike the Sun, these fields are apparently stable
(at least on human observation time scales),
and
their observed variability is interpreted as rotational modulation
with the magnetic axis inclined to the rotation axis {\citep{stibbs1950}}.
These fields could originate from interstellar fields that
were locked into the stars during the formation process {\citep{1982ARA&A..20..191B}}.
Or they might alternatively come from the
dynamo process in the stars {\citep{stepien2000}}.

We now give an overview of the surface fields of pulsating main-sequence stars with low and intermediate mass.
The first example of magnetic pulsators is the rapidly oscillating Ap (roAp) stars \citep{1982MNRAS.200..807K, Kurtz:1990aa},
which show high-order acoustic modes with a typical periods around 10\,min under significant influence of the surface field of the order of $10^3\,\mathrm{G}$.
Although the roAp stars are found in the same region in the HR diagram as $\delta$ Sct stars, which show only low to intermediate order modes, the large-scale strong surface magnetic field has long been thought to be incompatible with low-order oscillation modes.

This situation changed 
when surface fields were detected in a few $\delta$ Sct stars \citep{Thomson-Paressant:2023aa},
and
when \citet{Murphy:2020aa} discovered the first example of $\delta$ Sct--roAp hybrid stars, \kic{11296437}, with the surface field of $2.8\pm 0.5\,\mathrm{kG}$.
Furthermore, \citeauthor{Murphy:2020aa} argued, based on theoretical analysis,  that strong surface fields could inhibit high-order gravity modes.
In fact,
there has so far been no detection of a large-scale surface magnetic field in $\gamma$ Dor stars and
$\delta$ Sct--$\gamma$ Dor hybrid pulsators
\citep[e.g.][]{Thomson-Paressant:2023aa,Hubrig:2023aa}.
However, the hypothesis of suppression of high-order gravity modes by strong magnetic fields must be examined quantitatively because surface magnetic fields and
high-order gravity modes do coexist in 
slowly pulsating B (SPB) and
SPB--$\beta$ Cep hybrid stars
\citep[e.g.][]{Neiner:2003aa,Briquet:2013aa},
which are also main-sequence stars, but with
higher masses
($3$--$9\,\mathrm{M}_{\sun}$)
than $\gamma$ Dor stars.
In this context, we note that
the low-frequency peak found in the spectrum of some $\gamma$ Dor stars may be caused by surface spots, which are associated with small-scale surface magnetic activity
\citep{Henriksen:2023aa,Henriksen:2023ab,Antoci:2025aa}.
In summary, thanks to improvements in observational techniques, the number of confirmed main-sequence magnetic pulsators is steadily increasing.  This naturally leads to a stronger motivation to understand these stars also from a theoretical point of view.

The surface magnetic fields are detected not only in
main-sequence stars,
but also in stars in the 
initial (pre-main-sequence) stage of their lives
(T Tauri stars)
and
those in the final stage
(white dwarfs and neutron stars)
\citep[e.g.][]{Bagnulo:2021aa}.
The magnetic fields thus play significant roles 
throughout the entire life of stars
affecting many physical processes,
including
star formation, 
rotation, mass accretion, flares and winds
\citep[e.g.][]{mestel2012stellar}.

While surface magnetic fields are measured
by the Zeeman effect of spectral lines,
asteroseismology provides a unique method
to constrain the internal magnetic fields
through their effect on stellar oscillations.
Among others,
\citet{Li:2022aa}
have carefully examined the oscillation frequencies
of three red-giant stars,
which were observed by the {\em Kepler} spacecraft,
to deduce the detection
of fields of $30 - 100$\,kG in the core.
This analysis has been extended to
a larger sample of stars
by
\citet{Deheuvels:2023aa},
\citet{Li:2023aa}
and
\citet{2024MNRAS.534.1060H}.

Turning to main-sequence stars,
\citet{2022MNRAS.512L..16L}
estimated an upper limit of
the magnetic-field strength
in the near-core region of
the main-sequence B star HD 43317
to be
of order $500$\,kG.
%assuming that the field has a dipolar configuration.
This was based on the picture that
the observed suppression of gravity modes
in the low-frequency range
is due to the conversion of constituent waves
from the internal gravity waves
to resonant Alfv\'en waves
as a result of significant interaction
with the magnetic field
in the layer of the steep gradient of chemical composition
just outside the convective core
\citep{2017MNRAS.466.2181L}.
The essential part of this picture was originally presented by
\citet{Fuller:2015aa}
to explain unusually low amplitudes of dipolar modes
in a fraction of red-giant stars observed by the Kepler space telescope \citep{Mosser:2012aa,Garcia:2014aa}.
However,
\citet{Mosser:2017aa}
{contradicted this idea by demonstrating}
that
the low-amplitude dipolar modes of red giants are {formed by the coupling between core and envelope oscillations, which implies that the waves transmitted from the envelope to the core come back to the envelope (at least partially).}
{This is} not expected by the mechanism of
\citet{Fuller:2015aa},
at least in its original form.
%More studies are clearly needed 
Continuing efforts have been made
to understand
the mechanism of mode suppression and its relation
to the magnetic field
from both observational aspects
\cite[e.g][]{Stello:2016aa,2024A&A...690A.324C}
and theoretical ones
\citep[e.g.][]{2020MNRAS.493.5726L,2023MNRAS.523..582R}.

% \citet{Burssens:2023aa}

Following recent work on red-giant stars,
we report in this paper
the detection of a magnetic field 
in the deep interior of
a main-sequence F star, \TheStar,
which can be classified as a $\delta$\,Sct--$\gamma$\,Dor hybrid pulsator \citep[][{hereafter \citetalias{Saio:2015aa}}]{Saio:2015aa}.
The structure of this paper is as follows:
the main analysis is presented in Section~\ref{sec:analysis} with the details given in Appendices;
Section~\ref{sec:discussion} is devoted to discussions;
we finally give conclusions in Section~\ref{sec:conclusions}.

\section{Asymmetry of frequency splittings}
\label{sec:analysis}

\subsection{Target}
\label{subsec:target}

\TheStar\ has
a \Kepler\ magnitude of $\mathrm{Kp} = 14$
\citep{Brown:2011ut}
and
a spectral type of F0
\citep{Nemec:2017wm}.
There is no observational evidence
that the star belongs to
a binary (or multiple) system
\citep{Murphy:2018aa}.
The properties of the star are summarised in
Table \ref{tab:stellar_params}.

\citetalias{Saio:2015aa}
analysed the \Kepler\ long cadence data
of the star
in quarters 1 to 17
to find
rich frequency spectra of 
both of gravity and acoustic modes.
Using the \MESA\ stellar evolution
code
\citep{2011ApJS..192....3P,2013ApJS..208....4P,2015ApJS..220...15P,2018ApJS..234...34P,2019ApJS..243...10P,Jermyn:2023aa},
they then constructed
evolutionary models,
which reproduce
the observed properties well.
The best model has
a mass of $1.45\,\mathrm{M}_{\sun}$,
an
initial metal abundance of
$Z_0=0.01$
and
an age of $1.9\,\mathrm{Gyr}$,
which implies
the late phase
of the main-sequence stage
(see Table~\ref{tab:models}).
From
the observed 
frequency splittings,
they estimated 
the rotation periods of 
the core and the envelope
to be
$63.51\pm 0.28\,\mathrm{d}$
and
$66.18\pm 0.58\,\mathrm{d}$,
respectively.

The star can be classified as
a $\delta$ Sct--$\gamma$ Dor hybrid pulsator,
though its rotation period
is much longer than the typical value
of $\sim 1$ d
of $\gamma$ Dor stars
\citep{Li:2020aa}. {Most stars rotating this slowly with this $T_{\mathrm{eff}}$, $\log g$, and age show Am chemical peculiarities, or Ap peculiarities if there is a surface magnetic field. A high-resolution spectroscopic study of \TheStar\ would be useful to examine its surface abundances.}

\begin{table}
    \caption{Parameters for \TheStar\ from
    {\citet{Brown:2011ut} (B11),}
    \citet{Huber:2014wh} (H14)
    and \citet{Nemec:2017wm} (N17).}
    \centering
    \begin{tabular}{lcc}
    \hline
    Parameter & Value & Reference\\
    \hline
    {Kepler magnitude (mag)} & {$13.998$} & {B11}
    \\
%    $\log T_{\mathrm{eff}}$
%         &
%    $3.839 \pm 0.018$ &
%    \\
%    & &
%    $3.845 \pm ?$ &
%    \\
    $T_{\mathrm{eff}}$ (K) &
    $6900 \pm 292$ & H14
    \\ & 
    $7550 \pm 100$ & N17
    \\
    $\log g$ (cgs) &
    $3.52 \pm 0.40$ & H14
    \\
    & 
    $3.52 \pm 0.15$ & N17
    \\
    $v \sin i$ (km\,${\mathrm{s}}^{-1}$) &
    $< 6 \pm 1$ & N17
    \\
    $[\mathrm{Fe/H}]$ & 
    $-0.15 \pm 0.30$ & H14
    \\
    & 
    $+0.1\pm 0.3$ & N17
    \\
%    $\log L/\mathrm{L}_{\sun}$ & &
%    \\
    \hline
    \end{tabular}
    \label{tab:stellar_params}
\end{table}

\subsection{Measurement}

While the internal rotation
can be inferred from
the difference between
the lowest frequency and highest frequency
of each g-mode triplet
\citep[e.g.][]{Aerts:2010aa},
we may study magnetic fields and 
the second-order effect of rotation
based on
the asymmetry of triplets, which is defined by
\begin{equation}
\asymm_{n} \equiv
\frac{\nu_{n,1,1} + \nu_{n,1,-1}}{2}
- \nu_{n,1,0}
\;.
\label{eq:t_n_obs}
\end{equation}
Here,
$\nu_{n,\ell,m}$ generally represent 
mode frequencies
with radial order $n$,
spherical degree $\ell$
and azimuthal order $m$.
%It is clear from the definition
%that $\asymm_{n}$ measures
%the degree of non-constant frequency spacing of %each triplet.

\begin{figure}
    \centering
    \includegraphics[width=\columnwidth]{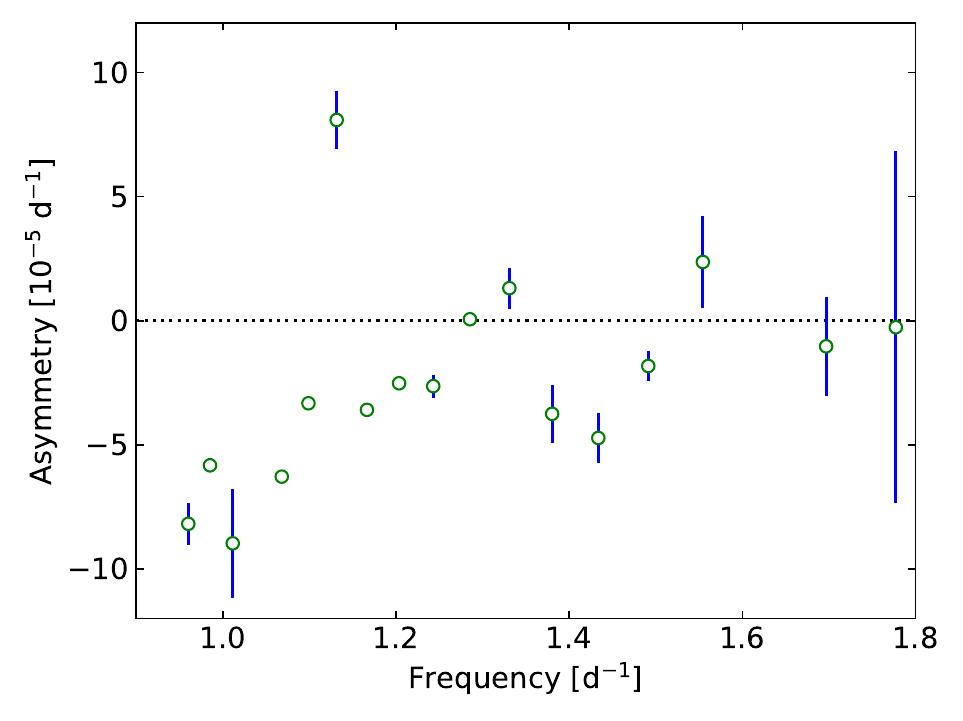}
    \caption{Asymmetry of
    frequency splittings
    $\asymm_n$
    (see equation \ref{eq:t_n_obs})
    for
    high-order gravity modes in \TheStar.
    Errors are smaller
    than the symbol size
    for the points without
    error bars.}
    \label{fig:asymmetry}
\end{figure}

In Fig.~\ref{fig:asymmetry}, we plot $\asymm_n$ as a function
of $\nu_{n,1,0}$
for all of the seventeen gravity-mode triplets 
between $0.9\,\mathrm{d}^{-1}$ and $1.8\,\mathrm{d}^{-1}$,
whose frequencies are
listed in Table 1
of \citetalias{Saio:2015aa}.
The first point we should note is that
all $\asymm_n$ 
except those at $1.29\,\mathrm{d}^{-1}$,
$1.70\,\mathrm{d}^{-1}$
and
$1.78\,\mathrm{d}^{-1}$
are statistically significantly different from zero,
although all $\left|\asymm_n\right|$ are smaller than
the frequency resolution,
\begin{align}
f_{\mathrm{res}}
& =
\frac{1}{4T_0}
=
1.7 \times 10^{-4}\,\mathrm{d}^{-1}
\label{eq:freq_res}
\end{align}
\citep{Kallinger:2008aa},
where $T_0 = 1459\,\mathrm{d}$ is the total observation time-span of the \Kepler\ primary mission Q1--17 data for \TheStar.
{The pulsation frequencies can be determined to higher precision than the frequency resolution, so long as there are no undetected, unresolved frequencies within the spectral window of the  mode frequency peaks. The consistency of our results for $\asymm_n$ indicates that this is the case.} 
Given that, we list the following properties of $\asymm_n$:
(1) a negative and decreasing trend below $1.25\,\mathrm{d}^{-1}$;
(2) 
{pseudo-sinusoidal behaviour, particularly for $\nu \gtrsim 1.3\,\mathrm{d}^{-1}$, with a wavelength of several data points};
(3) an outlier at $1.13\,\mathrm{d}^{-1}$.
The main question of this paper is
how we can understand these.

\subsection{Theory}
\label{sec:theory}

In the present analysis,
we assume
{that asymmetry  of the frequency splitting arises from the three effects,}
\begin{equation}
    \asymm_n = 
    \asymm_n^{(\mathrm{rot})}
    + \asymm_n^{(\mathrm{mag})}
    + \asymm_n^{(\mathrm{glitch})}
    \;,
\label{eq:tn_model}
\end{equation}
in which
$\asymm_n^{(\mathrm{rot})}$,
$\asymm_n^{(\mathrm{mag})}$
and
$\asymm_n^{(\mathrm{glitch})}$
indicate
the effects of rotation, a magnetic field
and a discontinuous structure (glitch),
respectively.
There are three remarks about equation (\ref{eq:tn_model}):
First, we assume that the first-order effect of rotation
is much larger than that of the magnetic field.
In fact, Table 1 of \citetalias{Saio:2015aa} shows that
the frequency splittings are all equal to about $8\times 10^{-3}\,\mathrm{d}^{-1}$,
which is larger by two orders of magnitude than the observed asymmetry in Fig.~\ref{fig:asymmetry}.
However, since 
{the rotational splitting is symmetric,}
the first-order effect of rotation
cannot influence $\asymm_n$ {and}
we need to analyse its second-order effect.
Second, in contrast, it is sufficient to consider the first-order effects for the magnetic field and the glitch.
Finally, we assume that
the deformation of the equilibrium structure
caused by the Lorentz force is negligible compared to that caused by the centrifugal force.
{We will discuss this assumption at the end of Section \ref{sec:mag_strength}.}

Each of the three terms on the right-hand-side of equation (\ref{eq:tn_model})
is discussed separately
in the following subsections.

\subsubsection{Rotation effect}
\label{sec:rotation}

For simplicity, we {restrict} ourselves to the case of
uniform rotation (without the magnetic field), which is a good approximation
for \TheStar\ 
(see Section~\ref{subsec:target}).
From a physical point of view,
the second-order effect of rotation is composed of two sources (in the Eulerian picture),
the second-order effect of the Coriolis force,
which directly affects the oscillations,
and the deformation of the equilibrium structure
due to the centrifugal force.

Although we need to rely on sophisticated methods
and stellar models
to quantify the corresponding
$\asymm_n^{(\mathrm{rot})}$ accurately,
it would be worth providing a model-independent
analytical formula for the order-of-magnitude estimates.
For high-order gravity modes, the Coriolis force becomes more important than the centrifugal force, because the former effect is inversely proportional to the frequency (the ratio between the Coriolis force and the acceleration of the oscillation).
Using equation (117) of \citet{Dziembowski:1992aa}
\citep[see also][]{Brassard:1989aa},
we estimate for dipolar modes ($\ell=1$),
\begin{equation}
     \asymm_n^{(\mathrm{rot, asymp})}
     \equiv
    \frac{\nurot^2}{40 \nu_{n,1}}
    \quad\mathrm{as}\quad
    \nu_{n,1} \rightarrow 0
    \;,
    \label{eq:t_n_C1_asymp}
\end{equation}
in which $\nurot$ means the (cyclic) rotation frequency.
Equation (\ref{eq:t_n_C1_asymp}) can be rewritten as
\begin{equation}
\asymm_n^{(\mathrm{rot, asymp})} = 6\times 10^{-6}
\left(\frac{\Prot}{64\,\mathrm{d}}\right)^{-2}
\left(\frac{\nu_{n,1}}{1\,\mathrm{d}^{-1}}\right)^{-1}
\,\mathrm{d}^{-1}
\;,
\label{eq:t_rot_asymp}
\end{equation}
where $\Prot = \nurot^{-1}$ means the rotation period.
The rotation effect is opposite in sign to
and smaller in amplitude
by an order {of magnitude} than
the observed asymmetry
in Fig.~\ref{fig:asymmetry}.
We conclude that there exist
some physical effects other than the rotation
in the star.

\subsubsection{Magnetic effect}
\label{sec:mag_effect}

\begin{figure}
\centering
\includegraphics[width=\columnwidth]{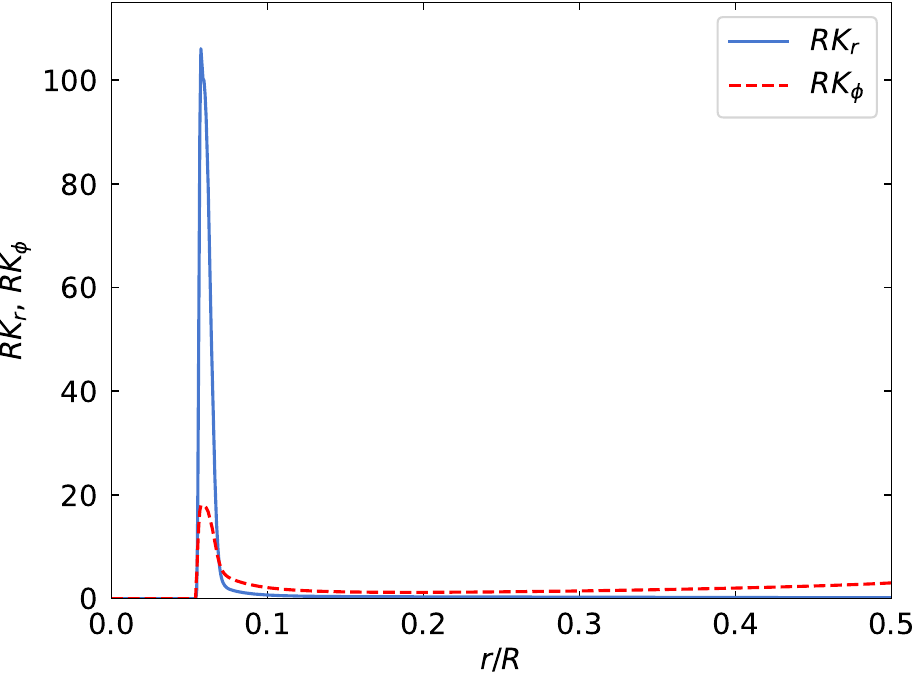}
\caption{Profiles
of $\Kr$ (solid curve) and $K_{\phi}$ (dashed curve)
multiplied by the total radius $R$
for the best evolutionary model 
(model A in Table \ref{tab:models})
with
the upper limit of
$\GB$
set at \xupestpercent\ per cent of
the total radius.
{%
The gradual increase of $R K_{\phi}$ towards larger $r/R$ actually complicates the interpretation,
which is treated in Section \ref{sec:mag_strength} in detail.
}
}
\label{fig:kernels}
\end{figure}

The frequency change
due to a weak magnetic field
is analysed
in the framework of
the regular perturbation theory
\citep[e.g.][]{Gough:1990tj,Shibahashi:1993vs}.
Analyses that
can be adapted to
the case of high-order and
low-degree gravity modes
have been developed in some recent papers
\citep[e.g.][]{Mathis:2021wn,Li:2022aa}.
We extend these previous works
under the assumption of
$|\Bphi| \gg |\Br|,\, |\Btheta|$,
where
$\Br$, $\Btheta$ and $\Bphi$
represent
$r$, $\theta$ and $\phi$ components
of the magnetic field
in the spherical coordinates
(with the rotation axis in the direction of $\theta = 0$),
respectively. 
The reason for dominant $\Bphi$
is that this component could easily be
increased by rotation
(${\Omega}$ effect). 
In order to define the $\phi$ component, it is necessary to identify a magnetic axis. Although we assume that this magnetic axis is aligned with the rotation axis, this does not necessarily mean that the field is axisymmetric. Namely, each component can generally depend on $\phi$.
{(Still, we may note that 
Gauss's law for magnetism ($\nabla\cdot\bB = 0$) leads to $\left|\partial \Bphi/\partial\phi\right| \ll \left|\Bphi\right|$, which implies approximate axisymmetry.)}
The motivation of this extension
comes from the fact
that the asymmetry in
Fig.~\ref{fig:asymmetry}
does not perfectly follow
the inverse-cube law of frequency
\citep[e.g.][]{Bugnet:2021aa}.
In fact,
we show in appendix
\ref{sec:mag_pert}
{(equation (\ref{eq:tmag_g}))}
that
the asymmetry of frequency splittings
is given 
%for dipolar modes
as a function
of the unperturbed frequency
$\nu_{n,1}$
by
\begin{equation}
    \asymm_n^{(\mathrm{mag})}
    =
    \frac{a}{\nu_{n,1}^3} 
    +
    \frac{b}{\nu_{n,1}}
    \;,
    \label{eq:t_n_mag}
\end{equation}
in which 
frequency-independent
parameters $a$ and $b$
are given by
\begin{equation}
    a = \Sr \mean{\Wrr \Br^2}
    \label{eq:a}
\end{equation}
and
\begin{equation}
    b = 
    \Sh
%    \left(
%    \mean{W_{\theta} \Btheta^2}
%    +
    \mean{W_{\phi} \Bphi^2}
%    \right)
    \;,
    \label{eq:b}
\end{equation}
respectively.
Here,
$\Sr$ and $\Sh$
represent
the sensitivity to
the equilibrium structure,
defined 
in terms of the density $\rho$
and
the Brunt-V\"ais\"al\"a frequency $N$
by
\begin{equation}
    \Sr \equiv
    \frac{3}{128\pi^5}
    \frac{%
    \int_{\GB} \frac{N^3}{\rho r^3} \mathrm{d}r}%
    {%}
    \int_{\G} \frac{N}{r} \mathrm{d}r}
    \;,
    \label{eq:Sr}
\end{equation}
and
\begin{equation}
    \Sh
    \equiv
    -\frac{9}{32\pi^3}
    \frac{\int_{\GB}\frac{N}{\rho r^3}\mathrm{d}r}{\int_{\G}\frac{N}{r}\mathrm{d}r}
    \;,
    \label{eq:Sh}
\end{equation}
respectively.
The domains of the radial integral 
$\G$
and
$\GB$
respectively
mean
the gravity-mode cavity
and
its subdomain 
where the magnetic field exists.
The gravity-mode cavity 
$\G$
extends
to almost the entire radiative region
in the case of
intermediate-mass main-sequence stars.
The angle brackets in equations (\ref{eq:a}) and (\ref{eq:b})
stand for
the volume average over $\GB$,
which is
defined by
\begin{equation}
\mean{W_{\alpha} B_{\alpha}^2}
    \equiv
    \int_{\GB}
    K_{\alpha}(r) 
    \overline{W_{\alpha} B_{\alpha}^2}
    \;
    \mathrm{d}r
    \quad
    \mbox{for $\alpha=\rmr$ and $\phi$}
    \;,
    \label{eq:W_avg}
\end{equation}
with the spherical average introduced by
\begin{equation}
    \overline{W_{\alpha} B_{\alpha}^2}
    \equiv
    \frac{1}{4\pi}
    \int_{4\pi}
    B_{\alpha}^2\left(r,\theta,\phi\right)
    W_{\alpha}\left(\cos\theta\right)
    \sin\theta \,
    \mathrm{d}\theta\, \mathrm{d}\phi
    \;.
    \label{eq:sph_av}
\end{equation}
The kernels in the radial direction are defined by
\begin{equation}
    \Kr (r)
    \equiv
    \left(
    \int_{\GB}
    \frac{N^3}{\rho r^3}
    \mathrm{d}r
    \right)^{-1}
    \frac{N^3}{\rho r^3} 
    \label{eq:K_r}
\end{equation}
and
\begin{equation}
%    K_{\theta}(r)
%    =
    K_{\phi}(r)
    \equiv
    \left(
    \int_{\GB}
    \frac{N}{\rho r^3}
    \mathrm{d}r
    \right)^{-1}
    \frac{N}{\rho r^3} 
    \;.
    \label{eq:K_phi}
\end{equation}
As examples,
the profiles of
$\Kr$ and $K_{\phi}$
are plotted
in Fig.~\ref{fig:kernels}
for the best model with
the upper limits of $\GB$ set at 50 per cent of the total radius.
The upper limits should be carefully fixed
to be consistent with the assumptions of the analysis
(see Section~\ref{sec:mag_strength}).
While the amplitude of $\Kr$ concentrates on the sharp peak around $r=0.06 R$, which corresponds to the layers of a steep gradient {in} mean molecular weight,
$K_{\phi}$ not only takes
a local maximum at the same position as $\Kr$,
but also has a long tail towards larger $r$.
The difference is because
$\Kr$ depends on {a} higher power of $N$ than $K_{\phi}$.
On the other hand,
the weight functions for the spherical averages are 
introduced by
\begin{equation}
    \Wrr\left(\cos\theta\right)
    \equiv
    P_2\left(\cos\theta\right)
    =
    \frac{3 \cos^2\theta - 1}{2}
    \;,
    \label{eq:P2}
\end{equation}
%\begin{equation}
%    W_{\theta}\left(\cos\theta\right)
%    = \frac{7 \cos^2\theta - 3}{4}
%\end{equation}
and
\begin{equation}
    W_{\phi}\left(\cos\theta\right)
    \equiv \frac{5 \cos^2\theta - 1}{4}
    \;.
    \label{eq:Wphi}
\end{equation}
In equation (\ref{eq:P2}),
function
$P_2$ is the Legendre polynomial of degree two.

We have made no assumption 
about the ratio between the two terms
on the right-hand side 
of equation (\ref{eq:t_n_mag}).
Therefore,
the second term 
can in principle dominate
over the first term,
when
$|\Bphi|$ is much larger than $|\Br|$.
In this case, the asymmetry would be
proportional to the inverse of frequency,
rather than the inverse cube.

The difference 
between the two terms
on the right-hand side 
of equation (\ref{eq:t_n_mag})
can be understood from a physical point of view.
Since the oscillation motion
is predominantly horizontal
for high-order gravity modes,
the magnetic field in the radial direction
would be significantly bent by 
the short-wavelength motions
to generate a strong
restoring Lorentz force.
On the other hand,
the field in the azimuthal direction
is nearly parallel to the motion
for $\ell=1$ and $m=\pm 1$ modes
near the equator, where the oscillation amplitude is largest,
so that there is little restoring force.
For the axisymmetric modes ($m=0$),
the motion is perpendicular to the azimuthal field, which means that we may expect
some Lorentz force to restore the motion.
However, since the wavelength in the horizontal direction is 
on the order of the stellar radius, which is
much larger than that in the radial direction, the size of the restoring force would be much smaller than in the case of the radial field.
Thus, the radial field influences
the high-order and low-degree gravity modes much more significantly than the azimuthal field.
This qualitatively explains 
why we observe
$|\Sr| \gg \nu^2 |\Sh|$ for $\nu=1\,\text{d}^{-1}$
{(see the last two lines of Table~\ref{tab:models})},
which implies that
the first term is much larger
than the second term
if $|\Br| \sim |\Bphi|$.
In addition,
as the radial wavelength gets shorter
for higher-order (smaller-frequency) gravity modes,
the radial field is accordingly
bent with the shorter scale,
which results in a larger 
restoring Lorentz force,
and hence a more significant
impact on the oscillation frequencies.
This is the reason for the strong frequency dependence of the first term.
In contrast,
the effect of the azimuthal field
would depend little on
the order of modes,
because
the bending scale of the field line
in the horizontal direction
is not affected by the radial wavelength.
This implies a weaker frequency dependence
of the second term.
In fact, the inverse dependence on $\nu_{n,1}$
originates from the perturbation to $\nu^2$,
so that we may regard that 
the impact of the azimuthal field
is essentially independent of the frequency
if it is measured by the perturbation to the squared frequency.

\subsubsection{Glitch effect}

{In Fig.~\ref{fig:asymmetry}, we observe that the data points are not distributed randomly, nor do they follow a linear trend, but instead they show correlation, which in the high-frequency range is pseudo-sinusoidal with a wavelength of several points.}
This is a typical signature of
discontinuous structure (glitch) in the star,
which disturbs the wave propagation.
Because $\asymm_{n}$ is not sensitive to any spherically
symmetric structure, which is usually assumed
in the literature about glitches, we formulate a framework
to analyse aspherical glitches
in Appendix \ref{sec:aspherical_glitches}.
The glitches generally induce oscillatory structures in the diagram of period and period difference with
a constant {wavelength}, but varying amplitude.
{Such a component can actually be identified in Fig.~\ref{fig:asymmetry} based on the detailed analysis in Sections~\ref{sec:glitch_type} and \ref{sec:fitting}.}
While the {wavelength} provides us with the information
about the location of each glitch,
the amplitude {modulation} depends on the type of discontinuity.
We demonstrate in Appendix \ref{sec:three_types}
that
the amplitude depends on the period
linearly, constantly or reciprocally,
if
the discontinuity is associated with
the density itself,
its first derivative
or its second derivative,
respectively.
We {will} carefully examine
{the} type of discontinuity during {our} data analysis {(Section~\ref{sec:glitch_type})}.

\subsection{Interpretation}
\label{subsec:interp}

\subsubsection{Evolutionary models}
\label{subsubsec:models}

We use
three evolutionary models
A, B and C
of \TheStar,
which are constructed by
\citetalias{Saio:2015aa},
to interpret the observed
asymmetry of frequency splittings.
The properties of the models
are summarised in
Table~\ref{tab:models}.
Their main differences
are found
in mass $M$,
initial metal abundance $Z_0$
and
parameter $h_{\mathrm{ov}}$
of the convective overshooting.
Model A corresponds to
the best model, which reproduces
the observed frequencies
most accurately,
while the other two models
also have 
quite close frequencies
\citepalias[see Fig.~11 of][]{Saio:2015aa}.

\begin{table}
    \centering
     \caption{Properties of
    evolutionary models
    by \citetalias{Saio:2015aa}.
    Symbols
    $X_{\mathrm{c}}$,
    $h_{\mathrm{ov}}$
    and
    $\Pi_1$
    stand
    for the central hydrogen
    abundance,
    the parameter
    of overshooting
    from the convective core
    and
    the period spacing of dipolar gravity modes,
    respectively,
    while $\Sr$ and $\Sh$ are defined by
    equations (\ref{eq:Sr})
    and (\ref{eq:Sh}), respectively.
    The upper limit of $\GB$ (the magnetic region in the gravity-mode cavity) is
    expressed by $\xup$,
    which is measured in units
    of the fractional radius.
    The other symbols
    have their usual meanings.}
    \label{tab:models}
    \begin{tabular}{lccc}
    \hline
    Model & A${}^*$ & B & C\\
    \hline
    $M/\mathrm{M}_{\sun}$
      & $1.45$
      & $1.50$
      & $1.54$
    \\
    $T_{\mathrm{eff}}$ (K)
      & $6625$
      & $6748$
      & $7221$
    \\
    $\log L/\mathrm{L}_{\sun}$
      & $0.854$
      & $0.907$
      & $1.050$
    \\
    $\log R/\mathrm{R}_{\sun}$
      & $0.309$
      & $0.319$
      & $0.331$
    \\
    $\log g$ (cgs)
      & $3.982$
      & $3.977$
      & $3.962$
    \\
    Age ($\mathrm{Gyr}$)
      & $1.9$
      & $1.7$
      & $1.4$
    \\
    $X_{\mathrm{c}}$
      & $0.149$
      & $0.142$
      & $0.111$
    \\
    $X_0$
      & $0.724$
      & $0.724$
      & $0.727$
    \\
    $Y_0$
      & $0.266$
      & $0.266$
      & $0.266$
    \\
    $Z_0$
      & $0.010$
      & $0.010$
      & $0.007$
    \\
    $h_{\mathrm{ov}}$
      & $0.005$
      & $0.000$
      & $0.005$
    \\
    $\Pi_1$ (s)
      & 2349
      & 2335
      & 2306
    \\
    \hline
    for $\xup = \xupest$
    \\
        $\Sr$
    $(10^{-31}\,\mathrm{cm}\,
    \mathrm{g}^{-1}\,
    \mathrm{s}^{-2})$
        & $3.0$
        & $2.8$
        & $2.9$
    \\
    $\Sh$
    $(10^{-24}\,\mathrm{cm}\,
    \mathrm{g}^{-1})$
    & $-2.5$
    & $-2.4$
    & $-2.4$
    \\
    \hline
    \end{tabular}
    \\
    {\it Note:} A${}^*$ indicates
    our best model \citepalias[see][]{Saio:2015aa}.
\end{table}

\subsubsection{Rotation effect}
\label{sec:int_rot_effect}

There exist two different methods to analyse
the second-order rotation effect.
One is based on perturbation theory
\citep[e.g.][]{Saio:1981aa,Gough:1990tj,Dziembowski:1992aa},
while the other relies on two-dimensional numerical computation
\citep[e.g.][]{Lee:1995wz}.
We try both methods.

\begin{figure}
	\includegraphics[width=\columnwidth]{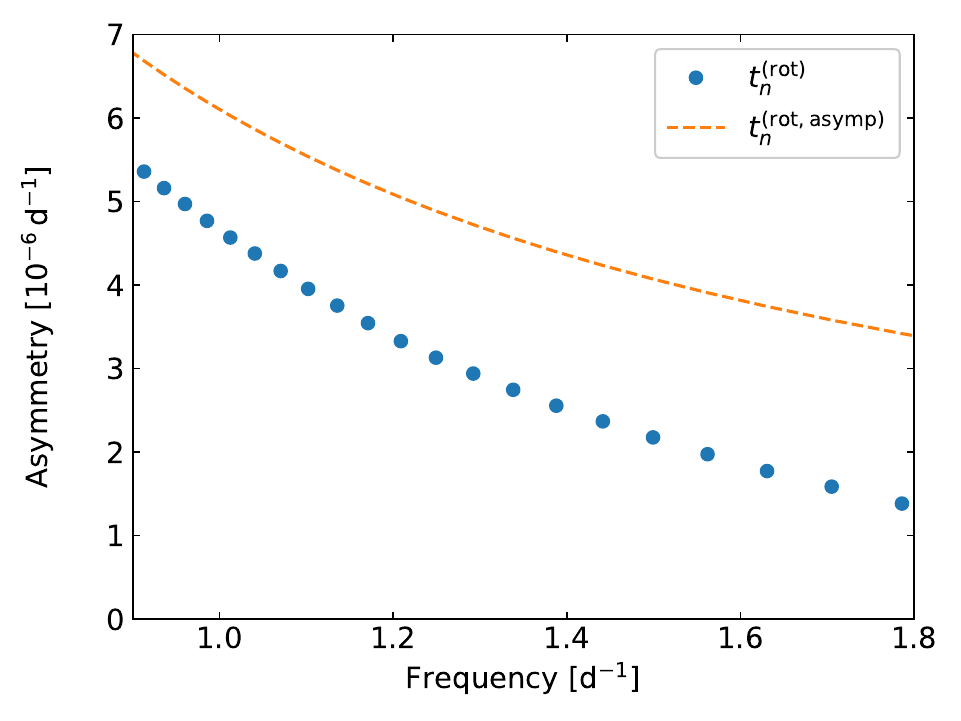}
    \caption{Asymmetry of
    frequency splittings
    caused by the second-order
    effect of rotation
    for the modes of
    our best evolutionary model (model A
    in Table \ref{tab:models}).
    The total effects
    $\asymm_n^{(\mathrm{rot})}$
    and 
    the asymptotic estimates
    $\asymm_n^{(\mathrm{rot,asymp})}$
    (see equation (\ref{eq:t_n_C1_asymp}))
    are shown
    by the filled dots and the dashed curve, respectively.
    }
    \label{fig:rot_asym}
\end{figure}

We first calculate
the effect
for the modes
of the best evolutionary
model
by \citetalias{Saio:2015aa}
(Section~\ref{subsec:target}).
We set the rotation period
to $\Prot = 64$\,d, which corresponds to
$\nurot=
0.0156
\,\mathrm{d}^{-1}$.
Fig.~\ref{fig:rot_asym} shows
$\asymm_n^{(\mathrm{rot})}$
and
$\asymm_n^{(\mathrm{rot,asymp})}$
(equation (\ref{eq:t_n_C1_asymp}))
for the modes in the
observed frequency range
in Fig.~\ref{fig:asymmetry}.
We have checked
that 
the values of
$\asymm_n^{(\mathrm{rot})}$,
which are estimated based on perturbation theory
\citep{Saio:1981aa},
are consistent with
a two-dimensional calculation
by the program
of \cite{Lee:1995wz}
within one per cent.
This reconfirms the conclusion
of Section \ref{sec:rotation}
that
the observed asymmetry cannot be explained
only by the rotation.
Fig.~\ref{fig:rot_asym} also demonstrates that
the asymptotic formula
(equation (\ref{eq:t_n_C1_asymp}))
overestimates
the true values by only a factor of three at most,
which is acceptable for order-of-magnitude estimates.

We also calculate
$\asymm_n^{(\mathrm{rot})}$
for two other evolutionary models
with masses of
$1.50\,\mathrm{M}_{\sun}$
and
$1.54\,\mathrm{M}_{\sun}$,
which are shown in
Fig.~13 of \citetalias{Saio:2015aa},
and confirm that there is no essential difference
from the case of the best model.

\subsubsection{Type of the glitch}
\label{sec:glitch_type}

Given the estimates
for $\asymm_n^{(\mathrm{rot})}$
based on the evolutionary models,
we may interpret the remaining contribution to
the observed asymmetry
$\asymm_n$
as the combined effects of a magnetic field
and a glitch.
While the magnetic effects can be described by
equation (\ref{eq:t_n_mag}),
we first need to {decide} which formula to use for the glitch signature.

\begin{figure}
\centering
    \includegraphics[width=\columnwidth]{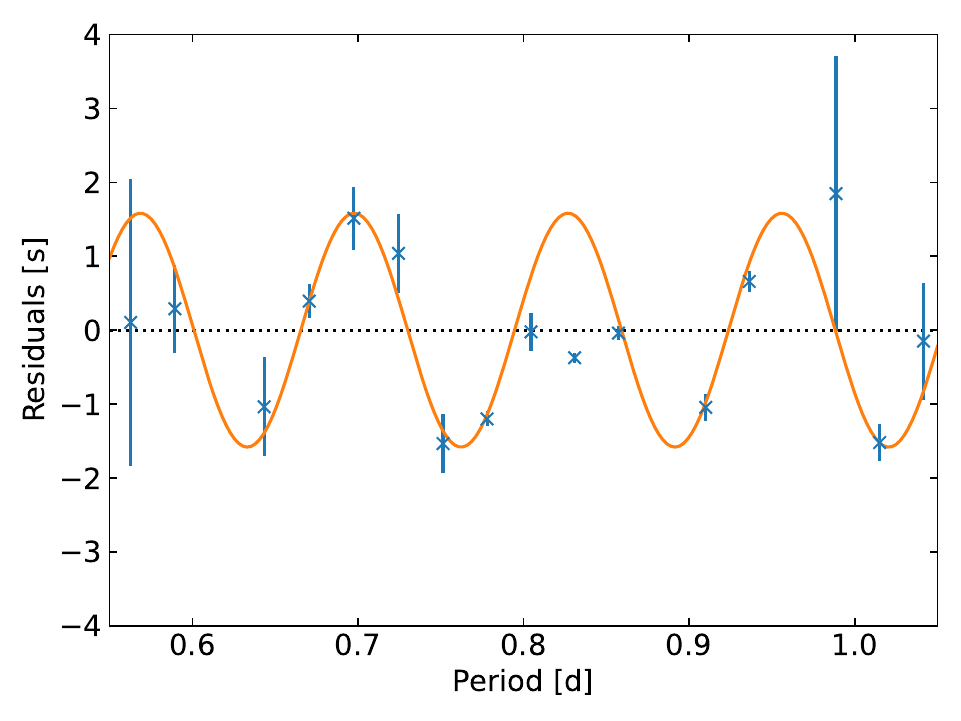}
\caption{Residuals 
of the fitting in period,
$-\nu_{n,1}^{-2}\left(\asymm_n - \asymm_n^{(\mathrm{rot})} - \asymm_n^{(\mathrm{mag})}\right)$,
as a function of the mode period.
The fitting is performed based on equation (\ref{eq:t_n_mag})
for model A (see Table~\ref{tab:fitting})
using only the eight modes with periods longer than $0.8\,\mathrm{d}$.
The residuals are computed for not only those modes
but also the modes with shorter periods.
The solid curve represents
the best fit to the residuals
by a sinusoidal function of constant amplitude,
neglecting the data point
at $0.83\,\mathrm{d}$.
{There is another rejected point at $0.88\,\mathrm{d}$, whose ordinate is outside the plot range. This corresponds to the outlier at $1.13\,\mathrm{d}^{-1}$ in Fig.~\ref{fig:asymmetry}
(see Section~\ref{subsec:outlier}).}
}
\label{fig:residuals}
\end{figure}

For this purpose,
we first fit equation (\ref{eq:t_n_mag}) to $\asymm_n - \asymm_n^{(\mathrm{rot})}$ in the low-frequency range between $0.95\,\mathrm{d}^{-1}$ and $1.25\,\mathrm{d}^{-1}$
(without the outlier at $1.13\,\mathrm{d}^{-1}$), where the {pseudo-sinusoidal} component has only a small amplitude (see Fig.~\ref{fig:asymmetry}).
Then,
using the fitted parameters $a$ and $b$,
we extract the magnetic contribution from $\asymm_n - \asymm_n^{(\mathrm{rot})}$
in the whole range between $0.95\,\mathrm{d}^{-1}$ and $1.8\,\mathrm{d}^{-1}$.
The residuals are shown as a function of period
in Fig.~\ref{fig:residuals}.
As shown by the solid curve,
we confirm that
these residuals can be explained by
a sinusoidal function with a constant amplitude,
except for one data point at $0.83\,\mathrm{d}$.
{The sinusoidal variation}
is expected for a glitch
associated with the discontinuity in the first derivative of density
(see Appendix \ref{sec:three_types}).

After the preliminary steps to determine
the type of the glitch,
global fitting 
in the whole frequency range
should be performed 
using simultaneously
equation (\ref{eq:t_n_mag})
and
\begin{equation}
\asymm_{n}^{(\mathrm{glitch)}}
=
\gla
\nu_{n,1}^2
\sin\left[
2\pi\left(
\frac{\glb}{\nu_{n,1}} - \glc
\right)
\right]
\label{eq:tn_glitch}
\end{equation}
(see equation (\ref{eq:a2k_H})),
in which $\gla$, $\glb$ and $\glc$ are constant parameters
to be fixed.

Equation (\ref{eq:tn_glitch}) suffers from two types of degeneracy.
First, the expression is invariant under the transformation of
\begin{align}
    \left(\gla,\glc\right)
    &\rightarrow
    \left(-\gla,\glc+\frac{1}{2}+k\right)
    \;,
\label{eq:glitch_deg_1}
\end{align}
in which $k$ is an arbitrary integer.
%This trivial degeneracy can be removed by constraining %the range of $\glc$ as, e.g.
%\begin{align}
%0 & \le \glc < \frac{1}{2}
%\;.
%\end{align}
Second,
since the mode periods $P_{n,1}=\nu_{n,1}^{-1}$ 
{of high order gravity modes have an almost constant spacing}, $\Pi_1$,
the sampling theorem tells us that
the frequency $\glb$ cannot be distinguished from 
its mirror image with respect to
the Nyquist frequency, $\left(2\Pi_1\right)^{-1}$. 
This implies {that} the expression
{is invariant}
under the transformation,
\begin{align}
\left(\glb,\glc\right)
&\rightarrow
\left(
\Pi_{1}^{-1} - \glb,
\varepsilon + \frac{1}{2} - \glc
\right)
\;,
\label{eq:mirror_sym}
\end{align}
if the mode periods follow
\begin{align}
P_{n,1} & = \left(n + \varepsilon \right) \Pi_1
\;,
\end{align}
in which $\varepsilon$ is a constant that corresponds to the total phase offset
introduced at the inner and outer turning points of the gravity-mode cavity. 
We do not need to consider the other Nyquist aliases
because $0 < \glb < \Pi_1^{-1}$, which will become clear later from equation (\ref{eq:glitch_pos_relation}).
This type of degeneracy 
{is identical to}
the core/envelope mirror symmetry 
discussed by \citet{Montgomery:2003aa}
(see Section~\ref{sec:glitch_prop}).
Strictly speaking,
this degeneracy is approximate because the period spacing is not exactly constant in reality. 
It is still possible that the two sets of parameters
$(\glb,\glc)$, which are approximately related to each other by equation (\ref{eq:mirror_sym}),
give equally good fits to the data.

\subsubsection{Combined effects of the magnetic field and the glitch}
\label{sec:fitting}

We fit in total five parameters,
two ($a$ and $b$) in equation (\ref{eq:t_n_mag})
and
three ($\gla$, $\glb$ and $\glc$) in equation (\ref{eq:tn_glitch}),
to the difference
$\asymm_n - \asymm_n^{(\mathrm{rot})}$.
%
%which is computed based on
%the mode identification
%in Table 7 of \citetalias{Saio:2015aa}.
%
We exclude from the fitting
the two data points
%with $\asymm_n=8.1\times 10^{-5}\,\mathrm{d}^{-1}$
at $\nu_{n,1,0}=1.13\,\mathrm{d}^{-1}$
and $1.20\,\mathrm{d}^{-1}$.
The former clearly follows a different trend
from the others,
while the latter cannot be explained by
the assumed form of the glitch signature
in equation (\ref{eq:tn_glitch}).
In fact,
inclusion of these points makes the fitting much worse.
Possible origins
for the two points
are discussed in
Section~\ref{subsec:outlier}.
We use
the {\tt 
curve\_fit} function
in the {\tt SciPy} library
of {\tt Python}
to fit the remaining
fifteen data points.
We adopt as a set of initial guesses
of the five parameters
the values obtained during
the preliminary steps
to determine the type of the glitch
in Section \ref{sec:glitch_type}.
We make another set of initial guesses
using equation (\ref{eq:mirror_sym}).
The results originating from the first and second sets
are referred to as cases 1 and 2, respectively.

\begin{figure*}
    \centering
    \begin{tabular}{@{}cc@{}}
    case 1 & case 2\\
    \includegraphics[width=\columnwidth]{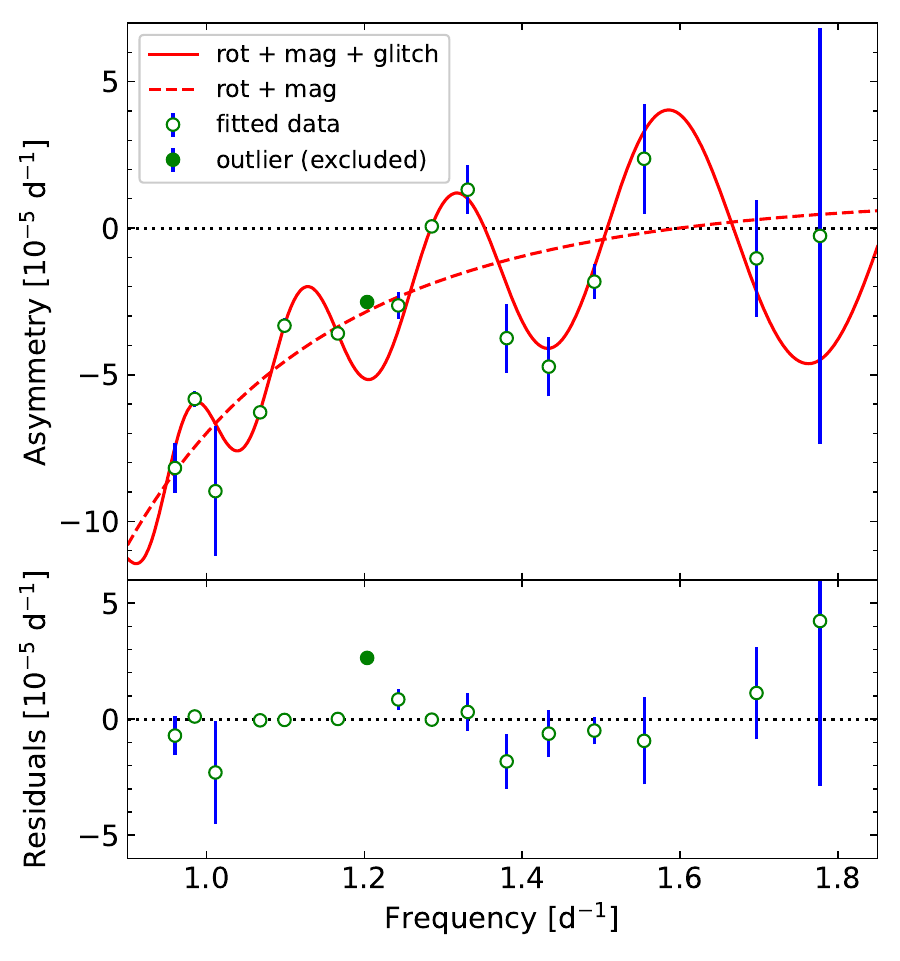}
    &
    \includegraphics[width=\columnwidth]{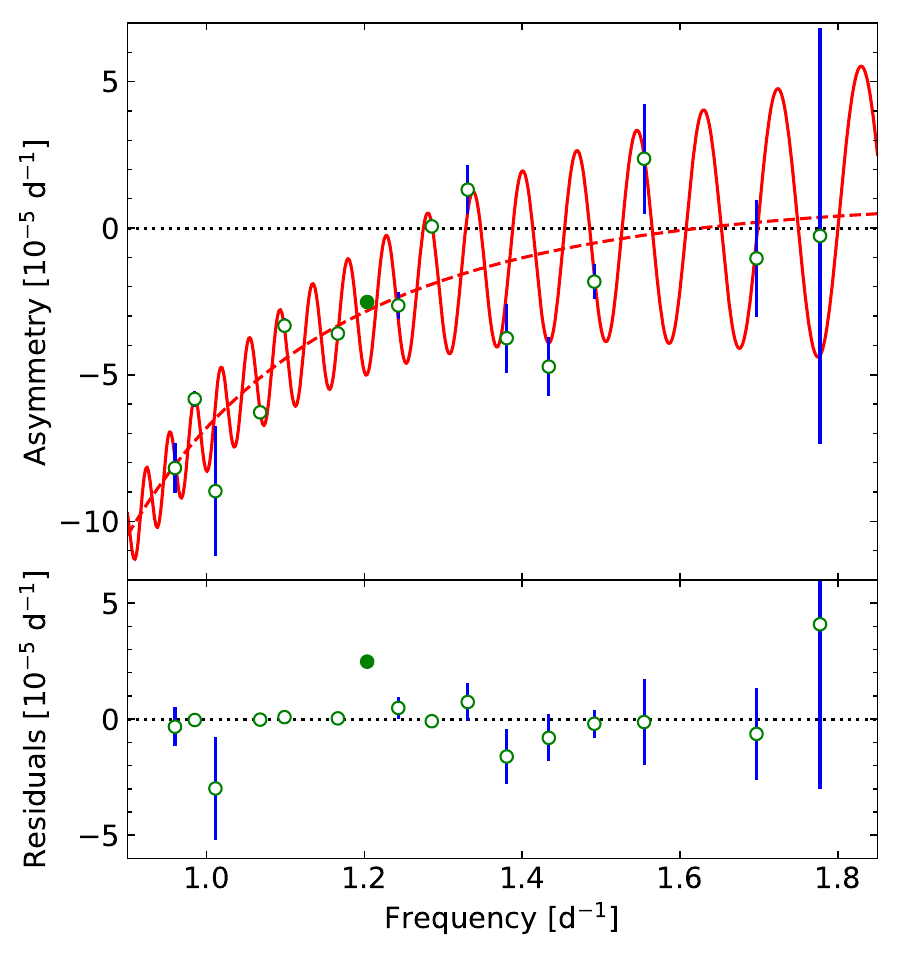}
    \end{tabular}
    \caption{%
    Asymmetry of
    frequency splittings $\asymm_n$
    of \TheStar\
    fitted with
    the model that takes account of
    rotation, a magnetic field and a glitch
    (upper part of each panel)
    and the residuals (lower part).
    The results for the two cases, 1 and 2
    (see Table~\ref{tab:best_fit_params}), 
    with different positions of the glitch
    (see Fig.~\ref{fig:N_XH}),
    are presented in the left and right panels, respectively.
The rotation effect is estimated 
    based on our
    best evolutionary model
    (model A in Table \ref{tab:models}).
{In each panel, there are two data points excluded from the fitting,
one at $1.20\,\mathrm{d}^{-1}$ indicated by the filled circle and the other at $1.13\,\mathrm{d}^{-1}$, whose ordinate is outside the plot range (Fig.~\ref{fig:asymmetry}).
See Section~\ref{subsec:outlier} for the possible origins of these points.
}
    }
    \label{fig:fitting}
\end{figure*}

The results of the fitting are presented in
Table~\ref{tab:best_fit_params} for model A (see Table~\ref{tab:models}).
We find that the fitting is good in both cases 1 and 2
since $\chi^2/\mbox{df}$ is close to one,
although the value in case 2 is slightly smaller.
The fitted values of $\asymm_n$
and the residuals
are shown
in Fig.~\ref{fig:fitting}.

Since $\asymm_n^{(\mathrm{rot})}$ weakly depends
on the evolutionary models,
we repeat the fitting for the two other evolutionary models 
(models B and C in Table~\ref{tab:models})
and find little difference in all of the five parameters.

%
%The chi square
%over the number of
%degrees of freedom
%is given by
%$\chi^2/(8 - 2) = 15.5$
%for all the models.

\begin{table}
\centering
\caption{
The best fit values of the five parameters
in equations (\ref{eq:t_n_mag}) and (\ref{eq:tn_glitch})
to explain the observed asymmetry in the frequency splittings $\asymm_n$ in Fig.~\ref{fig:asymmetry}
after subtracting the contribution of rotation that is estimated based on our best evolutionary model A in Table~\ref{tab:models}.
Two cases, 1 and 2, are considered
because of the degeneracy about the glitch signature described by equation
(\ref{eq:mirror_sym}).
In each case,
the chi-squared per degrees of freedom
($\chi^2/\mbox{df}$) is given
in the last row.
}
\label{tab:best_fit_params}
\begin{tabular}{lcc}
\hline
& case 1 & case 2
\\
\hline
\multicolumn{3}{l}{
parameters of the magnetic effect}
\\
\hline
    $a$ $(10^{-4}\,\mathrm{d}^{-4})$
        & $-1.18\pm 0.08$
        & $-1.13\pm 0.07$
    \\
    $b$ $(10^{-5}\,\mathrm{d}^{-2})$
        & $4.3\pm 0.6$
        & $4.0\pm 0.6$
    \\
\hline
\multicolumn{3}{l}{
parameters of the glitch effect}
\\
\hline
    $\gla$ $(10^{-5}\,\mathrm{d})$
        & $-1.6\pm 0.1$
        & $-1.51\pm 0.09$
    \\
    $\glb$ $(\mathrm{d}^{-1})$
        & $7.7\pm 0.1$
        & $29.9\pm 0.1$
    \\
    $\glc$ 
        & $0.14\pm 0.09$
        & $0.57\pm 0.09$
    \\
\hline
$\chi^2/\mbox{df}$
& $0.997$
& $0.732$
\\
\hline
\end{tabular}
\end{table}

%\subsubsection{Configuration of the field}

We can make a simple remark about the configuration
of the inferred magnetic field.
The negative sign of $a$
in Table~\ref{tab:best_fit_params}
implies that $\left|\Br\right|$ is larger on average near the equatorial region than in the polar region because of equation (\ref{eq:P2}).
Similarly,
the positive sign of $b$
means that
$\left|\Bphi\right|$ is more confined to the equator than the poles
(see equation (\ref{eq:Wphi})).

\subsubsection{Properties of the glitch}
\label{sec:glitch_prop}

\begin{figure}
\centering
\includegraphics[width=\columnwidth]{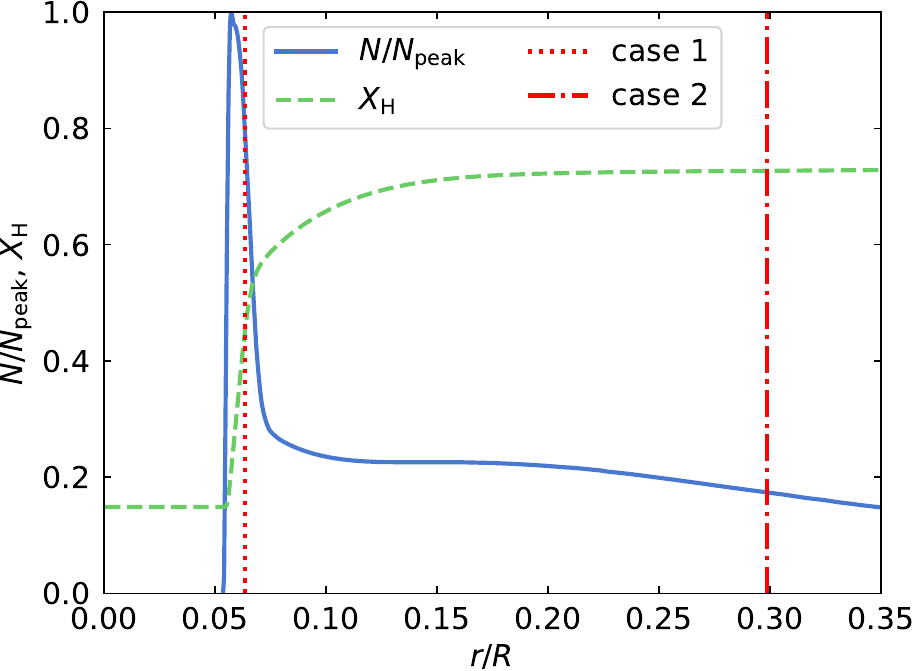}
\caption{Two possible positions of the aspherical buoyancy glitch in fractional radius ($r/R$) indicated by the vertical dotted and dashed lines.  The solid curve represents the \BV\ frequency ($N$) normalised by the peak value ($N_{\mathrm{peak}}$) at $r/R=0.06$, while the dashed curve stands for the hydrogen mass fraction ($X_{\mathrm{H}}$).
This plot is based on the structure of model A
in Table~\ref{tab:models}.}
\label{fig:N_XH}
\end{figure}

Here we interpret the three parameters, $\gla$, $\glb$ and $\glc$, about the glitch.
The fitted value of $\glb$
in Table~\ref{tab:best_fit_params} can be used to estimate the position of the glitch ($r_*$)
based on the relation,
\begin{align}
\int_{\rin}^{r_*}
\frac{N}{r}\;\dr
&=
\sqrt{2}\pi^2 \glb
\;,
\label{eq:glitch_pos_relation}
\end{align}
{which can be obtained by comparing equation (\ref{eq:tn_glitch}) with equation (\ref{eq:a2k_H}).}
Here, $\rin$ means the inner edge of the gravity-mode cavity, which almost coincides with the outer boundary of the convective core.
We use model A in Table~\ref{tab:models} to obtain
\begin{align}
\frac{r_*}{R} & =
\begin{cases}
0.0636 \pm 0.0002 & \mbox{in case 1}\;,
\\
0.298 \pm 0.003 & \mbox{in case 2}\;.
\end{cases}
\label{eq:gpos1}
\end{align}
The two possible positions are shown in Fig.~\ref{fig:N_XH},
together with the \BV\ frequency
and the hydrogen mass fraction.
The position in case 1 
means that the discontinuity is located in the layer of the steep gradient of the hydrogen profile,
which is created when the convective core shrinks in mass during evolution of the star.
It is plausible that some mixing processes
near the boundary between the convective core and the radiative envelope
could generate the discontinuity in the first derivative of the hydrogen profile.
On the other hand, the position in case 2
implies that the glitch is located deep in the radiative region,
where it is not clear how to make a discontinuous structure in general. 
Because of this,
{case 1 is preferable to case 2 from a physical point of view}.

The amplitude of the glitch signature $\gla$
is related to that of the step function $\mathfrak{D}_1^{(1)}$ that describes the discontinuity in the quadrupole component of the \BV\ frequency
(see equation (\ref{eq:dN2k_step})).
The relation is given by
\begin{align}
\mathfrak{D}_1^{(1)} & =
-\frac{8\sqrt{5\pi^3}}{3} \frac{\gla}{\Pi_{1}}
\;,
\label{eq:D11_rel}
\end{align}
in which $\Pi_{1}$ is the period spacing defined by
equation (\ref{eq:Piell_def}).
Using the fitted value in Table~\ref{tab:best_fit_params}
and the structure of model A in Table~\ref{tab:models},
we obtain
\begin{align}
\mathfrak{D}_1^{(1)}
&
=
\begin{cases}
0.020
\pm
0.001
& \mbox{in case 1}\;,
\\
0.018
\pm
0.001
& \mbox{in case 2}\;.
\end{cases}
\label{eq:D11_est}
\end{align}
The constant factor in equation (\ref{eq:D11_rel}) is equal to 33.2, which implies that
the size of the glitch signature $\gla$ is sensitive only weakly to the amplitude of the discontinuity.
The positive sign of $\mathfrak{D}_1^{(1)}$ means that
the quadrupole component of the \BV\ frequency decreases in the radial direction at the glitch, which in turn implies that
(the quadrupole component of)
the density gradient becomes less steep.
If the \BV\ frequency is dominated by the composition gradient,
the gradient of the mean molecular weight also becomes less steep across the discontinuity.
However, there is a warning at this point.
The sign of $\mathfrak{D}_1^{(1)}$ could be opposite
because of the degeneracy given by equation (\ref{eq:glitch_deg_1}).

We now turn to $\glc$,
which is related to
the phase lag ($\varphi_{\mathrm{in}}$)
at the inner turning point of the gravity-mode cavity
(see equation (\ref{eq:xih})),
by
$\glc = \varphi_{\mathrm{in}}/\pi$.
We can compare $\glc$
in Table~\ref{tab:best_fit_params} 
with its theoretical value
of $0.25$, which corresponds to $\varphi_{\mathrm{in}} = \pi/4$ \citep{UOASS1989}
to conclude that they are consistent with each other
in case 1
because the difference is only $1.2$ times greater than
the uncertainty.
The corresponding factor in case 2 increases to at most $1.9$
if we accept the change,
\begin{align}
\left(\gla,\glc\right)
&=
\left(-1.51, 0.57\right)
\rightarrow
\left(1.51, 0.07\right)
\;,
\end{align}
based on equation (\ref{eq:glitch_deg_1}).
%
%This consistency (at least in case 1) further supports %our interpretation based on the buoyancy glitch model.

\subsection{Strengths of the magnetic field}
\label{sec:mag_strength}

We can infer the properties of
the internal magnetic field based on equations
(\ref{eq:a})
and
(\ref{eq:b}).
However, these expressions are in the asymptotic limit, which is less accurate in the outer layers of the star.  We carefully examine this problem and establish more robust estimates of the field strengths.

\subsubsection{Asymptotic expressions for the lower bounds to the strengths of the field}
\label{sec:field_strength_asymp}

We first illustrate the principle
of estimating the lower bounds of the strengths of the field.
The parameter $a$
can be interpreted
by equation
(\ref{eq:a}).
Using
$\Sr$
of
the evolutionary models,
we can estimate
$\mean{\Wrr \Br^2}$,
which in turn can be used to
constrain the lower bound
to the root-mean-square
of the radial component
of the magnetic field
$\RMSr$.
For negative values
of $\mean{\Wrr \Br^2}$,
we can utilise
the relation
\begin{equation}
    \mean{\Br^2}
    >
    -2\mean{\Wrr \Br^2}
    = \frac{2(-a)}{\Sr}
    \label{eq:Br_est_asymp}
\end{equation}
\citep[see][]{Li:2022aa}.
Similarly,
in equation (\ref{eq:b})
we may use 
$b$ in Table~\ref{tab:best_fit_params}
and $S_{\text{h}}$ of the models
to estimate $\mean{W_{\phi} \Bphi^2}$,
which imposes a constraint on the lower bound to the root-mean-square of the azimuthal component, $\RMSphi$, by the relation,
\begin{equation}
\mean{\Bphi^2}
    >
    -4 \mean{W_{\phi} \Bphi^2}
    =
    \frac{4b}{(-\Sh)}
    \;.
    \label{eq:Bphi_est_asymp}
\end{equation}
In order to use equations (\ref{eq:Br_est_asymp}) and (\ref{eq:Bphi_est_asymp}), we need to assume the radius $\rup$, outside which no magnetic field exists at all.
However,
there is no direct observational constraint on $\rup$.
Instead, we may set $\rup$ to the outer edge of the gravity-mode cavity,
{but} this approach also has a problem
because equations (\ref{eq:Br_est_asymp}) and (\ref{eq:Bphi_est_asymp}) are valid only in the asymptotic limit, which is not realised very well in the outer layers.
Therefore,
we consider it {improper} to apply
equations (\ref{eq:Br_est_asymp}) and (\ref{eq:Bphi_est_asymp})
as they are.
In the following sections,
we {carefully} examine the problems of these relations to revise them.

\subsubsection{Problem of the asymptotic expressions}

\begin{figure*}
    \centering
    \includegraphics[width=\columnwidth]{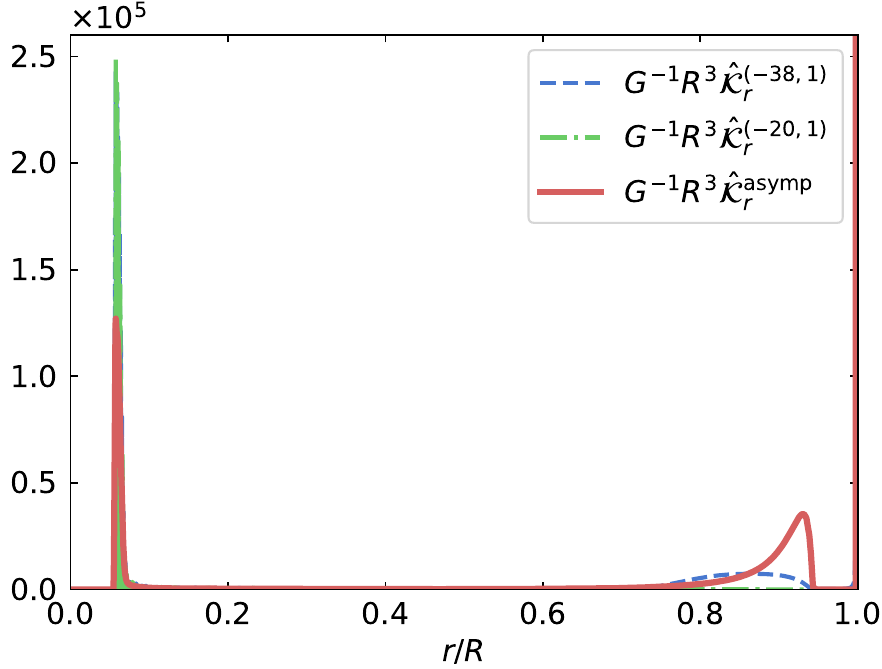}
    \includegraphics[width=\columnwidth]{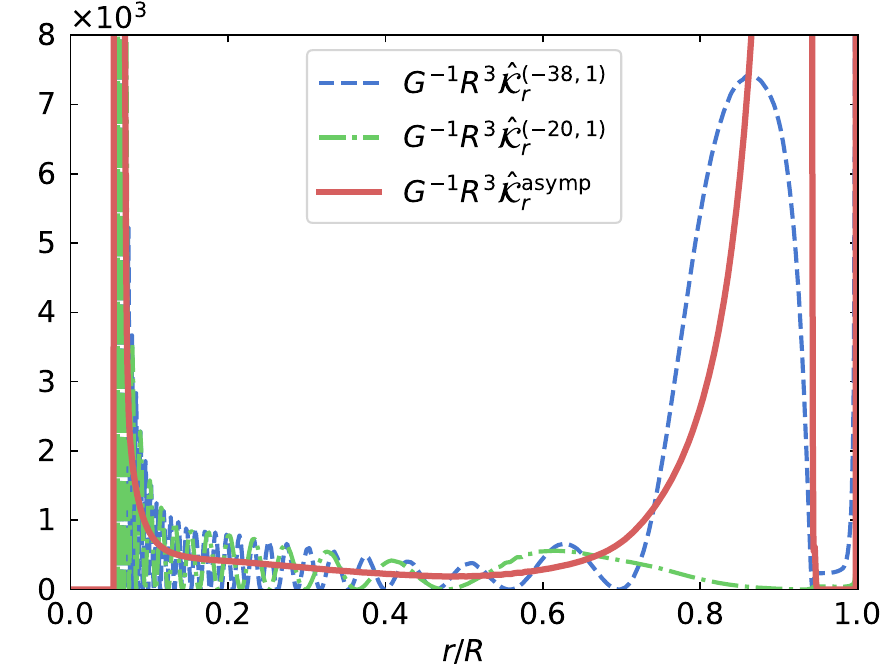}
    \includegraphics[width=\columnwidth]{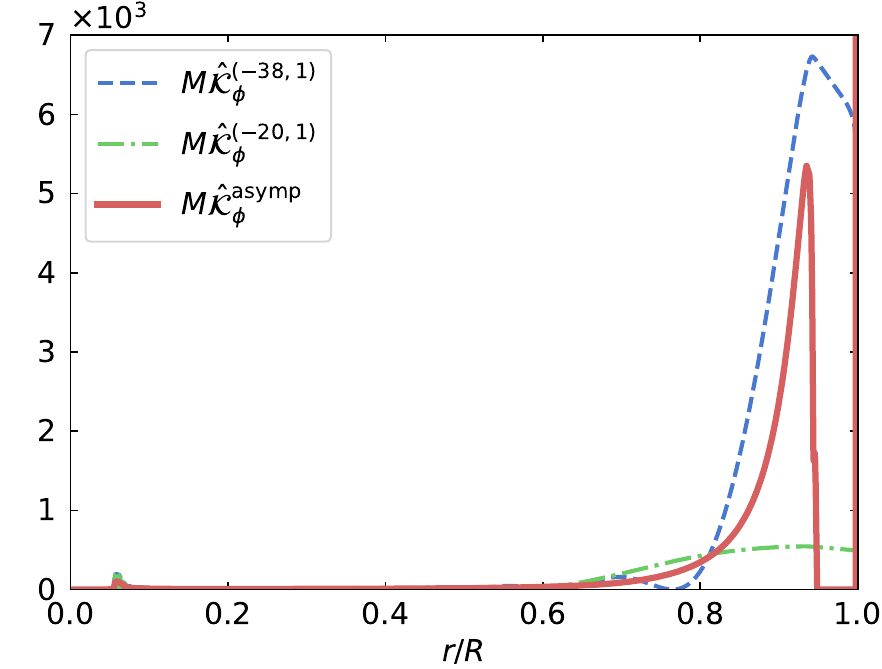}
    \includegraphics[width=\columnwidth]{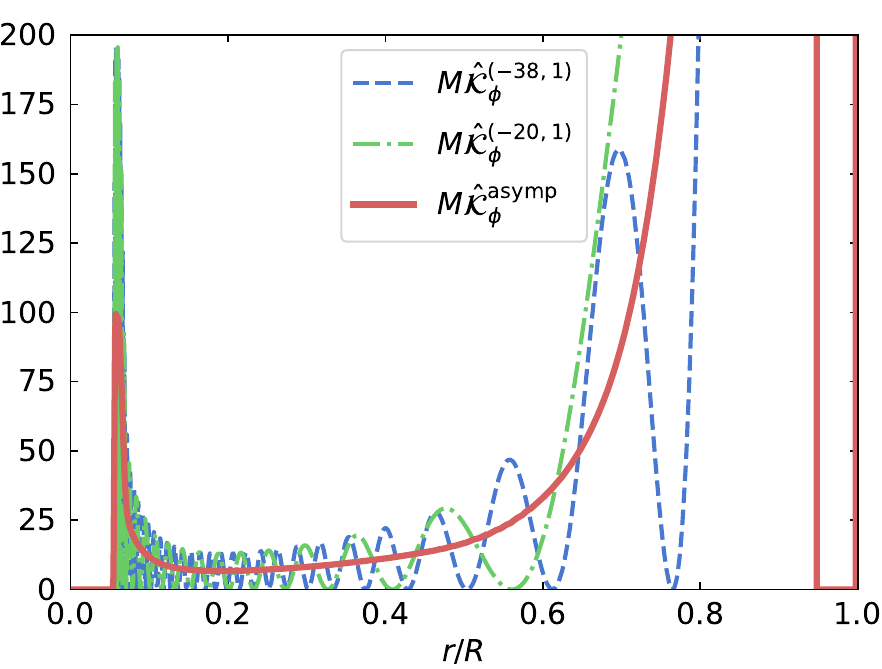}
    \caption{Profiles of 
    $\hat{\mathcal{K}}_{\rmr}^{(n,1)}$
    (with $n=\nlow,\,\nhigh$)
    and
    $\hat{\mathcal{K}}^{\mathrm{asymp}}_{\rmr}$
    normalised by $G R^{-3}$
    (upper panels)
    and
    $\hat{\mathcal{K}}_{\phi}^{(n,1)}$
    (with $n=\nlow,\,\nhigh$)
    and
    $\hat{\mathcal{K}}^{\mathrm{asymp}}_{\phi}$
    normalised by $M$
    (lower panels)
    as functions of the fractional radius $r/R$
    for model A in Table~\ref{tab:models}.
    The frequencies of the modes with
    $n=\nlow$ and $n=\nhigh$ are
    $0.96\,\mathrm{d}^{-1}$
    and
    $1.79\,\mathrm{d}^{-1}$, respectively.
    The left panels are the plots to see the overall structure,
    while the right panels 
    {have} smaller {ordinate} scales
    to resolve the oscillatory behaviour
    of $\hat{\mathcal{K}}_{\rmr}^{(n,1)}$
    and
    $\hat{\mathcal{K}}_{\phi}^{(n,1)}$.}
    \label{fig:Khat}
\end{figure*}

An essential point of the arguments
in Section~\ref{sec:field_strength_asymp}
is that $a$ and $b$ are independent of the mode frequencies.
This is correct only in the asymptotic limit,
while their general expressions are given by
\begin{align}
a_n & \equiv
\frac{3}{128\pi^5}
\int_0^R
\hat{\mathcal{K}}_{\rmr}^{(n,1)}
\overline{\Wrr \Br^2}
\;\dr
\end{align}
and
\begin{align}
b_n & \equiv
-\frac{9}{32\pi^3}
\int_0^R 
\hat{\mathcal{K}}_{\phi}^{(n,1)}
\overline{W_{\phi} \Bphi^2}
\;\dr
\;,
\label{eq:bn_def}
\end{align}
in which $\hat{\mathcal{K}}_{\rmr}^{(n,1)}$
and $\hat{\mathcal{K}}_{\phi}^{(n,1)}$ are defined by
\begin{align}
\hat{\mathcal{K}}_{\rmr}^{(n,1)}
&
\equiv
\frac{
\left(
2\pi\nu_{n,1}
\right)^2
r^2 \left(\tdr{\xih{n,1}}\right)^2
}{%
\int_0^R \left( \xir{n,1}^2 + 2 \xih{n,1}^2 \right)
\rho r^2\;\dr
}
\end{align}
and
\begin{align}
\hat{\mathcal{K}}_{\phi}^{(n,1)}
&
\equiv
\frac{
2 \xih{n,1}^2
}{%
\int_0^R \left( \xir{n,1}^2 + 2 \xih{n,1}^2 \right)
\rho r^2\;\dr
}
\;,
\end{align}
respectively
(see Appendix~\ref{sec:dom_terms}).
Here, $\xir{n,\ell}$ and $\xih{n,\ell}$ represent
the radial and horizontal displacements,
respectively,
with radial order $n$ and spherical degree $\ell$.
The asymptotic expressions for
$\hat{\mathcal{K}}_{\rmr}^{(n,1)}$
and $\hat{\mathcal{K}}_{\phi}^{(n,1)}$ are
provided by
\begin{align}
\hat{\mathcal{K}}^{\mathrm{asymp}}_{\rmr}
&
\equiv
\frac{%
\frac{N^3}{\rho r^3}
}{%
\int_G \frac{N}{r}
\;\dr}
\end{align}
and
\begin{align}
\hat{\mathcal{K}}^{\mathrm{asymp}}_{\phi}
&
\equiv
\frac{%
\frac{N}{\rho r^3}
}{%
\int_G \frac{N}{r}
\;\dr
}\;.
\end{align}
Fig.~\ref{fig:Khat} shows
the profiles of
$\hat{\mathcal{K}}_{\rmr}^{(n,1)}$ (with $n=\nlow,\,\nhigh$%
\footnote{We follow \citet{Takata2006b}
for the definition of the radial order $n$.})
and
$\hat{\mathcal{K}}_{\rmr}^{\mathrm{asymp}}$
in the upper panels
and
$\hat{\mathcal{K}}_{\phi}^{(n,1)}$ (with $n=\nlow,\,\nhigh$)
and
$\hat{\mathcal{K}}_{\phi}^{\mathrm{asymp}}$
in the lower panels
for model A in Table~\ref{tab:models}.
We observe in the upper panels that
the profiles of $\hat{\mathcal{K}}_{\rmr}^{(n,1)}$
are highly concentrated around $r/R\approx 0.06$ and
so oscillatory in the inner layers
($r/R \lesssim 0.5$)
that their average behaviour can be described by
$\hat{\mathcal{K}}_{\rmr}^{\mathrm{asymp}}$ very well.
The high peak is located in the layers
of the steep gradient of chemical compositions
just outside the convective core.
The oscillatory behaviour occurs because the radial wave number $\kr$ is large enough.
On the other hand,
in the outer layers ($r/R \gtrsim 0.5$), we find
that (1) the profiles of
$\hat{\mathcal{K}}_{\rmr}^{(n,1)}$ become rapidly
less oscillatory, that (2) the
amplitude of $\hat{\mathcal{K}}_{\rmr}^{\mathrm{asymp}}$
increases steeply and that (3)
the amplitude of $\hat{\mathcal{K}}_{\rmr}^{(n,1)}$
differs significantly between the two modes.
The reason for point (1) is that
$\kr$ becomes smaller due to the decrease in the Lamb frequency rather than the \BV\ frequency
\citepalias[see Fig.~12 of][]{Saio:2015aa}.
In fact, 
unlike the higher-order mode (with $n=\nlow$),
the outer turning point of the lower-order mode 
(with $n=\nhigh$)
is not fixed by the \BV\ frequency but the Lamb frequency,
which explains point (3).
Point (2) is due to the lower density in the near-surface layers.
All of the three points demonstrate that
the asymptotic expression
$\hat{\mathcal{K}}_{\rmr}^{\mathrm{asymp}}$
becomes inaccurate in the outer layers.

In the lower panels of Fig.~\ref{fig:Khat},
$\hat{\mathcal{K}}_{\phi}^{(n,1)}$ and
$\hat{\mathcal{K}}_{\phi}^{\mathrm{asymp}}$ 
show similar structures
to $\hat{\mathcal{K}}_{\rmr}^{(n,1)}$ and
$\hat{\mathcal{K}}_{\rmr}^{\mathrm{asymp}}$,
respectively,
although the peak 
of $\hat{\mathcal{K}}_{\phi}^{\mathrm{asymp}}$ 
in the innermost radiative region
is much smaller than that
in the outermost radiative region
around $r/R\approx 0.93$.
The much higher weight in the outer layers
is because
$\hat{\mathcal{K}}_{\phi}^{\mathrm{asymp}}$
is proportional to a lower power of $N$ than
$\hat{\mathcal{K}}_{\rmr}^{\mathrm{asymp}}$.
The largest amplitude of
$\hat{\mathcal{K}}_{\phi}^{(n,1)}$
in the outer layers
appears to imply that these layers contribute significantly to the integral in equation (\ref{eq:bn_def}).
{However}, we argue in the next section that this is not true.

\subsubsection{Maximum field strengths of the analysis at each radius}

Here, we recall the two assumptions given
in Section~\ref{sec:theory}.
The first one is that
the first-order rotation effect
is much larger than the direct effect of the magnetic field.
The former and latter
can be estimated by the dimensionless parameters,
\begin{align}
s & \equiv \frac{2\nurot}{\nu}
\;,
\end{align}
and
\begin{align}
s_{\mathrm{m}}
& \equiv
\left(
\frac{\Br}{\Brup}
\right)^2
\;,
\end{align}
respectively,
in which $\Brup$ is defined by
equation (\ref{eq:Brup}).
{Here, $s$ is called the spin parameter and provides the ratio of the Coriolis force and the inertial term, whereas $s_{\mathrm{m}}$ is the ratio of the horizontal component of the Lorentz force and that of the pressure gradient.}
The condition of $s_{\mathrm{m}} \ll s$ 
(for $\ell=1$) yields
\begin{align}
\left|\Br\right| \ll
\Brmax \equiv
\sqrt{64\pi^5 \nurot \nu^3 \rho} \frac{r}{N}
\;.
\label{eq:Brmax_cond}
\end{align}

\begin{figure}
    \centering
    \includegraphics[width=\columnwidth]{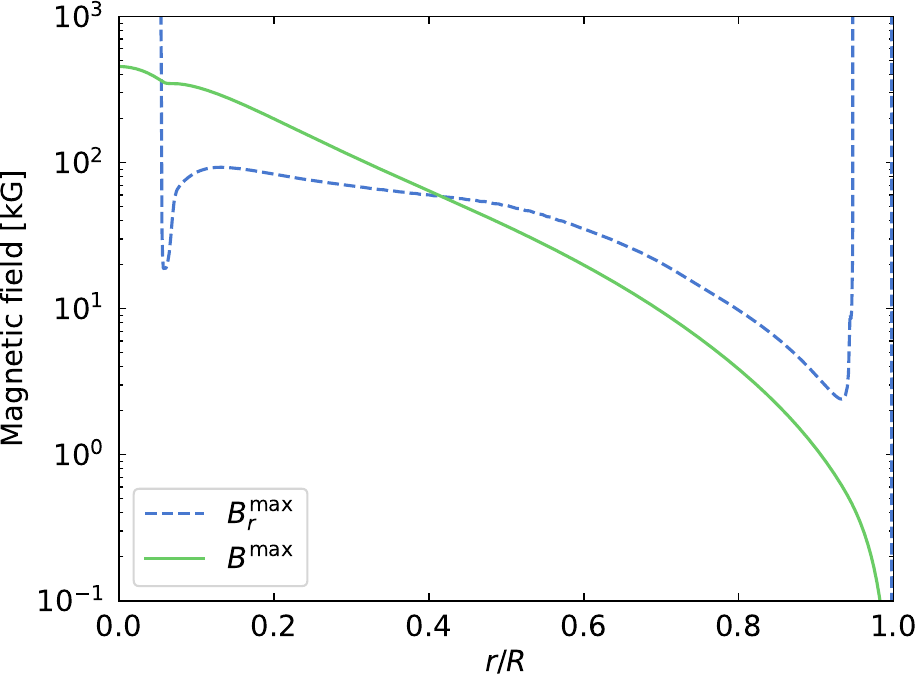}
    \caption{The maximum strengths
    of the radial component of the magnetic field ($\Brmax$)
    and the total magnetic field ($\Bmax$)
    that come from the assumptions of the present analysis.
    {The condition of $\left|\Br\right| < \Brmax$ means that the Coriolis force contributes to the restoring force of the oscillation more importantly than the Lorentz force, while $\left|B\right| < \Bmax$ means that the equilibrium structure is deformed by the centrifugal force more significantly than the Lorentz force.}
    Model A in Table~\ref{tab:models} is used.
    The rotation period and the oscillation frequency are assumed to be $64\,\mathrm{d}$ and
    $1\,\mathrm{d}^{-1}$, respectively.
    }
    \label{fig:Bmax}
\end{figure}

The second assumption is
that
the magnetic deformation of the equilibrium structure
is smaller than the rotational deformation.
The effect of the magnetic deformation can be estimated by the ratio between the magnetic and gas pressure ($p$),
\begin{align}
\mathfrak{r}_{\mathrm{mag}}
& \equiv \frac{B^2}{8\pi p}
\;,
\end{align}
whereas that of the rotational deformation can be estimated by
the ratio between the centrifugal force at the equator and the gravity,
\begin{align}
\mathfrak{r}_{\mathrm{rot}}
& \equiv \frac{r \left(2\pi\nurot\right)^2}{g}
\;,
\end{align}
where $g$ is the gravitational acceleration.
Then, the condition of $\mathfrak{r}_{\mathrm{mag}} \ll \mathfrak{r}_{\mathrm{rot}}$ means
\begin{align}
B \ll
\Bmax
\equiv
\sqrt{
\frac{32\pi^3\nurot^2 r p}{g}}
\;.
\label{eq:Bmax_cond}
\end{align}
The profiles of $\Bmax$ and $\Brmax$ are shown in Fig.~\ref{fig:Bmax}.
We observe that
$\Bmax$ decreases monotonically and rapidly
because $p$ does so towards the surface.
Although the amplitude of $\hat{K}_{\phi}^{(n,1)}$
roughly increases in proportion to the inverse of $\rho$ in the outer layers of the star,
$\hat{K}_{\phi}^{(n,1)} (\Bmax)^2$
becomes smaller for larger radii
because $p/\rho$ decreases. 
We therefore understand that the contribution of the outer layers to the integral in equation (\ref{eq:bn_def}) is small.

We may stress the meanings of these maximum strengths.
The estimates for the radial and total magnetic fields at each radius must not be larger than $\Brmax$ and $\Bmax$, respectively,
for the analysis to be self-consistent.
If this is not the case,
the fundamental relation of the present analysis, equation (\ref{eq:tn_model}),
cannot be justified.
The fact that equation (\ref{eq:tn_model}) provides
good fits to the data suggests, but does not prove, that
these conditions are actually satisfied.
Since $\Bmax < \Brmax$ for $r/R > 0.42$, we concentrate on $\Bmax$ to discuss the contribution from the near-surface layers to the integrals in equation (\ref{eq:bn_def}).

\subsubsection{Estimates for the root-mean-square of the strengths of the field}

We modify
equations
(\ref{eq:Br_est_asymp})
and
(\ref{eq:Bphi_est_asymp}),
taking into account the two problems,
the poor asymptotic expressions in the outer layers
and
the maximum strengths of the field in the analysis.
Since the problems are more severe for $\Bphi$ than $\Br$,
we first discuss $\Bphi$.

Since 
%$b > 0$ in Table~\ref{tab:best_fit_params} and 
$W_{\phi} \ge -1/4$
(see equation (\ref{eq:Wphi})),
we find from
equation (\ref{eq:bn_def})
\begin{align}
\frac{128\pi^3}{9} b_n
& \le
\int_{0}^{R}
\hat{\mathcal{K}}_{\phi}^{(n,1)}
\overline{\Bphi^2}
\;\dr
=
I_{n,\mathrm{in}} + I_{n,\mathrm{out}}
\;,
\label{eq:KBphi_sep}
\end{align}
where we have separated the integral 
in two parts at radius $\rup$ by introducing
\begin{align}
I_{n,\mathrm{in}} & \equiv
\int_0^{\rup} 
\hat{\mathcal{K}}_{\phi}^{(n,1)}
\overline{\Bphi^2}
\;\dr
\label{eq:I_n_in}
\end{align}
and
\begin{align}
I_{n,\mathrm{out}} & \equiv
\int_{\rup}^R 
\hat{\mathcal{K}}_{\phi}^{(n,1)}
\overline{\Bphi^2}
\;\dr
\;.
\end{align}
The idea is to choose $\rup$ such that the asymptotic expression $\hat{K}_{\phi}^{\mathrm{asymp}}$ is accurate enough for $r \le \rup$ while keeping $I_{n,\mathrm{out}}$ 
small enough.

We may substitute
$\hat{K}_{\phi}^{\mathrm{asymp}}$
into $\hat{\mathcal{K}}_{\phi}^{(n,1)}$
in $I_{n,\mathrm{in}}$.
The error of this approximation can be estimated as
\begin{align}
\frac{9}{128 \pi^3 b_n}
\left|
I_{n,\mathrm{in}}
-
\int_{\rin}^{\rup}
\hat{K}_{\phi}^{\mathrm{asymp}}
\overline{\Bphi^2}
\;\dr
\right|
&
\le
\kc_n
\;,
\label{eq:cn_limit}
\end{align}
in which we have defined
\begin{align}
\kc_n
& \equiv
\frac{9 \nu_{n,1}}{128\sqrt{2}\pi^2 b}
\left(\int_G \frac{N}{r}\;\dr\right)^{-1}
\left[
\frac{(\Bmax)^2}{\rho r^2}
\right]_{r=\rup}
\;.
\label{eq:dn_def}
\end{align}
{Note that $\rin$ in equation (\ref{eq:cn_limit}),
which has been introduced in equation (\ref{eq:glitch_pos_relation}),
means
the inner edge of the gravity-mode cavity.}
In deriving equation (\ref{eq:dn_def}),
we have substituted equation (\ref{eq:xih})
into equation (\ref{eq:I_n_in}),
and performed integration by parts.

On the other hand,
we may constrain the upper limit of the relative contribution of $I_{n,\mathrm{out}}$
to be
\begin{align}
\frac{9 I_{n,\mathrm{out}}}{128\pi^3 b}
&
\le
\kd_n
\equiv
\frac{9}{128\pi^3 b}
\int_{\rup}^{R}
\hat{\mathcal{K}}_{\phi}^{(n,1)}
\left(\Bmax\right)^2
\;\dr
\;.
\label{eq:dn_limit}
\end{align}

The relative errors $\kc_n$ and $\kd_n$
can be evaluated using the structure and the eigenfunctions of the equilibrium models
in Table~\ref{tab:models}
together with the fitted values of $b$
given in Table~\ref{tab:best_fit_params}.
We confirm 
for all the modes in the analysis of all the models
in Table~\ref{tab:models}
and for the both cases of the glitch positions
(cases 1 and 2)
\begin{align}
\kc_n & < \overline{\kc} = 0.012
\label{eq:cn_est}
\end{align}
and
\begin{align}
\kd_n & < \overline{\kd} = 0.1
\label{eq:dn_est}
\end{align}
for 
\begin{align}
\xup & =
\frac{\rup}{R} 
= 0.5
\;.
\label{eq:xup_choice}
\end{align}
Although we may choose other values of $\xup$,
we regard the value in equation (\ref{eq:xup_choice}) as fiducial.
{Note that this corresponds to the radial coordinate of $r = 1.02\,\mathrm{R}_{\sun}$ and the mass coordinate of $\Mr = 1.42\,\mathrm{M}_{\sun}$ (98 per cent of the total mass) in model A.}

Using equations (\ref{eq:KBphi_sep}), (\ref{eq:cn_limit})
and (\ref{eq:dn_limit})
{with $b_n = b$}, we can revise equation {(\ref{eq:Bphi_est_asymp})}
as
\begin{align}
\BphiRMS
    >
    \RMSphi
    \equiv
    \left[
    \frac{4b}{(-\Sh)}
    \left(
    1 
    - \overline{\kd}
    \right)
    \right]^{1/2}
    \;,
    \label{eq:Bphi_est_asymp_mod}
\end{align}
in which we have neglected $\overline{\kc}$
because it is smaller than $\overline{\kd}$ by an order-of-magnitude
{(see equations (\ref{eq:cn_est}) and (\ref{eq:dn_est}))}.
Here,
the integral domain $G_B$ that appears
in the definition of $\Sh$ 
(see equation (\ref{eq:Sh}))
is between $r=\rin$ and $r=\rup$.

In order to obtain the corresponding expression for
$\Br$,
for simplicity we make an additional assumption,
\begin{align}
\left|
\frac{\int_{\rup}^R \hat{\mathcal{K}}_{\rmr}^{(n,1)}\overline{\Wrr\Br^2}\;\dr}{\int_0^R \hat{\mathcal{K}}_{\rmr}^{(n,1)}
\overline{\Wrr \Br^2}\;\dr}
\right|
\le
{\frac{1}{4}}
\left|
\frac{\int_{\rup}^R \hat{\mathcal{K}}_{\phi}^{(n,1)}\overline{W_{\phi}\Bphi^2}\;\dr}{\int_0^R \hat{\mathcal{K}}_{\phi}^{(n,1)}
\overline{W_{\phi} \Bphi^2}\;\dr}
\right|
\;,
\end{align}
which roughly means that
$\Br$ is concentrated in the core region
to the degree similar to or {more} than $\Bphi$.
Then, 
also assuming that
the error in the asymptotic expression is negligible for $r \le \rup$,
we can revise equation (\ref{eq:Br_est_asymp}) as
\begin{align}
\BrRMS
    >
    \RMSr
    \equiv
    \left[
    \frac{2(-a)}{\Sr}\left( 1 - \overline{\kd}\right)
    \right]^{1/2}
    \;.
    \label{eq:Br_est_asymp_mod}
\end{align}

Table~\ref{tab:fitting} provides
$\RMSr$ and $\RMSphi$ for the three evolutionary models in Table~\ref{tab:models} (A, B and C)
and the two positions of the aspherical glitch
(cases 1 and 2)
with $\xup=\xupest$.
The results for model A and case 1,
$\RMSr=\Brest\,\mathrm{kG}$ and $\RMSphi=\Bphiest\,\mathrm{kG}$,
are completely consistent with the other cases with
different combinations of
evolutionary models
and glitch positions.
The estimates for $\RMSr$ and $\RMSphi$
are quite insensitive to
the evolutionary models
and the positions of the glitch.
The fact that
$\RMSphi$ is larger than
$B^{\max}$ at $r/R=0.5$,
which is equal to $37\,\mathrm{kG}$ for model A
(see Fig.~\ref{fig:Bmax}),
means that
the field distribution within $r/R \le 0.5$
is biased to the core.
We also confirm that
$\RMSr$ is well below $\Brmax$ in Fig.~\ref{fig:Bmax}.
{We finally note that if $\Br=\RMSr$ and $\Bphi=\RMSphi$ for $r/R\le 0.5$, the ratio of the magnetic pressure to the centrifugal force at the equator is equal to $0.04$--$0.7$ outside the convective core for model A and case 1.
This confirms one of the assumptions made in Section \ref{sec:theory}.}

\begin{table}
    \centering
    \caption{Estimates
    of 
    the lower bounds
    to root-mean-squares
    of $\Br$ and $\Bphi$.
    The assumed models are those in Table~\ref{tab:models},
    while cases 1 and 2 are different from each other in
    the position of the glitch (see Fig.~\ref{fig:N_XH}).
    The upper limit of the integrals
    in the numerators
    of equations (\ref{eq:Sr})
    and (\ref{eq:Sh})
    is set to $\xup = \xupest$ in unit
    of the fractional radius.
    }
    \label{tab:fitting}
    \begin{tabular}{lccc}
    \hline
    Model &  A${}^*$ & B & C\\
    \hline
    $\RMSr$
    $(\mathrm{kG})$
    \\
    case 1
        & $\Brest$
        & $3.7\pm 0.1$
        & $3.6\pm 0.1$
    \\
    case 2
    & $3.4\pm 0.1$
    & $3.6\pm 0.1$
    & $3.5\pm 0.1$
    \\
    \hline
    $\RMSphi$
    $(\mathrm{kG})$
    \\
    case 1
    & $\Bphiest$
    & $92 \pm 7$
    & $93 \pm 7$
    \\
    case 2
    & $88\pm 7$
    & $89\pm 7$
    & $89\pm 7$
    \\
    \hline
    \end{tabular}
        \\
    {\it Note.}${}^*$
    the best model \citepalias[see][]{Saio:2015aa}.
\end{table}
%

%\section{Search for other cases}
%
%other stars
%
%(on-line material?)
%
%no detection
%
%
%slow rotation is the key for detection
%

\section{Discussion}
\label{sec:discussion}

We have shown that
the observed asymmetry
of frequency splittings
in \TheStar\
can be explained
by a model with an internal magnetic field,
whose azimuthal component is much stronger
than the radial component.
Historically, it is well established that
a small fraction (about ten per cent) of
intermediate-mass main-sequence stars
have
strong and large-scale
magnetic fields at the surface,
which are associated with chemical peculiarities.
Because of this, much attention has been paid to the surface of these stars.
The result of this study enables us to move our focus to the interior of the stars with observational constraints.
%In fact, it is possible that the internal magnetic fields are common for early-type main-sequence stars.
%, which generally have the convective core.
%
%Namely, the problem of magnetic fields
%could be relevant not only in the small fraction of early-type stars, but also in any intermediate and high-mass main-sequence stars.
%
From the asteroseismic point of view,
we have so far been able to constrain only the radial component of the internal magnetic field in red giants,
which allows us to study {only} limited aspects of the magnetic problems in stars.
On the other hand, the predominantly toroidal configuration in the main-sequence star
revealed by the present study
has opened the possibility of investigating the origin, the impact on the angular momentum transport, and the evolutionary change of the field from a new angle.
Here, we discuss several topics concerning this result.

\subsection{Excluded data points}
\label{subsec:outlier}

We exclude
from the fitting
two data points
of the asymmetry of the frequency splittings,
(1) $\asymm_{n} = 8.1\times 10^{-5}\,\mathrm{d}^{-1}$
at $\nu_{n,1}=1.13\,\mathrm{d}^{-1}$
(see Fig.~\ref{fig:asymmetry})
and
(2) $\asymm_{n} = -2.5\times 10^{-5}\,\mathrm{d}^{-1}$
at $\nu_{n,1}=1.20\,\mathrm{d}^{-1}$
(see Fig.~\ref{fig:fitting}).
We speculate on the origin of these points.

First, we point out a potential problem in the frequency analysis that the frequency determination of a mode
can be perturbed
by
another mode with
a very close frequency
within the resolution
{
and an amplitude above the noise level}.
%{\red (any reference?) [DWK: We are not concerned with unresolved peaks at, or below, the noise level. The noise peaks themselves perturb the determined frequencies, but that effect is included in the error analysis. Your concern here is unresolved real peaks with amplitudes above the noise level. In some cases we are somewhat aware of those because the spectral window is visibly perturbed. In other cases it might not show and we will not know that we have this problem until we can get higher frequency resolution. Given the Kepler 4-yr data set, the latter is not going to happen for this star in the foreseeable future. And no, I do not know of any reference for this. Simon: Do you?][MT: thank you for the detailed comments. I modified the last sentence.]}

Actually, the problem of the close frequency
influences not only the data analysis
but also the physics of stellar oscillations.
If there are two modes with very close frequencies
(close degeneracy),
those frequencies could be perturbed
as a result of mode interaction
due to nonlinearity or rotation or magnetic fields.
This might possibly cause a considerable effect
in the frequency spectrum as a result of an avoided crossing.

In fact,
checking the frequencies of model A,
we find
a quadrupolar mode with
$\nu_{-56,2}=1.1347\,\mathrm{d}^{-1}$,
which is close to 
the dipolar mode with
$\nu_{-32,1}=1.1356\,\mathrm{d}^{-1}$,
which corresponds to data point (1).
Therefore, data point (1) could undergo
a significant effect of close degeneracy.
In addition,
although the frequency difference between the two modes of $9\times 10^{-4}\,\mathrm{d}^{-1}$ is larger than the frequency resolution 
given by equation (\ref{eq:freq_res}),
it is possible that
a small error in the evolutionary model shifts the theoretical frequencies to agree with each other
within the resolution.
{Only one other mode has a closer $\ell=2$ mode frequency, which is the one at $1.7\,\mathrm{d}^{-1}$. However, this mode has a much larger measurement uncertainty, so the influence of any mode interaction is difficult to discern.}
On the other hand, in the case of data point (2),
there do not exist for $2 \le \ell \le 4$
modes with such close frequencies to the dipolar one with
$\nu_{-30,1}=1.2088\,\mathrm{d}^{-1}$.
Still,
there is an $\ell=5$ mode with
$\nu_{-118,5}=1.2066\,\mathrm{d}^{-1}$,
which differs by
$2\times 10^{-3}\,\mathrm{d}^{-1}$ from the dipolar mode.
This is itself not surprising because the period spacing becomes smaller (the spectrum becomes denser) for larger $\ell$,
which implies a {higher} chance {of finding} a close frequency.
Since we do not expect strong interaction between
the $\ell=1$ and $\ell=5$ modes,
the effect of the close degeneracy for data point (2)
would be much smaller than for data point (1).
Although this is qualitatively consistent with
the fact that
data point (2) is much closer to the best-fit curves
than data point (1),
we postpone detailed quantitative analysis to future work.

\subsection{
Origin of
the detected magnetic field
}
\label{sec:origins}

We may consider at least four possible origins of the detected magnetic field:
(1) an interstellar field that is locked into the star
(fossil field)
{(Section \ref{sec:fossil})};
(2)
a field generated by
the dynamo process
in the convective core
and
moved or
left in the radiative region
{(Section \ref{sec:core_dynamo})};
%as the convective core recedes along with evolution;
(3)
a field generated and maintained by
a dynamo process of Tayler--Spruit type
in the radiative region {(Section \ref{sec:TS_dynamo})};
(4)
a field generated by
the magneto-rotational instability (MRI)
during a merger process {(Section \ref{sec:merger})}.
{We also discuss the stability of
the field (Section \ref{sec:stability}).}

\subsubsection{Fossil field}
\label{sec:fossil}

During formation of an intermediate-mass star,
weak magnetic fields that are {embedded in} the interstellar medium
can {become} concentrated in the star
as the medium collapses into it.
Unless the fields are completely destroyed
during the pre-main-sequence wholly convective phase,
they are eventually locked in the radiative region of the star.
If these fields {relax to} a stable configuration,
they can survive for the timescale of magnetic diffusion,
which is on the order of $10^{10}\,\mathrm{yr}$
(longer than the lifetime of the main-sequence stage for a star with a mass $\gtrsim 1.5\,\mathrm{M}_{\sun}$).
{While the problem of stability is discussed separately in Section \ref{sec:stability},
there have been many theoretical studies of the equilibrium structure of magnetic fields and its stability in the stellar radiative interior \citep[e.g.][]{Prendergast:1956aa,Woltjer:1960aa,
Braithwaite:2004aa,Braithwaite:2006aa,
Braithwaite:2008aa,
Duez:2010ab,Duez:2010ac}.
There have also been studies about the effect of
internal fields on stellar structure and evolution \citep[e.g.][]{Mestel:1977aa,Duez:2010aa}.
It {certainly will be} interesting to compare the result of the present analysis with these works in detail,
which is postponed to future work.
Instead, we concentrate on {the following} issue here.}

If the field originates from the interstellar field,
we may expect it to extend outside the surface.
This is not likely because our model with no surface field can explain the observed data quite well.
%
%\textcolor{red}{no evidence from p modes to support the oblique pulsator model.}
In addition,
if there were a large-scale surface magnetic field,
which is generally inclined to the rotation axis,
the oblique pulsator model \citep{Kurtz:1990aa}
would predict that all acoustic modes should show a multiplet structure in the frequency spectrum, whose components are equally split by the rotation frequency.
This phenomenon is not detected in \TheStar\ 
\citepalias[see Table 3 of][]{Saio:2015aa}.
Furthermore, there has been no confirmed detection of {any} large-scale magnetic field at the surface of any $\gamma$ Dor 
(or $\delta$ Sct--$\gamma$ Dor hybrid)
stars
\citep[e.g.][]{Thomson-Paressant:2023aa,Hubrig:2023aa}.

{One possible solution to this problem
could come from the hypothesis of \citet{Jermyn:2020aa},
which {was} proposed to explain
the observed bimodal distribution
of the surface magnetic {fields} of early-type
main-sequence stars \citep[e.g.][]{Auriere:2007aa,Lignieres:2014aa}.
A large-scale surface magnetic field
could be destroyed by
convective motions near the surface unless it is {sufficiently} strong.
Then, a dynamo process {would work} in the subsurface convective zone to generate a much weaker small-scale field.
We may examine this idea using simplified expressions.
A sufficient condition for a magnetic field
to suppress convection
is given by 
\begin{align}
 \frac{\Br^2}{\Br^2 + 8\pi\Gamma_1 p}
 > \nabla - \nabla_{\mathrm{ad}}
 \label{eq:GT_cond}
\end{align}
\citep{Gough:1966aa}.
Here, $\Gamma_1$ is the first adiabatic index,
while $\nabla$ and $\nabla_{\mathrm{ad}}$
mean the temperature gradient,
$\rd\ln T/\rd\ln p$, and its adiabatic value,
{$\left(\partial \ln T/\partial \ln p\right)_{S}$,}
respectively, where $S$ represents entropy.
If the radial component of the fossil field, $\Br$,
is strong enough to satisfy
equation (\ref{eq:GT_cond})
for $\nabla = \nabla_{\mathrm{rad}}$,
where $\nabla_{\mathrm{rad}}$ is the
radiative temperature gradient,
then 
the convection is suppressed and
the field keeps its large-scale structure on the surface of the star.
Otherwise, the convective motions significantly modify the field to generate small-scale structures,
which would lead to unstable configurations,
and hence decay of the field.

Our best model of \TheStar\ (model A) 
with $1.45\,\mathrm{M}_{\sun}$
has only one subsurface convective zone
in the outermost five per cent in radius
($5\times 10^{-6}$ in fractional mass),
where hydrogen (H\,I), He\,I and He\,II undergo ionisation.
An essential point is that the
opacity in this zone is dominated by that of the ionisation of hydrogen,
which results in very large values of $\nabla_{\mathrm{rad}}$
with the maximum value of $\approx 400$.
Since the left-hand side of equation (\ref{eq:GT_cond}) is always smaller than one,
this condition can never be satisfied in most of the convective zone,
which implies that the magnetic field would not suppress the subsurface convection.
However, the convection associated with the ionisation of hydrogen
is so efficient that the strength of the dynamo-generated small-scale field should range between $0.4$\,kG (near the top of the zone) and $1.3$\,kG (near the bottom), {values that} are obtained by assuming the equipartition of energy between the magnetic field and the convective motion.
This means that the field strength at the photosphere is about a few hundred {Gauss},
which has not been detected in any $\gamma$ Dor stars so far.
In summary, if we adopt the picture of \citet{Jermyn:2020aa}, 
the large-scale fossil field could indeed be erased by the subsurface convection, but it would be replaced with a small-scale dynamo-generated field with considerable strength, which does not have any observational support.
We therefore need to revise the picture or switch to some other mechanism 
\citep[see e.g.][]{Braithwaite:2017aa}
to explain the detected internal field of \TheStar\ as a fossil one.
In particular, since the assumption of the equipartition of energy could be too crude in the above discussion,
it is highly desirable to perform
more detailed analysis about the interaction
between the fossil magnetic field and the near-surface convection, which is outside the scope of this paper.
We should also note that the present argument does not {contradict} \citet{Jermyn:2020aa}, who {considered} only main-sequence models above $2\,\mathrm{M}_{\sun}$.
In such higher-mass structures, the main focus is on the subsurface convection zones associated with the ionisation of He\,II and iron-group elements,
which are much less efficient than the one with the ionisation of hydrogen,
and hence would generate a much weaker field by dynamo.
}

\subsubsection{Stability of the field}
\label{sec:stability}

\begin{figure}
\centering\includegraphics[width=\columnwidth]{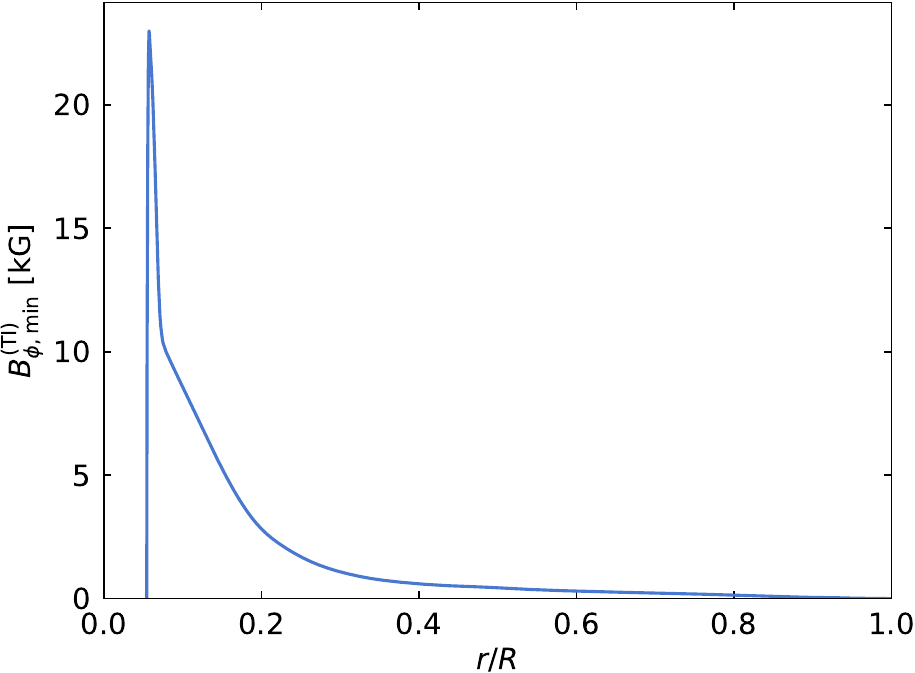}
\caption{Profile
of the minimum strength of the toroidal component
of the magnetic field, which is defined by
equation (\ref{eq;Bphi_min}),
for the Tayler instability to operate
based on model A in Table~\ref{tab:models}
with the rotation period of $64$\,d.}
\label{fig:Bphi_TI_min}
\end{figure}

Apart from its origin,
it is an interesting and important question whether
the predominantly toroidal configuration found in the present analysis is stable or not
\citep[e.g.][]{Braithwaite:2017aa}.
On the one hand,
it is well known from a theoretical point of view that
purely toroidal magnetic fields are unstable
against the Tayler instability \citep{Tayler:1973aa}.
In fact, we may estimate the minimum strength of the toroidal component that is required for the Tayler instability to operate by overcoming
the magnetic diffusion
%based on equation (8) of \citet{Spruit:2002aa} 
as
\begin{align}
B_{\phi,\min}^{(\mathrm{TI})}
\equiv
\left( 4 \pi r \rho {\Neff} \right)^{1/2}
\left( \eta \Omega \right)^{1/4}
\;,
\label{eq;Bphi_min}
\end{align}
in which $\eta$ and $\Omega$ stand for the magnetic diffusivity and the angular rotation rate, respectively.
Equation (\ref{eq;Bphi_min}) can be obtained by replacing $N$ by $\Neff$ in equation (8) of \citet{Spruit:2002aa}.
Here, $\Neff$ is the effective \BV\ frequency defined by
\begin{align}
\Neff^2
& \equiv
\frac{\eta}{\kappa} N_T^2 + N_{\mu}^2
\;,
\end{align}
in which $\kappa$ means the thermal diffusivity and $N_T$ and $N_{\mu}$ represent
the thermal and compositional part of the \BV\ frequency, respectively.
This replacement approximately takes into account the fact that the stabilising effect of thermal stratification becomes weaker due to thermal diffusion on the small spatial scales on which the Tayler instability occurs
\citep[see][]{Spruit:2002aa}.
The profile of 
$B_{\phi,\min}^{(\mathrm{TI})}$
is plotted in
Fig.~\ref{fig:Bphi_TI_min}.
It takes a maximum of $\sim 30\,\mathrm{kG}$ around
$r/R \sim 0.06$, which is below $\RMSphi$ given in Table~\ref{tab:fitting}.
This means that the detected field would be unstable against
the Tayler instability if we neglect the poloidal component.
On the other hand, from an observational point of view,
no detectable change in the oscillation frequencies
is found in \TheStar\ over the nearly four-year period of the \Kepler\ observation,
which is much longer than the Alfv\'en timescale (the characteristic scale of the system $l_{\mathrm{s}}$ divided by the Alfv\'en velocity) of $\sim 200\,\mathrm{d}$ for $B = 100\,\mathrm{kG}$, $\rho = 30\,\mathrm{g}\,\mathrm{cm}^{-3}$ and $l_{\mathrm{s}} = 0.1\,\times 2\pi R$.

These considerations lead to the following question:
can a purely toroidal field be stabilised by a small contribution of the poloidal component?
In fact, the theoretical study of \citet{Braithwaite:2009aa} gives
an affirmative answer to this question
because
{%
the poloidal component gives a restoring force
through bending of the field line
against the horizontal motion,
which is the main component of the Tayler instability.
That study}
provides the stability condition for axisymmetric magnetic fields as 
\begin{align}
\Bwa\,\frac{E}{U} &< \frac{E_{\mathrm{p}}}{E} \lesssim 0.8
\;,
\label{eq:stability_cond}
\end{align}
in which $E$, $U$ and $E_{\mathrm{p}}$ are the total magnetic energy, the gravitational energy and the energy of the poloidal component, respectively.
The parameter $\Bwa$ is on the order of $10$ for main-sequence stars.
Although the condition assumes a purely radiative structure with no convective zone, 
we nonetheless apply it to \TheStar, 
which has a small convective core (eight per cent in mass in model A), to check the field stability.
Using 
the results in Table~\ref{tab:fitting}
and
the structure of model A in Table~\ref{tab:models},
we obtain the following estimates:
\begin{align}
\frac{E}{U} & \sim 10^{-7}
\end{align}
and
\begin{align}
\frac{E_{\mathrm{p}}}{E} & \sim 10^{-3}
\;,
\end{align}
which clearly satisfy equation (\ref{eq:stability_cond}).
This means that the detected field in \TheStar\ is stable, if it is axisymmetric.

\subsubsection{Convective dynamo}
\label{sec:core_dynamo}

As for the second possibility,
three-dimensional numerical simulations by
\citet{2005ApJ...629..461B} and \citet{Hidalgo:2024aa}
have demonstrated 
that
{a} dynamo can {operate}
in the convective {cores} of A stars.
Although these simulations assume
a few (or more) times faster rotation rates
than that of \TheStar,
they show that
the magnetic energy can reach
at least the same order as the kinetic energy,
implying {an} average field strength
of several tens of kG in the convective core.
This number is slightly smaller than, but is still on the same order as our inference of $\RMSphi = \Bphiest\,\mathrm{kG}$ in the inner radiative region.
However, 
if we assume a crude estimate of 
\begin{align}
\left|
\frac{\Bphi}{\Br}
\right|
\sim
\frac{\RMSphi}{\RMSr}
\approx
30
\label{eq:Bphi_Br}
\end{align}
(see Table~\ref{tab:fitting}),
this large value is not realised in the convective core in the {cited} simulations.
The reason could possibly be given as follows.

If the field originates from the convective core,
we need some mechanism to move it to the radiative region.
Here, we may list two possibilities{:}
direct transport by overshooting 
%\textcolor{red}{or magnetic pumping}
at the top of the convective core,
or
the shrinking of the convective core with stellar evolution leaving the {generated} magnetic field
{in the radiative region}
near the outer edge of the convective core.
In either case,
if there exists rotational shear (radial differential rotation) at the convective/radiative boundary,
we may expect that the ${\Omega}$ effect operates to convert the radial component into the toroidal component in the radiative region.
Although \citetalias{Saio:2015aa} {estimated} that the degree of radial differential rotation of \TheStar\ between the core and the envelope is only a few per cent, this does not necessarily mean that the ${\Omega}$ effect is negligible because it is cumulative after many rotations, and the degree of radial differential rotation could have been higher in the past.

Interestingly,
the picture of the enhanced toroidal component
due to rotational shear
is supported by
the recent numerical simulations of
\citet{Ratnasingam:2024aa}
for a 7-$\mathrm{M}_{\sun}$ main-sequence star
with a rotation period of $4.04\,\mathrm{d}$
and a seed dipolar field of $\sim 1\,\mathrm{G}$.
These simulations with a higher mass and a shorter rotation period than \TheStar, which has the mass of $\sim 1.5\,\mathrm{M}_{\sun}$ and a rotation period of $64\,\mathrm{d}$,
provide a ratio of the toroidal field energy to the poloidal field energy that is comparable to that implied by equation (\ref{eq:Bphi_Br}),
although the asteroseismic analysis of \TheStar\ rejects the presence of such strong rotational shear 
as found in the simulations
in the near-core layers where the \BV\ frequency has a sharp peak.

Because the steep gradient of chemical composition
at the convective/radiative boundary
is generated by
the shrinking of the convective core along with evolution,
the layers of the steep gradient are those once in the convective core,
where the active dynamo process {was} in operation.
In addition,
since the steep gradient makes it difficult for the magnetic field to migrate to the outer region,
it is possible that the field is totally confined in those layers.
In this case,
we may set $\xup = 0.1$ and $\overline{\kd} = 0$
in equations (\ref{eq:Bphi_est_asymp_mod}) and (\ref{eq:Br_est_asymp_mod}) to obtain larger values of the lower bounds,
\begin{align}
\RMSphi & =
\Bphiestone
\,\mathrm{kG}
\end{align}
and
\begin{align}
\RMSr & =
\Brestone
\,\mathrm{kG}
\;,
\end{align}
respectively,
for model A with the glitch position of case 1.
These lead to an even larger ratio of
\begin{align}
\left|\frac{\Bphi}{\Br}\right|
\sim
\frac{\RMSphi}{\RMSr}
\approx
40
\;,
\end{align}
which means that the energy of the toroidal component is three orders of magnitude larger than that of the radial component.
It is obvious that a more detailed comparison of the asteroseismic result with three-dimensional simulations with the appropriate parameters for \TheStar\ is highly desirable.

%No signature of the periodic change
%of the field
%(at least, that below the four-year timescale
%by the Kepler data)?
%Stability of the oscillation frequencies
%can be used to constrain the periodicity
%of the magnetic activity.
%Highly time-dependent nature of the core dynamo
%in the numerical simulations
%is inconsistent with this.

\subsubsection{Radiative dynamo}
\label{sec:TS_dynamo}

A promising mechanism for the dynamo process
in the radiative region
is the one proposed by \citet{Spruit:2002aa},
which can be understood in the classical picture of the $\alpha$--${\Omega}$ dynamo. 
{Under} radial differential rotation,
the toroidal component of the magnetic field can be generated by stretching the (initially small) seed poloidal field (the ${\Omega}$ effect).
If the toroidal component becomes strong enough,
the Tayler instability sets in,
which essentially provides the $\alpha$ effect
to make the poloidal component from the toroidal component.
Since the predominantly toroidal field is still unstable against the Tayler instability, the process continues.
This mechanism (Tayler--Spruit dynamo) has received much attention, particularly because the magnetic field can transport angular momentum very efficiently in the radiative region of the stars \citep[e.g.][]{Cantiello:2014aa,Moyano:2023aa,Moyano:2024aa}.
However, the details of the mechanism are still under active debate, and more studies are clearly needed to understand its whole picture
\citep[e.g.][]{Zahn:2007aa,Fuller:2019aa,Petitdemange:2023aa,Petitdemange:2024aa}.

Having said all this, the first question to ask is whether the current structure of \TheStar\ satisfies the condition for the Tayler--Spruit dynamo or not.
\begin{figure}
\centering\includegraphics[width=\columnwidth]{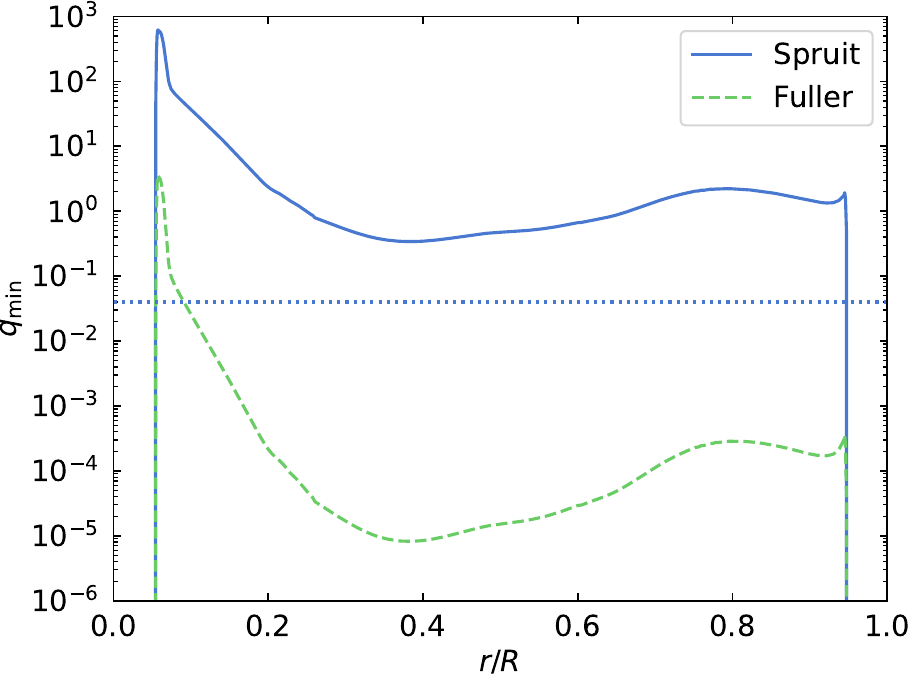}
\caption{The minimum values of the gradient of the rotation rate, $q$, defined by equation (\ref{eq:q_min}),
for the Tayler--Spruit dynamo to operate.
The structure of model A in Table~\ref{tab:models} is assumed with the rotation period of $64\,\mathrm{d}$.
Two cases are considered: One proposed by \citet{Spruit:2002aa} as in equation (\ref{eq:q_min_Spruit}) (solid line)
and the other by \citet{Fuller:2019aa} as in equation (\ref{eq:q_min_Fuller}) (dashed line).
The horizontal dotted line indicates the asteroseismic
estimate by \citetalias{Saio:2015aa}
(equation (\ref{eq:q_obs})).
}
\label{fig:q_min}
\end{figure}
In order to check this, we plot in Fig.~\ref{fig:q_min} the minimum values of the degree of differential rotation, $q$,
defined by
\begin{align}
q & \equiv
\left|
\frac{\mathrm{d}\ln\Omega}{\mathrm{d}\ln r}
\right|
\;,
\label{eq:q_min}
\end{align}
which is required to sustain the mechanism.
The original argument of \citet{Spruit:2002aa} provides
\begin{align}
q_{\min}^{\mathrm{Spruit}}
& =
\left( \frac{\Neff}{\Omega}\right)^{7/4}
\left( \frac{\eta}{r^2 \Neff} \right)^{1/4}
\;,
\label{eq:q_min_Spruit}
\end{align}
whereas \citet{Fuller:2019aa} reconsider the saturation mechanism 
{based on turbulent dissipation of
the perturbed field rather than the background toroidal field}
to present\footnote{For simplicity, we set $\alpha=1$ in equation (36) of \citet{Fuller:2019aa}.}
\begin{align}
q_{\min}^{\mathrm{Fuller}}
&
=
\left( \frac{\Neff}{\Omega}\right)^{5/2}
\left( \frac{\eta}{r^2 \Omega} \right)^{3/4}
\;.
\label{eq:q_min_Fuller}
\end{align}
On the other hand,
\citetalias{Saio:2015aa} constrain the difference between the core and envelope rotation rates, which implies
\begin{align}
q \approx 0.04
\;.
\label{eq:q_obs}
\end{align}
We thus find $q < q_{\min}^{\mathrm{Spruit}}$ in the entire radiative region, which means that the Tayler--Spruit dynamo does not work in \TheStar\ at least in its original form proposed by \citet{Spruit:2002aa}.
We also observe $q < q_{\min}^{\mathrm{Fuller}}$ near the peak around $r/R = 0.06$, which means that the Fuller-type mechanism can work only in the radiative region above the layer of the steep gradient of chemical compositions.

\begin{figure}
\centering
\includegraphics[width=\columnwidth]{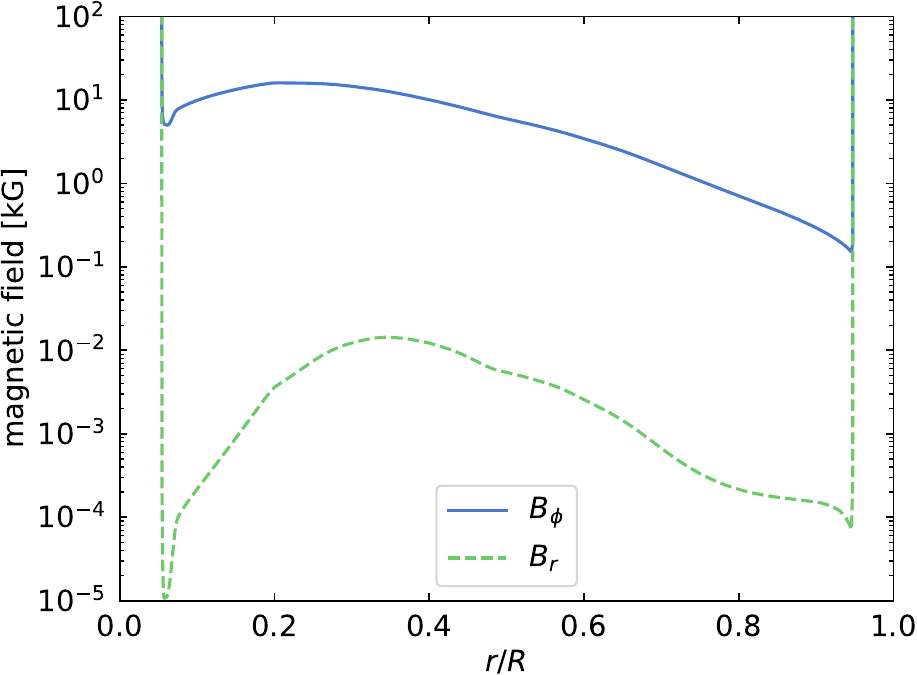}
\caption{%
Profiles of the radial ($\Br$) and azimuthal ($\Bphi$) components
of the magnetic field by the dynamo mechanism of \citet{Fuller:2019aa}.
The structure of model A in Table~\ref{tab:models} is assumed with the rotation period of $64\,\mathrm{d}$
and equation (\ref{eq:q_obs}) for the gradient of the rotation rate.}
\label{fig:B_Fuller}
\end{figure}

The next point to check is whether the field strengths detected are consistent with the prediction of the theory by \citet{Fuller:2019aa}.
We therefore plot in Fig.~\ref{fig:B_Fuller} the radial and azimuthal components of the field based on equations (23) and (22) of \citet{Fuller:2019aa},
assuming equation (\ref{eq:q_obs}) and the rotation period of $64\,\mathrm{d}$.
We confirm that
the predicted azimuthal component has its maximum of $16\,\mathrm{kG}$ around $r/R={0.20}$
{for $0.06 \le r/R \le 0.94$},
while the maximum of the radial component is found to be ${0.014}\,\mathrm{kG}$ around $r/R=0.34$ {in the same range}.
For both components,
the predicted values are well below the seismically estimated lower bound given in Table~\ref{tab:fitting}.
We neglect the divergent trends of $\Bphi$ and $\Br$ near the inner and outer boundaries of the radiative region.
This is justified by the fact that the trends come from the assumption of a finite differential rotation (equation (\ref{eq:q_obs})) even near the boundaries,
which is unrealistic because the angular momentum transport is so efficient as $\Neff\rightarrow 0$ that the differential rotation would disappear quickly.
We therefore conclude that the Fuller-type mechanism is not operative in \TheStar, either.

One remaining possibility is that
the star used to have strong differential rotation in the past, which was removed by efficient angular momentum transport by the magnetic field generated by the Tayler--Spruit dynamo.
After the differential rotation {subsided}, 
the field suffered from the Tayler instability, which could substantially modify the configuration to settle in with the detected strengths of the field components.
However, unless the structure of the star changes significantly, 
this scenario has difficulty explaining
the implied significant increase of the total magnetic energy as a result of the Tayler instability.
{Such significant structural change cannot come from the central condensation that occurs with single-star evolution. Only significant mass transfer or a merger would suffice, and the former is essentially ruled out by the non-detection of a companion (or remnant) via pulsation timing \citep{Murphy:2018aa} and the absence of eclipses or ellipsoidal variation in the light curve \citepalias{Saio:2015aa}.}
The final conclusion of this section is that
the dynamo mechanism in the radiative region is unlikely to be the origin of the detected field in \TheStar.

\subsubsection{Stellar merger}
\label{sec:merger}

One of the remarkable properties of \TheStar\ is that
its rotation period of $\sim 64\,\mathrm{d}$ is much longer than the typical period of $1\,\mathrm{d}$ of $\gamma$ Dor stars.
One possible explanation 
is that the star is a product of stellar merger.
In fact,
\kic{11145123},
which is another slowly rotating $\delta$ Sct--$\gamma$ Dor hybrid pulsator
with a similar mass and a rotation period of $\sim 100\,\mathrm{d}$,
is suspected to be a blue straggler
\citep{Kurtz:2014aa,Takada-Hidai:2017aa,Hatta:2021um}.
One reason for this is that
the best evolutionary model constructed by \citet{Kurtz:2014aa}
has an initial helium abundance of $0.36$, which is unusually high for a single-star model.
Although this is not the case for the models of \TheStar\  (see Table~\ref{tab:models}),
it would still be worth examining the merger or mass-transfer processes, which are possible mechanisms to form blue stragglers.

\citet{Schneider:2019aa} have numerically simulated such a merger process
to successfully explain the properties of the blue straggler $\tau$ Sco, which has a rotation period of $41\,\mathrm{d}$
and a surface magnetic field of a few hundred gauss.
They find that a strong internal magnetic field is generated by MRI during the merger,
and that
the merger product experiences
significant spin-down during the subsequent (thermal relaxation) phase,
before reaching the main-sequence position on the HR diagram.
Although the mass of their target ($\sim 15\,\mathrm{M}_{\sun}$) is much larger than \TheStar,
we may consider the possibility that the same physical processes are at work in both systems.
This hypothesis {can simultaneously explain (qualitatively)} 
the internal magnetic field
and the slow (and uniform) rotation,
{but it also implies}
that the star should have
a strong surface magnetic field, which is not supported by the present analysis
(as we discuss in Section~\ref{sec:fossil}).
In this context, we may note that
no surface magnetic field has been detected in \kic{11145123} by high-resolution spectroscopy \citep{Takada-Hidai:2017aa},
while \citet{Gizon:2016aa} conjecture the presence of a surface field much weaker than the Sun at its activity maximum based on the asymmetry of the frequency splittings of acoustic modes.
{
In order for the merger hypothesis to hold,
we need to explain how to confine the magnetic field in the stellar interior
(see Section \ref{sec:fossil})}.

\subsection{Comparison with the case of red giants}

%\subsubsection{Difference in the analysis}

The present result of asteroseismic detection of the internal magnetic field in \TheStar\ follows 
{those of} red-giant stars \citep{Li:2022aa,Deheuvels:2023aa,2024MNRAS.534.1060H}.
We compare the two cases in this section.

{
\citet{Li:2022aa} estimate that the detected $\BrRMS$ of 30 to 100 kG at the RGB stage should originate from that
between 3 and 5 kG at the main-sequence stage,
assuming magnetic flux conservation.
The fact that this range is consistent with our estimate of $\RMSr$ in Table~\ref{tab:fitting} supports the idea that the detected fields in \TheStar\ and red giants have the same physical origin.
}

{If we compare the analysis between the two cases,}
the main difference is that the present result can detect not only the radial component but also the azimuthal component of the field, whereas only the radial component is so far constrained in the red-giant case.
This is mainly because the gravity radial orders of $\sim 30$ of the detected modes in \TheStar\ are lower than those of $\gtrsim 100$ in stars at the red-giant branch (RGB) stage (except during the very early phase).
While mode frequencies are generally more sensitive to the radial component than to the azimuthal component
(Section~\ref{sec:mag_effect}),
the corresponding longer radial wavelengths (measured relative to the size of the gravity-mode cavity) lessen this effect in \TheStar.
As a result, the frequency perturbation due to the azimuthal component can be detected more easily in \TheStar.

On the other hand,
in the red-giant cases,
we can estimate the average field strength, $\BrRMS$,
although the analysis for \TheStar\ can provide only the lower bound, $\RMSr$.
This is because the red-giant analysis relies on the asymptotic frequency formula to fit individual frequencies, while the current analysis concentrates on the asymmetry of the frequency splitting.
In fact, a similar fitting of individual frequencies is not straightforward for \TheStar\ because of the significant contribution {of the glitch} (see Section~ \ref{sec:sph_glitch}).

\subsection{Implications on angular momentum transport}

The field strengths detected in \TheStar, which are given in Table~\ref{tab:fitting}, imply that the torque via Maxwell stresses, which is proportional to $\Br \Bphi$,
is larger by three orders of magnitude than predicted by the mechanism of \citet{Fuller:2019aa}.
The number would be even larger for the original Tayler--Spruit dynamo by \citet{Spruit:2002aa}.
Therefore, the angular momentum transport in \TheStar\ is much more efficient than described by the dynamo mechanisms associated with the Tayler instability in the radiative zone.
We may quantify this point based on the time scale.
Under the assumption that $\Omega$ depends only on the radius (shellular rotation),
the angular momentum transport in the radial direction due to the magnetic field can be formulated as a diffusion process \citep[e.g.][]{Maeder:2009aa}.
The corresponding diffusion coefficient (effective viscosity) is given by
\begin{align}
D_{\mathrm{mag}}
& \equiv
\frac{\left|\Br\Bphi\right|}{4\pi \rho q \Omega}
\;.
\end{align}
Then,
the time scale of the angular momentum transport for a characteristic length scale $l_{\mathrm{s}}$ can be estimated as
\begin{align}
& \tau_{\mathrm{mag}} \equiv \frac{l^2_{\mathrm{s}}}{D_{\mathrm{mag}}}
=
\frac{8\pi^2 \rho q l^2_{\mathrm{s}}}{\Prot\left|\Br\Bphi\right|}
\nonumber\\&
= 681 q
\left(\frac{\rho}{10^2\,\mathrm{g}\,\mathrm{cm}^{-3}}\right)
\left(\frac{\Prot}{64\,\mathrm{d}}\right)^{-1}
\left(\frac{l_{\mathrm{s}}}{\mathrm{R}_{\sun}}\right)^2
\left|\frac{\Br}{3.5\,\mathrm{kG}}\right|^{-1}
\left|\frac{\Bphi}{92\,\mathrm{kG}}\right|^{-1}
\mathrm{yr}\;,
\end{align}
in which $\Prot = 2\pi/\Omega$ means the rotation period
(as in equation (\ref{eq:t_rot_asymp}))
and $\rho=10^2\,\mathrm{g}\,\mathrm{cm}^{-3}$ is a good estimate of the density at the top of the convective core.
The length scale of $l_{\mathrm{s}} \approx \mathrm{R}_{\sun}$ corresponds to fifty per cent of the total radius of model A.
It is clear that, for \TheStar, $\tau_{\mathrm{mag}}$ is much shorter than the evolutionary time scale of $\sim 10^9\,\mathrm{yr}$.
This means that uniform rotation ($q=0$) is established very quickly in the layers with the magnetic field.

We may use this argument to reject, in a different way from that in Section \ref{sec:fossil}, the possibility that the magnetic field extends to the surface layers.
In this case, we may set $\xup=0.95$ (the base of the near-surface convective zone) to obtain $\RMSr = 2\,\mathrm{kG}$ and $\RMSphi = 13\,\mathrm{kG}$ (for model A).
Even with these lower values of the field strengths,
$\tau_{\mathrm{mag}}$ is still so short that the observed value of $q=0.04$ cannot be retained for the evolutionary time scale.
More precisely, 
assuming the two-zone structure,
\citetalias{Saio:2015aa} estimate that
the rotation rate in the inner forty percent in radius
is four per cent higher than that in the outer sixty percent.
The lower rotation rate in the outer region should have a significant contribution by the non-magnetic layers,
where there is no efficient transport of angular momentum.

Apart from the angular momentum transport inside the star, the slow rotation of \TheStar\ (with the period of $\sim 64\,\mathrm{d}$) needs to be explained separately
because the total angular momentum of the star is much smaller than typical $\gamma$ Dor stars,
which have rotation periods of $\sim 1\,\mathrm{d}$.
Since we interpret that there is no large-scale magnetic field at the surface, there is no chance for magnetic braking to operate.  Other possibilities include binary interaction and/or mass transfers, which are all speculative at this stage.

\subsection{Spherical counterpart of the aspherical glitch}
\label{sec:sph_glitch}

\begin{figure}
    \centering
    \includegraphics[width=\columnwidth]{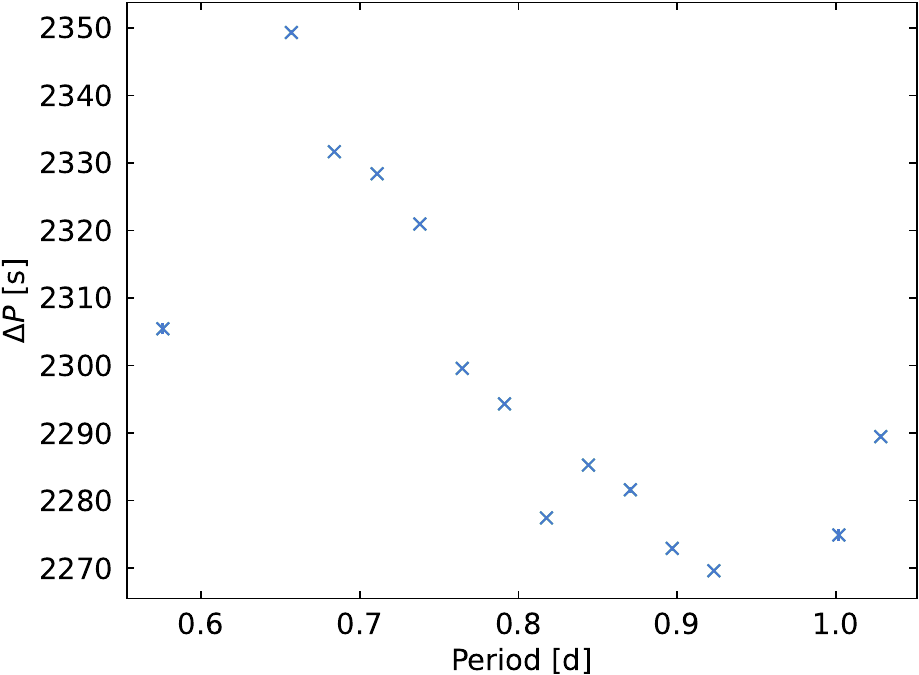}
    \caption{Diagram of period versus period spacing for high-order gravity modes of \TheStar.}
    \label{fig:P_dP}
\end{figure}

We may expect to identify
the spherical counterpart
of the aspherical glitch
in the period versus period spacing diagram,
which is reproduced in Fig.~\ref{fig:P_dP}.
%\citepalias[Fig.~9 of][]{Saio:2015aa}.
%
From equation (\ref{eq:a2k_H}), we can derive
the expression
for the period spacing as
\begin{align}
P_{n+1}-P_{n}
&
=
\Pi_1
+
\gla^{(0)}
\cos\left[
2\pi\left(
\glb
\frac{P_{n}+P_{n+1}}{2}
- \glc
\right)
\right]
\;,
\label{eq:delta_P}
\end{align}
in which the amplitude of
the oscillatory component is given by
\begin{align}
\gla^{(0)}
& \equiv
-\frac{\Pi_1 \mathfrak{D}_{0}^{(1)}}{4\sqrt{\pi^3}}
\sin\left(\pi\glb\Pi_1\right)
\;.
\label{eq:gla0}
\end{align}
Assuming that the amplitude of the spherical component of the glitch
is on the same order as the quadrupole component,
$\left|\mathcal{D}_{0}^{(1)} \right|\sim \left|\mathcal{D}_{1}^{(1)}\right|$,
we may use
$\mathcal{D}_{1}^{(1)}$ in equation (\ref{eq:D11_est}),
$\Pi_1$ in Table~\ref{tab:models}
and
$\glb$ in Table~\ref{tab:best_fit_params}
to obtain
\begin{align}
\left|\gla^{(0)}\right|
\sim 1\,\mathrm{s}
\;.
\label{eq:a0_est}
\end{align}
This estimate does not depend on the evolutionary model (models A or B or C)
or the glitch position (cases 1 or 2).

However, our preliminary analysis finds it
difficult to judge quantitatively
whether
the data contain the sinusoidal component given by the second term on the right-hand side of equation (\ref{eq:delta_P})
with the expected amplitude given by
equation (\ref{eq:a0_est}) or not.
This is because there are clearly multiple components with different amplitudes and periods.
In fact, the dominant oscillatory component has a much larger amplitude of $\sim 50$\,s
and a longer period of $\sim 0.5$\,d (see Fig.~\ref{fig:P_dP}).
This means that 
the nonlinearity 
(deviation from the sinusoidal function)
of the dominant component
is not negligible
when we discuss a period difference on the order of seconds.
Apart from the most dominant component,
we may observe at most that the second most dominant component shows an oscillatory behaviour with an amplitude on the same order as that given by equation (\ref{eq:a0_est})
and a period of $\sim 0.1\,\mathrm{d}$, which is consistent with $\glb^{
-1}=0.13\,\mathrm{d}$ for case 1 in Table~\ref{tab:fitting}.
Since more careful treatment is necessary,
we defer the detailed analysis to future work.

\section{Conclusions}
\label{sec:conclusions}

We have detected
an internal magnetic field
of \TheStar.
The lower bound
to the root-mean-squares
of the radial and azimuthal components
(with respect to the rotation axis)
within \xupestpercent\ per cent in radius
are estimated to be
$\Brest$\,kG
and
$\Bphiest$\,kG,
respectively.
The radial and azimuthal components
are more confined to the equator than to the poles.
The lower bound to the radial component
is clearly incompatible with
the prediction of the Taylor--Spruit dynamo,
suggesting
that the field could 
originate from other mechanisms,
such as convective-core dynamos,
fossil fields, and stellar mergers.
We have also identified a signature of an
aspherical discontinuous structure,
which is located in the layers of steep chemical composition {gradient}
just outside the convective core.
We suppose that
the structure is generated by some mixing processes at the boundary between the convective core and the radiative region.
%
%The search for 
%a similar signature
%in a sample of other ?? stars
%has ended in failure,
%which calls for
%increasing the size of the sample
%in the future study.
%
%This study has reconfirmed the potential of %asteroseismology.
%
The discovery of the predominantly toroidal magnetic field
has revealed a new aspect of magnetic fields in stars in general,
generating many questions
about their structure, origin and evolution.

\section*{Acknowledgements}

We thank NASA and the \Kepler\ team for their revolutionary data. 
We also acknowledge the referee for constructive comments.
MT thanks 
Ludovic Petitdemange,
Facundo D.~Moyano
and
Jim Fuller
%
% Rhita, Chloe, Coralie
%
for their instructive and insightful comments about the present work.
He is grateful
to
Rhita-Maria Ouazzani, Benoit Mosser and
their colleagues
at LIRA (formerly LESIA) of the Paris Observatory
and
Charly Pin\c{c}on 
at Universit\'e Paris-Saclay, Institut d'Astrophysique Spatiale
for a lot of active scientific discussions under their warm hospitality
and
financial support
based on
the visiting fellow programme of the Paris Observatory
in 2023 and 2025.
This research was also supported by the Munich Institute for Astro-, Particle and BioPhysics (MIAPbP) which is funded by the Deutsche Forschungsgemeinschaft (DFG, German Research Foundation) under Germany's Excellence Strategy -- EXC-2094 -- 390783311. 
MT was partially supported by JSPS KAKENHI Grant Numbers 
JP18K03695 and JP22K03672.
{SJM was supported by the Australian Research Council through Future Fellowship FT210100485.}

%%%%%%%%%%%%%%%%%%%%%%%%%%%%%%%%%%%%%%%%%%%%%%%%%%
\section*{Data Availability}

The data underlying this article will be shared on reasonable request to the corresponding author.



\bibliographystyle{mnras}
\bibliography{references} % if your bibtex file is called example.bib

%\begin{thebibliography}{99}
%\end{thebibliography}

%%%%%%%%%%%%%%%%%%%%%%%%%%%%%%%%%%%%%%%%%%%%%%%%%%

%%%%%%%%%%%%%%%%% APPENDICES %%%%%%%%%%%%%%%%%%%%%

\appendix

\section{Asymmetric frequency splittings of high-order gravity modes caused by the magnetic field}
\label{sec:mag_pert}

\subsection{Guidelines of the analysis}
\label{sec:guidelines}

We develop a regular perturbation analysis
to obtain an expression
for the frequency change
of high-order and low-degree gravity modes
due to slow rotation and a weak magnetic field that is confined in the stellar interior.
Following \citet{Li:2022aa},
the present analysis is designed to be
independent of the field configuration.
One of our fundamental assumptions is that
the rotation effect dominates
over the magnetic effect.
In this case, we may consider the rotation effect first,
and then take the magnetic effect into account in the second step.
An advantage of this approach is that the rotation perturbation lifts the degeneracy among 
the eigenmodes in the
spherically symmetric case
with respect to the azimuthal order, $m$,
so that we may apply the non-degenerate perturbation theory in the second step.
Then, the main point of the analysis
is to calculate the matrix element of the magnetic operator.
Since the expression for the matrix element
can be highly complicated,
we adopt its symmetric form
\citep[see][]{Kovetz:1966tr,Glampedakis:2007aa}.
We list two advantages of this approach:
(1) the variational principle can be applied;
(2) the frequency perturbations
are clearly real.

In general,
the magnetic effect
consists of
two aspects.
Firstly, the equilibrium structure
is deformed by
the Lorentz force.
This structure change contributes to the frequency change.
Then, oscillations occur
about the deformed structure
under the direct influence
of the Lorentz force.
Following \citet{Gough:1990tj},
we refer to the former and the latter as the indirect and direct effects, respectively.

\subsection{Oscillation equations in the presence of the magnetic field}

We start from the equation of motion
of a uniformly rotating magnetized fluid
in the rotating frame,
\begin{equation}
    \rho
\frac{\rd\boldsymbol{v}}{\rd t}
=
-\nabla p
-\rho \nabla\Phi
+ \frac{1}{\lspeed}
\bj\times\bB
- \rho
\left[
2 \bOmega\times\boldsymbol{v}
+ \bOmega\times\left(
\bOmega\times\boldsymbol{r}\right)
\right]
\;,
\label{eq:eq_of_motion}
\end{equation}
in which 
$\rd/\rd t$ is the Lagrangian time derivative,
$\boldsymbol{v}$ the velocity,
$p$ the pressure,
$\Phi$ the gravitational potential,
$\lspeed$ the speed of light in vacuum, 
$\bj$ the electric current density,
$\bB$ the magnetic field,
$\bOmega$ the rotation vector
and
$\boldsymbol{r}$ the position vector.
The third term on the right-hand side
of equation (\ref{eq:eq_of_motion})
stands for the Lorentz force,
while
the fourth and fifth terms 
represent the Coriolis and centrifugal forces,
respectively.
We assume that
$\bj$ is related to $\bB$
by
Ampere's law
(or the
Ampere--Maxwell equation 
neglecting
the displacement current),
\begin{equation}
    \nabla\times\bB
    =
    \frac{4\pi}{\lspeed}
    \bj
    \;.
    \label{eq:Ampere}
\end{equation}
Equation (\ref{eq:eq_of_motion}) can be recast as
\begin{align}
   \rho
\frac{\rd v_i}{\rd t}
=
-\partial_j \Pi_{j,\,i}
-\rho \partial_i \Psi
- 2\rho
\left( \bOmega\times\boldsymbol{v} \right)_i
\;,
\label{eq:eq_motion_alt}
\end{align}
where
subscripts $i$ and $j$ mean
the $i$-th and $j$-th components of the Cartesian coordinates, respectively.
We follow the Einstein summation convention
for repeated indices.
In equation (\ref{eq:eq_motion_alt}),
tensor $\Pi_{i,j}$ and scalar $\Psi$ are introduced by
\begin{equation}
\Pi_{i,j}
\equiv
\delta_{i,j}
\left(
p + \frac{\bB^2}{8\pi}
\right)
-
\frac{B_i B_j}{4\pi}
\label{eq:Pi_def}
\end{equation}
and
\begin{equation}
\Psi = \Phi - \frac{1}{2}
\left|\bOmega\times\br\right|^2
\;.
\end{equation}
In equation (\ref{eq:Pi_def}), $\delta_{i,j}$ is the Kronecker delta.
Assuming that
there is no velocity field in the equilibrium structure
(in the rotating frame),
we can linearize
equation (\ref{eq:eq_motion_alt})
to derive the oscillation equation
with respect to
a small displacement vector $\bxi$, 
\begin{equation}
\sigma^2 \rho \xi_i = 
\sigma C_{i,j} \xi_j
+
T_{i,j} \xi_j
\;,
\label{eq:osc_eq}
\end{equation}
in which 
we have assumed that $\bxi$
depends on time $t$ as
$\exp\left(-\text{i}\sigma t\right)$
with an angular frequency $\sigma$.
The tensorial operators $T_{i,j}$
and $C_{i,j}$
in equation (\ref{eq:osc_eq})
are defined by
\begin{equation}
T_{i,j}\xi_j
\equiv
\rho
\delta\left(\frac{1}{\rho}
\partial_j \Pi_{j,i}\right)
+
\rho \delta\left(\partial_i\Psi\right)
\label{eq:Tij_def}
\end{equation}
and
\begin{equation}
C_{i,j}\xi_j
\equiv
-2 \mathrm{i} \rho
\left(
\bOmega\times\bxi
\right)_i
\;.
\end{equation}
In equation (\ref{eq:Tij_def}),
$\delta$ is the operator for the Lagrangian perturbation.
In order to relate
the perturbed quantities 
in equation (\ref{eq:Tij_def})
to $\bxi$,
we need to use the linearized versions of
the continuity equation,
the Poisson equation
and
the adiabatic relation
between the Lagrangian pressure and density
perturbations.
In addition, we also use
the linearized induction equation of 
(ideal) magnetohydrodynamics,
\begin{equation}
\delta\bB\left[\bxi\right]
=
\left(\bB\cdot\nabla\right)\bxi
-
\left(\nabla\cdot\bxi\right)\bB
\end{equation}
\citep[e.g.][]{Roberts:1967aa}.

\subsection{Matrix element}
It can be shown that
the operator $T_{i,j}$ is symmetric in the sense,
\begin{equation}
\mathcal{T}
\left(\hbxi,
\bxi\right)
=
\left[
\mathcal{T}
\left(
\bxi,
\hbxi
\right)
\right]^*
\;,
\label{eq:calT_symmetry}
\end{equation}
where ${}^*$ indicates the complex conjugate.
Here, $\mathcal{T}$ is defined by
\begin{equation}
\mathcal{T}
\left(\hbxi,
\bxi\right)
\equiv
\int \hat{\xi}^*_i T_{i,j} \xi_j \,\dV
\;,
\label{eq:symmetry}
\end{equation}
in which
the domain of the integral is the entire stellar volume.
Equation (\ref{eq:calT_symmetry})
holds
for any $\bxi$ and $\hbxi$
that satisfy the proper boundary conditions.
The proof is almost the same as that
given by \citet{Kovetz:1966tr},
except that $\Phi$ should be replaced with $\Psi$.
Since
$C_{i,j}$ is also symmetric
\citep{LO1967},
the total operator 
$\sigma C_{i,j} + T_{i,j}$
on the right-hand side of
equation (\ref{eq:osc_eq}) is symmetric as a whole.
This property serves as a basis
of the variational principle.

The symmetric form of $\mathcal{T}$ is composed of three parts,
\begin{equation}
\mathcal{T}
\left(\hbxi,
\bxi\right)
=
\mathcal{L}
\left(\hbxi,
\bxi\right)
+
\mathcal{B}
\left(\hbxi,
\bxi\right)
+
\mathcal{R}
\left(\hbxi,
\bxi\right)
\;.
\label{eq:Texp}
\end{equation}
Here,
$\mathcal{L}$ and
$\mathcal{B}$
correspond to
the part that does not depend on $\bB$ directly,
and
that explicitly includes it,
respectively,
while $\mathcal{R}$ stands for
the contribution coming from the surface
and the exterior of the star.
They are further decomposed as
\begin{equation}
\mathcal{L}
\left(\hbxi,
\bxi\right)
=
\mathcal{S}_L
\left(\hbxi,
\bxi\right)
+
\mathcal{A}_L
\left(\hbxi,
\bxi\right)
+
\left[
\mathcal{A}_L
\left(\bxi,
\hbxi\right)
\right]^*
\;,
\label{eq:Lexp}
\end{equation}
\begin{equation}
\mathcal{B}
\left(\hbxi,
\bxi\right)
=
\mathcal{S}_B
\left(\hbxi,
\bxi\right)
+
\mathcal{A}_B
\left(\hbxi,
\bxi\right)
+
\left[
\mathcal{A}_B
\left(\bxi,
\hbxi\right)
\right]^*
\label{eq:Bexp}
\end{equation}
and
\begin{equation}
\mathcal{R}
\left(\hbxi,
\bxi\right)
=
\mathcal{S}_R
\left(\hbxi,
\bxi\right)
+
\mathcal{A}_R
\left(\hbxi,
\bxi\right)
+
\left[
\mathcal{A}_R
\left(\bxi,
\hbxi\right)
\right]^*
\;,
\label{eq:Rexp}
\end{equation}
where
$\mathcal{S}$
and
$\mathcal{A}$
mean
the symmetric and asymmetric parts, respectively. 
The definitions of
$\mathcal{S}_L$, 
$\mathcal{A}_L$ 
$\mathcal{S}_B$,
$\mathcal{A}_B$,
$\mathcal{S}_R$
and $\mathcal{A}_R$
are provided by
\begin{align}
\mathcal{S}_L
\left(\hbxi,
\bxi\right)
&\equiv
\int 
\Gamma_1 p
\left(\nabla\cdot\hbxi^*\right)
\left(\nabla\cdot\bxia\right)
\;
\dV
\nonumber\\&\phantom{=}\mbox{}
- G
\iint
\left[
\rho\left(\boldsymbol{r}_{\ra}\right)
\hbxi^*\left(\boldsymbol{r}_{\ra}\right)
\cdot
\nabla_{\ra}
\right]
\left[
\rho\left(\boldsymbol{r}_{\rb}\right)
\bxia\left(\boldsymbol{r}_{\rb}\right)
\cdot
\nabla_{\rb}
\right]
\nonumber\\&\phantom{=
- G
\iint
}\mbox{}
\times
\left(
\frac{1}{\left|
\boldsymbol{r}_{\ra}
-
\boldsymbol{r}_{\rb}
\right|}
\right)\;
\dV_{\ra}\,
\dV_{\rb}
\;,
\label{eq:S_part}
\end{align}
\begin{align}
\mathcal{A}_L
\left(\hbxi,
\bxi\right)
&\equiv
-
\int
\left(
\hbxi^*\cdot\nabla \Psi
\right)
\left(
\rho
\nabla\cdot\bxi
+
\frac{1}{2}\,
\bxi\cdot\nabla\rho
\right)
\;
\dV
\;,
\end{align}
\begin{align}
\mathcal{S}_B
\left(\hbxi,
\bxi\right)
&\equiv
\frac{1}{4\pi}
\int 
\delta\bB\left[\hbxi^*\right]
\cdot
\delta\bB\left[\bxia\right]
\;
\dV
\;,
\end{align}
\begin{align}
\mathcal{A}_B
\left(\hbxi,
\bxi\right)
&\equiv
\frac{1}{4\pi}
\int
\biggl[
\left(
\nabla\cdot\hbxi^*
\right)
\left(
\bxi\cdot
\left[
\left(
\bB\cdot\nabla
\right)
\bB
\right]
\right)
\nonumber\\&\phantom{=
\frac{1}{4\pi}
\int
\biggl[}\mbox{}
+
\frac{1}{2}\,
\hbxi^*\cdot
\left\{
\left(\bxi\cdot\nabla\right)
\left[
\left(\bB\cdot\nabla\right)\bB
\right]
\right\}
\biggr]
\;
\dV
\;,
\label{eq:AB_part}
\end{align}
\begin{align}
\mathcal{S}_R
\left(\hbxi,
\bxi\right)
&\equiv
\int
\left(\bn\cdot\hbxi^*\right)
\left(\bn\cdot\bxi\right)
\left(\bn\cdot\nabla\right)
\left[
\frac{\bBex^2}{8\pi}
-
\left(
p + \frac{\bB^2}{8\pi}
\right)
\right]
\;\dS
\nonumber\\&\phantom{=}\mbox{}
+
\int_{\Vex}
\frac{\bBex'\left[\hbxi^*\right]\cdot\bBex'\left[\bxia\right]}{4\pi}
\;\dV
\label{eq:S_R}
\end{align}
and
\begin{align}
\mathcal{A}_R
\left(\hbxi,
\bxi\right)
&\equiv
-\frac{1}{8\pi}
\int
\left(\bn\cdot\bB\right)
\hbxi^*\cdot\left[
\left(\bxi\cdot\nabla\right)\bBex
\right]
\;\dS
\;.
\label{eq:A_R}
\end{align}
In equation (\ref{eq:S_part}),
$\Gamma_1$ is the first adiabatic index,
while
$\bn$ in
equation (\ref{eq:S_R})
is
the unit normal vector of the stellar surface.
In equations (\ref{eq:S_R}) and (\ref{eq:A_R}),
$\bBex$ and $\bBex'\left[\bxi\right]$
mean the magnetic field in the exterior of the star
and its Eulerian perturbation
that is induced by displacement $\bxi$ in the interior.
The volume integral in
equation (\ref{eq:S_R}) is performed over
the whole region outside the star,
while the surface integrals in
equations (\ref{eq:S_R}) and (\ref{eq:A_R}) are
over the entire surface of the star.
When we derive
equations (\ref{eq:S_part})--(\ref{eq:AB_part}),
we have used the equilibrium relation,
\begin{align}
-\nabla 
\left(
p + \frac{\bB^2}{8\pi}
\right)
+ \frac{\left(\bB\cdot\nabla\right)\bB}{4\pi}
- \rho\nabla\Psi 
= 0
\;.
\label{eq:eq_cond}
\end{align}
Utilizing this relation, we can show that
equation (\ref{eq:Texp}) is equivalent to
equation (39) of \citet{Kovetz:1966tr}.%
\footnote{%
For comparison,
we need to replace $U$ with $-\Psi$
in the equation.
In addition, we believe that {in \citet{Kovetz:1966tr}}
the second term in the fourth line
has a sign error,
and that 
$B^2/(8\rho)$
in the sixth line
means
$B^2/(8\pi)$.
}

In addition to the absence of the magnetic field
at the surface and the exterior,
we also assume that the density is equal to zero
at the surface in the equilibrium structure.
Under these assumptions, we simply obtain
\begin{equation}
\mathcal{R}\left(\hbxi,\bxi\right) = 0
\;.
\end{equation}

\subsection{Frequency perturbation}

Equation (\ref{eq:osc_eq}) is a general expression
that assumes neither slow rotation nor a weak magnetic field.
The next step is to regard them
as small perturbations.
Although equation (\ref{eq:osc_eq})
concerns oscillations about
the (rotationally and magnetically) deformed equilibrium structure,
we may reinterpret this equation
as describing oscillations about
the non-rotating and non-magnetic structure
with two different kinds of perturbations,
the contributions to the restoring force and
the deformation.
Since we consider the major effect of rotation separately
in the first step,
we study here the frequency perturbation due only to the magnetic effect.
We can then separate
the quantities in
equation (\ref{eq:osc_eq})
into
the unperturbed part (with subscript $0$)
and
the perturbed part (with subscript $1$).
Thanks to the symmetric property of $T_{i,j}$,
we can derive an expression for the frequency perturbation as
\begin{equation}
\sigma_1
=
\frac{
\mathcal{T}_1\left(\bxi_0,\bxi_0\right)
-\sigma_0^2 
\int \rho_1 \left|\bxi_0\right|^2\;\dV
}{%
2 \sigma_0 \mathcal{I}\left(\bxi_0,\bxi_0\right)
}
\;,
\label{eq:sigma1}
\end{equation}
in which we have introduced
\begin{equation}
\mathcal{I}\left(\hbxi,\bxi\right)
\equiv
\int \rho_0 \hbxi^*\cdot\bxi\;\dV
\;.
\label{eq:calI}
\end{equation}
We implicitly assume in equation (\ref{eq:sigma1}) that the third- and higher-order effects of rotation
are negligible.
Strictly speaking,
as we explain in Appendix
\ref{sec:guidelines},
the unperturbed eigenfunctions should be those
after taking the rotation effect into account.
However,
since we analyse only the leading-order effect,
we may neglect the perturbation to $\bxi_0$
due to rotation, and
adopt as $\bxi_0$ the eigenfunctions
of the spherically symmetric structure.
In this case, $\bxi_0$ can be expressed 
in the spherical coordinates $\left(r,\theta,\phi\right)$
with the origin ($r=0$) set at the centre of the star
and the direction of $\theta = 0$ aligned with the rotation axis
as
\begin{equation}
\bxi_0 =
\xirn(r) \Ylm\left(\theta,\phi\right) \ber
+
\xihn(r) \nabYlm\left(\theta,\phi\right)
\;,
\label{eq:xi0}
\end{equation}
in which
$\Ylm\left(\theta,\phi\right)$
is the spherical harmonic
with the angular degree $\ell$
and the azimuthal order $m$.
While $\ber$ is the
unit vector in the radial direction,
the horizontal gradient operator $\nabperp$
is defined by
\begin{equation}
\nabperp 
=
\betheta\,\pdtheta{}
+
\bephi\,\frac{1}{\sin\theta}\pdphi{}
\;,
\end{equation}
where $\betheta$ and $\bephi$ are the unit vectors in the $\theta$ and $\phi$ directions, respectively.
The functions $\xirn$ and $\xihn$ depend not only on $r$ but also on $\ell$ and the radial order $n$.
The corresponding eigenfrequency $\sigma_0$ is also dependent on $n$ and $\ell$.

The matrix element
$\mathcal{T}_1\left(\bxi_0,\bxi_0\right)$
is obtained by perturbing equation
(\ref{eq:Texp}) as
\begin{equation}
    \mathcal{T}_1
\left(\bxi_0,
\bxi_0\right)
=
\mathcal{L}_1
\left(\bxi_0,
\bxi_0\right)
+
\mathcal{B}
\left(\bxi_0,
\bxi_0\right)
\;,
\label{eq:T1}
\end{equation}
in which 
$\mathcal{L}_1$
can in turn be derived by perturbing
equation (\ref{eq:Lexp}) as
\begin{equation}
\mathcal{L}_1
\left(\bxi_0,
\bxi_0\right)
=
\mathcal{S}_{L,1}
\left(\bxi_0,
\bxi_0\right)
+
2 \Re\left[
\mathcal{A}_{L,1}
\left(\bxi_0,
\bxi_0\right)
\right]
\;.
\label{eq:L1exp}
\end{equation}
Here, $\mathcal{S}_{L,1}$
and $\mathcal{A}_{L,1}$ are provided by
\begin{align}
&
\mathcal{S}_{L,1}
\left(\bxi_0,
\bxi_0\right)
\nonumber\\
&
\equiv
\int 
\left(\Gamma_1 p\right)_1
\left|\nabla\cdot\bxi_0\right|^2
\;
\dV
\nonumber\\&\phantom{=}\mbox{}
- 2 G
\Re\Biggl\{
\iint
\left[
\rho_0\left(\boldsymbol{r}_{\ra}\right)
\bxi^*_0\left(\boldsymbol{r}_{\ra}\right)
\cdot
\nabla_{\ra}
\right]
\left[
\rho_1\left(\boldsymbol{r}_{\rb}\right)
{\bxi_0}\left(\boldsymbol{r}_{\rb}\right)
\cdot
\nabla_{\rb}
\right]
\nonumber\\&\phantom{=
- G
\iint
}\mbox{}
\times
\left(
\frac{1}{\left|
\boldsymbol{r}_{\ra}
-
\boldsymbol{r}_{\rb}
\right|}
\right)\;
\dV_{\ra}\,
\dV_{\rb}
\Biggr\}
\label{eq:SL1}
\end{align}
and
\begin{align}
%&
\mathcal{A}_{L,1}
\left(\bxi_0,
\bxi_0\right)
%\nonumber\\
&\equiv
-
\int
\biggl\{
\left(\bxi^*_0\cdot\nabla\Phi_1\right)
\left(
\rho_0 \nabla\cdot\bxi_0
+
\frac{1}{2}\,\bxi_0\cdot\nabla\rho_0
\right)
\nonumber\\&\phantom{=\int\Bigl\{
}\mbox{}
+
\left(\bxi^*_0\cdot\nabla\Phi_0\right)
\left(
\rho_1 \nabla\cdot\bxi_0
+
\frac{1}{2}\,\bxi_0\cdot\nabla\rho_1
\right)
\biggr\}
\;
\dV
\;.
\end{align}
In equation (\ref{eq:T1}),
$\mathcal{B}$
corresponds to
the direct effect
of the magnetic field, 
while 
$\mathcal{L}_1$ 
and
the second term in the numerator of
equation (\ref{eq:sigma1})
describe the indirect effect.

\subsection{Dominant terms for high-order and
low-degree
gravity modes}

We evaluate equation (\ref{eq:sigma1})
for high-order and low-degree gravity modes.
A primary parameter is
\begin{equation}
\epsilon = \frac{\sigma_0}{N(r_*)} \ll 1
\;,
\end{equation}
in which $r_*$ is a representative radius
in the propagative region of gravity waves.
We also assume that
\begin{equation}
 \frac{\sigma_0 r_*}{c} \ll 1
 \;,
\end{equation}
in which $c$ is the sound speed. 
According to the asymptotic analysis
\citep[e.g.][]{UOASS1989},
the horizontal displacement $\xihn$
dominates over the radial displacement $\xirn$
by factor $\epsilon^{-1}$.
Because
the wavelength at radius $r$
is on the order of $\epsilon r$,
the first derivatives of the eigenfunctions
$\bxi$ with respect to $r$
are evaluated to be $\sim \epsilon^{-1} r^{-1} \bxi$.
Exceptionally, 
the divergence, $\nabla\cdot\bxi$, is only on the order of $r^{-1} \xirn$, which
reflects that
high-order gravity waves are almost uncompressed.

As for the field configuration,
we assume 
\begin{equation}
\epsilon \lesssim
\frac{\left|\Br\right|}
{\left|\Bphi\right|}
\ll
1
\label{eq:B_condition}
\end{equation}
and
$\left|\Btheta\right|\sim\left|\Bphi\right|$,
the latter of which will be changed at the last step.

Under these assumptions,
we retain
only the terms that can be
on the order of
$\epsilon^{-2} \xihn^2 \Br^2$ or
$\xihn^2 \Btheta^2$
or
$\xihn^2 \Bphi^2$
in the numerator 
on the right-hand side
of equation (\ref{eq:sigma1}).
We may thus identify the dominant terms
in $\mathcal{B}\left(\bxi_0, \bxi_0\right)$ as
\begin{align}
%&
\mathcal{B}\left(\bxi_0, \bxi_0\right)
%\nonumber\\
&
\approx
\frac{1}{4\pi}
\int 
\Bigl(
\left|
\left(\bB\cdot\nabla\right)
\bxi_0
\right|^2
\nonumber\\&\phantom{%
=\frac{1}{4\pi}\int}\mbox{}
+ 
\Re\left[
\bxi^*_0\cdot
\left\{
\left(\bxi_0\cdot\nabla\right)
\left[
\left(\bB\cdot\nabla\right)\bB\right]
\right\}
\right]
\Bigr)
\;
\dV
\;,
\label{eq:B_dom}
\end{align}
and confirm that
there is no contribution from
$\mathcal{L}_1\left(\bxi_0, \bxi_0\right)$
and
the second term
in the numerator of equation (\ref{eq:sigma1}).
This means that the indirect effects of the
magnetic field is not important for
high-order gravity modes
even in the order that we consider in the present analysis
\citep[see][]{Mathis:2023aa}.
As a result, equation (\ref{eq:sigma1}) is reduced to
\begin{equation}
\sigma_1
\approx
\frac{
\mathcal{B}\left(\bxi_0,\bxi_0\right)
}{%
2 \sigma_0 \mathcal{I}\left(\bxi_0,\bxi_0\right)
}
\;.
\label{eq:sigma1_red}
\end{equation}

We remark on the second term
on the right-hand side of equation
(\ref{eq:SL1}),
which appears to be on the order of $\xihn^2 \Bphi^2$.
However,
this integral includes the Eulerian perturbation to the gravitational potential, $\Phi'_{\bxin}$,
which is given by
\begin{equation}
\Phi'_{\bxin}\left(\br\right)
=
-G
\int
\rho\left(\br_{\ra}\right)
\bxi\left(\br_{\ra}\right)
\cdot\nabla_{\ra}
\left(
\frac{1}{\left|\br - \br_{\ra}\right|}
\right)
\;
\dV_{\ra}
\;.
\end{equation}
It is well established that
$\Phi'_{\bxin}$ is negligibly small
for high-order gravity modes
\citep{Cowling1941}.

\subsection{Evaluation of the dominant terms}
\label{sec:dom_terms}

\subsubsection{Angular integrals}
The main result of this section
is 
the expression for
the angular integral in equation
(\ref{eq:B_dom}),
which is provided by
\begin{align}
&
\int_{4\pi}
\left(
\left|
\left(\bB\cdot\nabla\right)
\bxi_0
\right|^2
%\nonumber\\&\phantom{=}\mbox{}
+ \Re\left[
\bxi^*_0\cdot
\left\{
\left(\bxi_0\cdot\nabla\right)\left[
\left(\bB\cdot\nabla\right)\bB
\right]
\right\}
\right]
\right)
\;
\dOmega
\nonumber\\
&
\approx
\left(\tdr{\xihn}\right)^2
\int_{4\pi}
\Br^2 \Dlm \;\dOmega
%\nonumber\\&\phantom{=}\mbox{}
+
\left(\frac{\xihn}{r}\right)^2
\int_{4\pi}
\left(
\Btheta^2 \Flm
+
\Bphi^2 \Glm
\right)
\dOmega
\;,
\label{eq:Bangint}
\end{align}
where
$\Dlm$, $\Flm$ and $\Glm$
are all functions of $\theta$.
{
In equation (\ref{eq:Bangint}),
$\dOmega$ means an infinitesimal element of a solid angle, given by $\dOmega = \sin\theta\,\rd\theta\,\rd\phi$.
}
Their derivation is lengthy, but straightforward
if we use
generalised spherical harmonics
\citep{Gelfand1956},
which are widely used in geophysics
\citep[e.g.][]{Phinney:1973aa,Dahlen:1999aa}
and have also been
adapted to helio- and asteroseismic problems
\citep[e.g.][]{Ritzwoller:1991aa,Hanasoge:2017aa,Kiefer:2017aa,Das:2020aa}.
The results are summarised as follows:
\begin{align}
 \Dlm
&=
\sum_{k=0}^{\ell}
U_{\ell,\,k}
\left(
K - 2 L
\right)
\oP{2k}{\ell}{m}
P_{2 k}\left(\cos\theta\right)
\;,
\label{eq:Dlm}
\\
 \frac{F_{\ell}^{m} + G_{\ell}^{m}}{2}
& =
\sum_{k=0}^{\ell}
U_{\ell,\,k}
\left[
-\left( K - 2 L \right)
\left( K + 1 \right)
- 2 L^2
\right]
\nonumber\\&\phantom{=}\mbox{}\times
\oP{2k}{\ell}{m}
%\sqrt{\frac{4\pi}{4k+1}}\GSH{2k}{0}{0} 
P_{2k}\left(\cos\theta\right)
\label{eq:FGsum}
\\
\noalign{\noindent\mbox{and}}
 \frac{F_{\ell}^{m} - G_{\ell}^{m}}{2}
& =
\sum_{k=1}^{\ell}
U_{\ell,\,k}\,
\frac{
K \left( K + 2 L - 1 \right)
- 4 L \left( L + 1 \right)
}{4\left(K - 1\right)}\,
\nonumber\\&\phantom{=}\mbox{}\times
\oP{2k}{\ell}{m}
P_{2k}^{2}\left(\cos\theta\right)
\;,
\label{eq:FGsub}
\end{align} 
in which we have introduced
\begin{align}
U_{\ell,\,k}
&
\equiv
\left(-1\right)^{k+1}
\frac{\left(4k+1\right)\left(2k\right)!\left(2\ell+1\right)!\left(\ell+k\right)!}%
{4\pi \ell \left(k!\right)^2
\left(2\ell+2k+1\right)!
\left(\ell-k\right)!}
\;,
\\
 K & \equiv  k \left( 2k + 1\right)
 \;,
\\
 L & \equiv \frac{\ell\left(\ell+1\right)}{2}
\\
\noalign{\noindent\mbox{and}}
\oP{j}{\ell}{m}
&\equiv
\left(-1\right)^{-\ell+m}
\frac{\ell
\sqrt{\left(2\ell-j\right)! \left(2\ell+j+1\right)!}
}{\left(2\ell\right)!}
\wtj{\ell}{\ell}{j}{m}{-m}{0}
\;.
\label{eq:Pjlm}
\end{align}
Note that
$P_{2k}$ 
in equations (\ref{eq:Dlm})
and (\ref{eq:FGsum})
are the Legendre polynomials,
while
$P_{2k}^2$ in equation (\ref{eq:FGsub})
are the associated Legendre functions.
The last factor on the right-hand side of
equation (\ref{eq:Pjlm}) is Wigner's 3-$j$ symbol.
The coefficients
$\oP{j}{\ell}{m}$
are polynomials in $m$ of degree $j$,
which form a basis to represent
the frequency splitting as
\begin{align}
\nu\left(n,\ell,m\right)
&
=
\sum_{j=0}^{2\ell}
a_j\left(n,\ell\right)
\oP{j}{\ell}{m}
\;.
\end{align}
Here, 
$a_j$ is called the a-coefficient of order $j$.
The asymmetry in the frequency splittings defined by
equation (\ref{eq:t_n_obs}) is related to $a_2$ as
\begin{align}
\asymm_n & = 3 a_2\left(n,1\right)
\;.
\end{align}
The coefficients
$\oP{j}{\ell}{m}$
%in equation (\ref{eq:Pjlm})
are first introduced by
\citet{Ritzwoller:1991aa}
to analyse frequency splittings
in the spectrum of solar oscillations,
while they are also used in asteroseismic analyses
\citep[e.g.][]{Benomar:2023aa}.
Here, we adopt the normlization
introduced by \citet{Schou:1994aa}.
It is worth noting the following relation:
\begin{equation}
U_{\ell,\,k} \oP{2k}{\ell}{m}
=
-\sqrt{\frac{4k+1}{4\pi}}
\int_{4\pi}
Y_{2k}^{0} \left|Y_{\ell}^{m}\right|^2
\;\dOmega
\;.
\label{eq:UPQ_replation}
\end{equation}
%in which $Y_{\ell}^{m}$ is generally
%a spherical harmonic function of degree $\ell$ and order %$m$.
The expressions for
$\Dlm$, $\Flm$ and $\Glm$
are given for
$\ell=1$ and $2$ as follows:
\begin{align}
 D_{1}^{m} & 
= \frac{1}{2\pi} + \frac{1}{8\pi} \,\oP{2}{1}{m} \left(3\cos^2\theta - 1\right)
\;,
\label{eq:D1m}
\\
 F_{1}^{m} & = 
\frac{3}{8\pi}\,
\oP{2}{1}{m}
\left(
3 - 7 \cos^2\theta
\right)
\;,
\\
 G_{1}^{m} & =
\frac{3}{8\pi}\,
\oP{2}{1}{m}
\left(
1 - 5 \cos^2\theta
\right)
\;,
\label{eq:G1m}
\\
 D_{2}^{m} & = 
\frac{3}{2\pi}
- \frac{15}{28 \pi}\, \oP{2}{2}{m} P_2\left(\cos\theta\right)
- \frac{3}{14 \pi}\, \oP{4}{2}{m} P_4\left(\cos\theta\right)
\;,
\end{align}
\begin{align}
\frac{F_{2}^{m}+G_{2}^{m}}{2} & 
=
\frac{3}{\pi}
- \frac{15}{14\pi}\,\oP{2}{2}{m} P_2\left(\cos\theta\right)
\nonumber\\&\phantom{=}\mbox{}
+ \frac{93}{28\pi}\,\oP{4}{2}{m} P_4\left(\cos\theta\right)
\\
\noalign{\noindent\mbox{and}}
\frac{F_{2}^{m}-G_{2}^{m}}{2} & 
=
- \frac{15}{28\pi}\,\oP{2}{2}{m} P_{2}^{2}\left(\cos\theta\right)
\nonumber\\&\phantom{=}\mbox{}
- 
\frac{17}{112\pi}\,\oP{4}{2}{m} P_{4}^{2}\left(\cos\theta\right)
\;,
\end{align}
in which
$\oP{j}{\ell}{m}$ coefficients are provided by
\begin{align}
 \oP{2}{1}{m} & = 3 m^2 - 2
\\
 \oP{2}{2}{m} & = m^2 - 2
\\
\noalign{\noindent\mbox{and}}
 \oP{4}{2}{m} & = \frac{35 m^4 - 155 m^2 + 72}{6}
\;.
\end{align}
Note that $\Dlm$, $\Flm$ and $\Glm$
depend on $m$ only through 
$\oP{2k}{\ell}{m}$,
which include only even powers of $m$.
The independence of the sign of $m$ comes from the symmetric property of the operator $\mathcal{B}$
(see equation (\ref{eq:Bexp})).

\subsubsection{Radial integrals}

In order to evaluate $\xihn$ in equation
(\ref{eq:Bangint}),
we use its asymptotic expression,
\begin{align}
 \xihn
&
\approx
A
\left(
\frac{N}{\sigma_0^3 \rho r^3}
\right)^{1/2}
\sin\left(
\int^r_{\rin} \kr\;\dr - 
%\frac{\pi}{4}
\varphi_{\mathrm{in}}
\right)
\;,
\label{eq:xih}
\end{align}
in which $A$ is a normalisation constant, 
$\rin$ the inner turning point of the gravity-mode cavity,
$\varphi_{\mathrm{in}}$
the phase lag introduced at $r=\rin$
and
$\kr$ defined by
\begin{align}
\kr
& \equiv
\frac{\sqrt{\ell\left(\ell+1\right)}N}{\sigma_0 r}
\end{align}
\citep[e.g.][]{UOASS1989}.
We accordingly obtain
\begin{align}
\tdr{\xihn}
&
\approx
A
\left(
\frac{\ell\left(\ell+1\right) N^3}{\sigma_0^5 \rho r^5}
\right)^{1/2}
\cos\left(
\int^r_{\rin} \kr\;\dr - 
%\frac{\pi}{4}
\varphi_{\mathrm{in}}
\right)
\;.
\end{align}
When we evaluate the radial integrals
that include
$\xihn^2$ or $(\rd\xihn/\rd r)^2$,
the highly oscillatory factors
proportional to
$\sin^2$ and $\cos^2$
can be replaced with $1/2$,
and the domain of integrals can be set to $\GB$.
We may thus calculate
equation
(\ref{eq:B_dom})
as
\begin{align}
\mathcal{B}\left(\bxi_0, \bxi_0\right)
&
\approx
\frac{A^2\ell\left(\ell+1\right)}{8\pi\sigma_0^5}
\int_{G_B}
\frac{N^3}{\rho r^3}
\int_{4\pi}
\Br^2 \Dlm \;\dOmega\,
\dr
\nonumber\\&\phantom{=}\mbox{}
+
\frac{A^2}{8\pi \sigma_0^3}
\int_{G_B}
\frac{N}{\rho r^3}
\int_{4\pi}
\left(
\Btheta^2 \Flm
+
\Bphi^2 \Glm
\right)
\dOmega
\,
\dr
\;.
\label{eq:Beval}
\end{align}
Similarly,
we may compute
$\mathcal{I}\left(\bxi_0^*,\bxi_0\right)$
in equation (\ref{eq:calI})
as
\begin{align}
\mathcal{I}\left(\bxi_0,\bxi_0\right)
& =
\int
\left[
\xirn^2 + \ell\left(\ell+1\right)\xihn^2
\right]
\rho r^2\;\dr
\approx
\ell\left(\ell+1\right)
\int
\rho r^2 \xihn^2
\;\dr
\nonumber\\
&
\approx
\frac{A^2 \ell\left(\ell+1\right)}{2\sigma_0^3}
\int_{\G}
\frac{N}{r}
\;\dr
\;.
\label{eq:Ieval}
\end{align}
Using equations (\ref{eq:Beval}) and (\ref{eq:Ieval}) in equation (\ref{eq:sigma1_red}),
we obtain 
the expression for
the magnetic perturbation
to cyclic frequency
as
\begin{align}
\nu_1
&=
\left(
\int_{\G}
\frac{N}{r}
\;\dr
\right)^{-1}
\Biggl[
\frac{1}{128 \pi^5 \nu_0^3}
\int_{G_B}
\frac{N^3}{\rho r^3}
\int_{4\pi}
\Br^2 \Dlm \;\dOmega\,
\dr
\nonumber\\&\phantom{=}\mbox{}
+
\frac{1}{32\pi^3 \ell\left(\ell+1\right)\nu_0}
\int_{G_B}
\frac{N}{\rho r^3}
\int_{4\pi}
\left(
\Btheta^2 \Flm
+
\Bphi^2 \Glm
\right)
\dOmega
\,
\dr
\Biggr]
\;.
\label{eq:nu1}
\end{align}

\subsection{Asymmetry of the frequency splittings of dipolar modes}

We derive the expression for
the asymmetry of the frequency splittings
for $\ell = 1$.
Using equations (\ref{eq:D1m})--(\ref{eq:G1m}),
we first obtain
\begin{align}
 D_1^1 - D_1^0 &
=
\frac{3}{8\pi} \left( 3 \cos^2\theta - 1 \right)
\equiv
\frac{3}{4\pi}\, \Wrr\left(\cos\theta\right)
\;,
\\
 F_1^1 - F_1^0 & 
= -\frac{9}{8\pi}\left( 7 \cos^2\theta - 3 \right)
\equiv
-\frac{9}{2\pi}\, W_{\theta}\left(\cos\theta\right)
\\
\noalign{\noindent\mbox{and}}
 G_1^1 - G_1^0 & 
= -\frac{9}{8\pi}\left( 5 \cos^2\theta - 1 \right)
\equiv
-\frac{9}{2\pi}\, W_{\phi}\left(\cos\theta\right)
\;,
\label{eq:G10}
\end{align}
where $W_{\alpha}\left(\cos\theta\right)$
with $\alpha=\rmr,\,\theta$ and $\phi$
are normalised so that
they take their maxima of
$W_{\alpha}=1$ at $\theta=0$.
Noting that
the asymmetry is equal to the difference
in frequency perturbation
between $m=1$ and $m=0$
with the same radial order $n$
and the spherical degree $\ell=1$,
we obtain
its expression
from equations
(\ref{eq:nu1})--(\ref{eq:G10})
as
\begin{align}
\asymm_n^{\text{(mag)}}
&=
\left(
\int_{\G}
\frac{N}{r}
\;\dr
\right)^{-1}
\biggl(
\frac{3}{128 \pi^5 \nu_{n,1}^3}
\int_{G_B}
\frac{N^3}{\rho r^3}
\,
\overline{\Wrr \Br^2}
\;
\dr
\nonumber\\&\phantom{=}\mbox{}
-
\frac{9}{32\pi^3 \nu_{n,1}}
\int_{G_B}
\frac{N}{\rho r^3}
\,
\overline{
W_{\theta} \Btheta^2
+
W_{\phi} \Bphi^2
}
\;
\dr
\biggr)
\;,
\label{eq:tmag_g}
\end{align}
in which
overlines
mean the spherical averages
(see equation (\ref{eq:sph_av})).
Equation (\ref{eq:tmag_g}) is
equivalent to equation (\ref{eq:t_n_mag}),
if we assume $|\Btheta| \ll |\Bphi|$.

%\subsection{Magnetic glitches}

\subsection{Validity condition of the perturbation analysis}
\label{sec:validity}

The perturbation analysis 
presented in this paper
can be justified
only if the magnetic effect is a small
perturbation.
This condition can be rephrased as
the perturbed Lorentz force being much smaller than
the total restoring force that exists in the absence of the magnetic field.
For high-order gravity modes,
the dominant force is the buoyancy force,
which is always in the radial direction,
whereas
only the Eulerien perturbation to the pressure gradient
($\nabla p'$)
contributes to the horizontal component of the total force.
We first derive the condition of
the horizontal component
of the Lorentz force $\boldsymbol{L}'$
being much smaller than
that of $\nabla p'$.
We then discuss the condition about the radial components.

The horizontal component of $\nabla p'$
can be estimated 
from the perturbed equation of motion
to be
\begin{equation}
\left(
\nabla p'
\right)_{\perp}
=
\frac{p'}{r} \nabYlm
\approx
\sigma^2 \rho \xihn \nabYlm
\;,
\label{eq:dp_h}
\end{equation}
in the Cowling approximation.
On the other hand,
the dominant horizontal component of
the Lorentz force is given 
for high-order gravity modes by
\begin{equation}
\boldsymbol{L}'_{\perp}
\approx
-\frac{1}{4\pi} \Br^2 \kr^2 \xihn \nabYlm
\label{eq:mag_h}
\end{equation}
(see equation (\ref{eq:B_dom})).
Here,
we do not assume equation (\ref{eq:B_condition}),
but consider that all of the components of $\bB$ 
are on the same order,
which does not seriously influence
the order-of-magnitude estimate 
in this section.

From equations (\ref{eq:dp_h}) and (\ref{eq:mag_h}),
we may introduce
a dimensionless parameter to measure
the importance of the Lorentz force by
\begin{equation}
s_{\mathrm{m}}
\equiv
\left(\frac{\Br}{\Brup}\right)^2
\;,
\label{eq:Br_cond}
\end{equation}
in which
$\Brup$ is defined by
\begin{equation}
\Brup
\equiv
\sqrt{\frac{4\pi\rho}{\ell\left(\ell+1\right)}}
\frac{\sigma^2 r}{N}
\;.
\label{eq:Brup}
\end{equation}
If we do not take the rotation into account,
the condition for the magnetic field to be weak enough
to apply the perturbation analysis
is given simply by
$s_{\mathrm{m}} \ll 1$, or equivalently
$\left|\Br\right| \ll \Brup$.
The upper limit given by
equation (\ref{eq:Brup})
is essentially the same as that derived 
by
\citet{Bugnet:2021aa}
in their equation (29)
and 
larger by factor two
than the critical field strength
in equation (3)
of \citet{Fuller:2015aa}
for $\ell=1$.

We now turn to the radial components.
Since $\boldsymbol{L}'$ is perpendicular to $\bB$,
the radial component of $\boldsymbol{L}'$
is different in amplitude from its horizontal component 
by at most factor 
\begin{equation}
f_L =\left|\frac{B_{\perp}}{\Br}\right|
\;,
\end{equation}
in which $B_{\perp}$ means the
amplitude of the horizontal component of $\bB$.
On the other hand, 
the buoyancy force $\rho N^2 \xirn$ is larger than
$\left|\left(\nabla p'\right)_{\perp}\right|$
by factor 
\begin{equation}
f_N =
\frac{N}{\sigma}
\end{equation}
for high-order gravity modes.
Therefore,
in the situations where
$f_L \lesssim f_N$,
the buoyancy force dominates over
the radial component of $\boldsymbol{L}'$
if equation (\ref{eq:Br_cond}) is satisfied.
In fact,
from Table~\ref{tab:fitting} of this paper
and Fig.~12 of \citetalias{Saio:2015aa},
we estimate
$\left(f_L,\,f_N\right) \sim 
\left(26,\,10\mbox{--}100\right)$
for $r/R \lesssim 0.5$.

\section{Aspherical buoyancy glitches}
\label{sec:aspherical_glitches}

\subsection{Background}

Physical processes in the stellar interior often
create a layer of rapid variation in chemical composition and other physical quantities
typically near the boundary of mixing regions.
If the scale height of the variation
is much shorter than the wavelength of waves
that go through the layer,
the structure is essentially considered as
a discontinuous surface,
which
generally disturbs the wave propagation,
and hence modifies the frequencies of modes
that consist of those waves.
This is called a glitch problem
in asteroseismology.
The problem is particularly important
for high-order modes
because the glitch
induces characteristic signatures in
the spectrum of mode frequencies (or periods)
depending on its structure.
Such signatures, which are frequently observed in real stars,
provide us with a unique probe into
the properties of
the physical processes that cause the glitch.

In this appendix,
we confine ourselves to
the glitch problem of
high-order gravity modes,
for which a considerable amount of
effort has already been made.
The present analysis extends it
to take account of the glitch structure that is not spherically symmetric.
For simplicity,
we assume that the discontinuity is so weak
that we may apply the variational principle
(or the perturbation theory
in the non-degenerate case)
to estimate the change in the mode frequencies (or periods).
This also implies another assumption that
the degeneracy among the modes
with the same radial order and the same spherical
degree, but with different azimuthal orders,
has been lifted by the effect of the rotation
before we consider a smaller effect of the glitch
(see appendix~\ref{sec:guidelines}).

\subsection{Framework for high-order gravity modes
with uniform rotation}

Under the assumptions that we have made,
our task is simply to evaluate
equation (\ref{eq:sigma1})
without taking the magnetic field into account.
In this case,
it is convenient to eliminate $\Psi$ in equation (\ref{eq:Texp}) 
using 
\begin{align}
\nabla p + \rho\nabla\Psi 
= 0
\label{eq:eq_cond_noB}
\end{align}
(see equation (\ref{eq:eq_cond}))
to obtain
\begin{align}
\mathcal{T}
\left(\hbxi,
\bxi\right)
& =
\mathcal{S}_T
\left(\hbxi,
\bxi\right)
+
\mathcal{A}_T
\left(\hbxi,
\bxi\right)
+
\left[
\mathcal{A}_T
\left(\bxi,\hbxi\right)
\right]^*
\;,
\end{align}
in which the symmetric part $\mathcal{S}_T$
and the asymmetric part $\mathcal{A}_T$ are defined by
\begin{align}
\mathcal{S}_T
\left(\hbxi,
\bxi\right)
& \equiv
\int 
\frac{1}{\Gamma_1 p}\,
p'\left[\hbxi^*\right]
p'\left[\bxia\right]
\;
\dV
\nonumber\\&\phantom{=}\mbox{}
- G
\iint
\left[
\rho\left(\boldsymbol{r}_{\ra}\right)
\hbxi^*\left(\boldsymbol{r}_{\ra}\right)
\cdot
\nabla_{\ra}
\right]
\left[
\rho\left(\boldsymbol{r}_{\rb}\right)
\bxia\left(\boldsymbol{r}_{\rb}\right)
\cdot
\nabla_{\rb}
\right]
\nonumber\\&\phantom{=
- G
\iint
}\mbox{}
\times
\left(
\frac{1}{\left|
\boldsymbol{r}_{\ra}
-
\boldsymbol{r}_{\rb}
\right|}
\right)\;
\dV_{\ra}\,
\dV_{\rb}
\;,
\label{eq:ST_exp}
\end{align}
and
\begin{align}
\mathcal{A}_T
\left(\hbxi,
\bxi\right)
& =
-\frac{1}{2}
\int
\left(
\hbxi^*\cdot\nabla p
\right)
\bxi\cdot\left(
\frac{\nabla p}{\Gamma_1 p}
- \frac{\nabla\rho}{\rho}
\right)
\;\dV
\;.
\end{align}
In equation (\ref{eq:ST_exp}),
$p'$ means the Eulerian perturbation to the pressure,
\begin{align}
p'\left[\bxi\right]
&
=
-\Gamma_1 \nabla\cdot\bxi
- \bxi\cdot\nabla p
\;.
\end{align}
For high-order gravity modes,
we may neglect $\mathcal{S}_T$
because
the first term corresponds to
the potential energy of the acoustic oscillations,
and the second term depends on
the perturbation to the gravitational potential
\citep{Cowling1941}.
In order to evaluate $\mathcal{A}_T$,
we note that
$p$ and $\rho$ are functions of only $\Psi$
in uniformly rotating stars
(see equation (\ref{eq:eq_cond_noB})).
We therefore find
\begin{align}
\mathcal{T}\left(\bxi,\bxi\right)
&
\approx
\int
\rho
\left|
\bxi\cdot\bepsi
\right|^2
\ABV^2
\;\dV
\;,
\end{align}
in which $\bepsi$ and $\ABV^2$ are defined by
\begin{align}
\bepsi & 
\equiv \frac{\nabla\Psi}{\left|\nabla\Psi\right|}
\end{align}
and
\begin{align}
\ABV^2
& \equiv
\frac{\left|\nabla\Psi\right|^2}{\rho}
\left(
-\tdpsi{p}
\right)
\left(
\frac{1}{\Gamma_1 p}\tdpsi{p}
- \frac{1}{\rho}\tdpsi{\rho}
\right)
\nonumber\\
& \approx
\frac{1}{\rho}
\left(
-\pdr{p}
\right)
\left(
\frac{1}{\Gamma_1 p}\pdr{p}
- \frac{1}{\rho}\pdr{\rho}
\right)
\;.
\label{eq:calN2}
\end{align}
It is obvious that $\ABV^2$ reduces to
the (squared) \BV\ frequency $N^2$
in spherically symmetric structures.
The approximate equality in equation (\ref{eq:calN2})
is because $\nabla\Psi$ is almost in the radial direction.

\subsection{General formula of the glitch signature}

In order to describe the analysis precisely,
we consider three different equilibrium structures
that are very close to each other.
The first (structure 0) is the spherically symmetric one
without rotation and any glitches.
The second (structure r) has a slow uniform rotation, but
without any glitches.
The third (structure g) rotates at the same rate as the second and has a glitch.
We distinguish the variables of these structures
by subscripts 0, r and g for
the first, second and third structures, respectively.
We derive the frequency difference between
structures g and r in the following way:
we separately compute
the difference between structures r and 0
and that between structures g and 0,
and then take the difference between the two differences.
Although structures {g and r} are both deformed by the centrifugal force,
its effect on the frequencies cancels out in the leading order when their differences between the two structures are computed.
We thus obtain
\begin{align}
\mathcal{T}_1\left(\bxi_0,\bxi_0\right)
&
\approx
\int
\rho_0
N_0^2
\xirn^2
\left|\Ylm\right|^2
\frac{\Delta\ABV^2}{N_0^2}
\;\dV
\;,
\label{eq:T1_glitch}
\end{align}
in which $\Delta\ABV^2$ is
the difference at the same position,
defined by
\begin{align}
\Delta\ABV^2 
& =
\ABV^2_{\mathrm{g}}
-
\ABV^2_{\rmr}
\;.
\end{align}
In equation (\ref{eq:T1_glitch}),
the difference in density
is neglected because
it has a much smaller impact on the induced
glitch signature in the frequency (or period) spectrum than that in $\ABV^2$,
which depends on the derivative of the density.
In addition,
we choose to neglect the difference
in $\bepsi$ between structures r and g
by assuming that $\Psi$ of structure g ($\Psi_{\mathrm{g}}$)
is a function of only $\Psi_{\rmr}$,
which implies that $\nabla\Psi_{\mathrm{g}}$ is in the same direction as $\nabla\Psi_{\rmr}$ at every point.
After adopting this assumption,
we further approximate $\bepsi$ by $\ber$
in equation (\ref{eq:T1_glitch}).

Using equation
(\ref{eq:xih})
and the corresponding expression
for $\xirn$,
\begin{align}
 \xirn
&
\approx
-A
\left(
\frac{\ell\left(\ell+1\right)}{\sigma_0 N_0 \rho_0 r^3}
\right)^{1/2}
\cos\left(
\int^r_{\rin} \kr\;\dr - 
%\frac{\pi}{4}
\varphi_{\mathrm{in}}
\right)
\;,
\end{align}
we obtain from equations (\ref{eq:sigma1}) and (\ref{eq:T1_glitch})
the oscillatory component of
the difference in the mode period as
\begin{align}
\Delta P\left(n,\ell,m\right)
&
=
\sum_{k=0}^{\ell}
\hat{a}^{\mathrm{g}}_{2k}\left(n,\ell\right)
\oP{2k}{\ell}{m}
\;,
\label{eq:dP_glitch}
\end{align}
in which 
the period a-coefficients $\hat{a}^{\mathrm{g}}_{2k}$ are defined by
\begin{align}
\hat{a}^{\mathrm{g}}_{2k}\left(n,\ell\right)
&
\equiv
\frac{
U_{\ell,\,k} \Pi_{\ell}
}{\sqrt{\left(4k+1\right)\pi}}\,
\int_{\mathrm{G}}
\frac{\Delta\ABV_{2k,0}^2}{N_0^2}
\cos\left(2\mathcal{K}\right)
\;\dK
\;.
\label{eq:ap_g}
\end{align}
Here, we have introduced the following definitions:
\begin{align}
\Pi_{\ell} & \equiv
\frac{2\pi^2}{\sqrt{\ell\left(\ell+1\right)}}
\left(
\int_{\mathrm{G}} \frac{N_0}{r}\;\dr
\right)^{-1}
\;,
\label{eq:Piell_def}
\end{align}
\begin{align}
\Delta\ABV_{2k,0}^2(r)
&
\equiv
\int_{4\pi}
Y_{2k}^{0}
\Delta\ABV^2
\;\dOmega
\end{align}
and
\begin{align}
\mathcal{K} & \equiv
\frac{\sqrt{\ell\left(\ell+1\right)}}{\sigma_0}
\int^r_{\rin} 
\frac{N_0}{r}
\;\dr - \varphi_{\mathrm{in}}
\;.
\label{eq:Kphase}
\end{align}
In equation (\ref{eq:ap_g}),
we neglect the contribution of the components that depend on the mode period only smoothly,
because it should be ascribed to the period difference between structures r and 0.
We retain this treatment throughout this section.

%(see equation (\ref{eq:UPQ_replation})),

\subsection{Three types of discontinuity}
\label{sec:three_types}

Equations (\ref{eq:dP_glitch}) and (\ref{eq:ap_g}) provide the fundamental formulae that describe the glitch signature in the period spectrum of high-order gravity modes.
The signature depends on the type of discontinuity in $\Delta\ABV^2$.
In the following,
we consider three different types, which are 
located at $r = r_*$
inside the gravity-mode cavity
\citep[see][]{Miglio:2008aa}.
The corresponding expressions
for $\hat{a}^{\mathrm{g}}_{2k}$ are derived
as functions of the mode period $P_0$
in each case.

If 
the density itself is discontinuous,
$\Delta\ABV^2$ follows a
Dirac delta function ($\delta$)
as
\begin{align}
\frac{\Delta\ABV_{2k,0}^2}{N_0^2}
&
=
\mathfrak{D}_{k}^{(0)} r_*
\delta\left( r - r_* \right)
\;,
\end{align}
in which $\mathfrak{D}_{k}^{(0)}$ is a
dimensionless constant.
Then,
the $\hat{a}^{\mathrm{g}}_{2k}$ is given by
\begin{align}
\hat{a}^{\mathrm{g}}_{2k}\left(n,\ell\right)
&
=
\mathfrak{E}^{(0)}_{\ell,k}
P_0\left(n,\ell\right)
\cos\left[
\frac{2 \pi P_0\left(n,\ell\right)}{\Pi_{\ell,*}}
- 2 \varphi_{\mathrm{in}}
\right]
\;,
\label{eq:a2k_D}
\end{align}
where $\mathfrak{E}^{(0)}_{\ell,k}$ and $\Pi_{\ell,*}$ are defined by
\begin{align}
\mathfrak{E}^{(0)}_{\ell,k}
&
\equiv
\sqrt{\frac{
\ell\left(\ell+1\right)
}{\left(4k+1\right)\pi}}
\frac{
U_{\ell,\,k} \Pi_{\ell} N_0\left(r_*\right)}{2\pi}
\,\mathfrak{D}_{k}^{(0)}
\end{align}
and
\begin{align}
\Pi_{\ell,*} & \equiv
\frac{2\pi^2}{\sqrt{\ell\left(\ell+1\right)}}
\left(
\int_{\rin}^{r_*} \frac{N_0}{r}\;\dr
\right)^{-1}
\;,
\end{align}
respectively.

If
the derivative of the density is discontinuous,
$\Delta\ABV^2$ is described by a
Heaviside step function ($H$)
with a dimensionless constant $\mathfrak{D}_{k}^{(1)}$
as
\begin{align}
\frac{\Delta\ABV_{2k,0}^2}{N_0^2}
&
=
\mathfrak{D}_{k}^{(1)}
H\left( r_* - r \right)
\;.
\label{eq:dN2k_step}
\end{align}
The corresponding $\hat{a}^{\mathrm{g}}_{2k}$
is provided by
\begin{align}
\hat{a}^{\mathrm{g}}_{2k}\left(n,\ell\right)
&
=
\mathfrak{E}_{\ell,k}^{(1)}
\sin\left[
\frac{2 \pi P_0\left(n,\ell\right)}{\Pi_{\ell,*}}
- 2 \varphi_{\mathrm{in}}
\right]
\;,
\label{eq:a2k_H}
\end{align}
in which $\mathfrak{E}_{\ell,k}^{(1)}$ is defined by
\begin{align}
\mathfrak{E}_{\ell,k}^{(1)}
&
=
\frac{
U_{\ell,\,k} \Pi_{\ell}
}{2\sqrt{\left(4k+1\right)\pi}}\,
\mathfrak{D}_{k}^{(1)}
\;.
\end{align}

In case where 
the second derivative of the density
is discontinuous,
$\Delta\ABV^2$ has a form
of the ramp function with a dimensionless constant $\mathfrak{D}_{k}^{(2)}$
as
\begin{align}
\frac{\Delta\ABV_{2k,0}^2}{N_0^2}
&
=
\begin{cases}
\mathfrak{D}_{k}^{(2)}
\left( 1 - \frac{r}{r_*} \right) & 
\mbox{for } r \le r_*
\;,
\\
0 & 
\mbox{for } r \ge r_*
\;.
\end{cases}
\end{align}
We then obtain
\begin{align}
\hat{a}^{\mathrm{g}}_{2k}\left(n,\ell\right)
&
=
\mathfrak{E}_{\ell,k}^{(2)}
P_0^{-1}\left(n,\ell\right)
\cos\left[
\frac{2 \pi P_0\left(n,\ell\right)}{\Pi_{\ell,*}}
- 2 \varphi_{\mathrm{in}}
\right]
\;,
\label{eq:a2k_r}
\end{align}
in which $\mathfrak{E}_{\ell,k}^{(2)}$ is given by
\begin{align}
\mathfrak{E}_{\ell,k}^{(2)}
&
\equiv
-
\sqrt{\frac{\pi}{\ell\left(\ell+1\right)\left(4k+1\right)}}
\frac{ U_{\ell,\,k} \Pi_{\ell}}{2 N_0(r_*)}\,
\mathfrak{D}_{k}^{(2)}
\;.
\end{align}

Equations (\ref{eq:a2k_D}), (\ref{eq:a2k_H}) and (\ref{eq:a2k_r}) demonstrate that,
for the three types of discontinuity,
the $\hat{a}^{\mathrm{g}}_{2k}$ coefficients
depend on
the sinusoidal functions of $P_0$
with the same period $\Pi_{\ell,*}$,
but that
their amplitude follows
a linear or constant or reciprocal function
of $P_0$
for the Dirac or Heaviside or ramp type, respectively.

%\section{Search for other cases}
%\label{sec:other_stars}

%%%%%%%%%%%%%%%%%%%%%%%%%%%%%%%%%%%%%%%%%%%%%%%%%%

% Don't change these lines
\bsp	% typesetting comment
\label{lastpage}
\end{document}